\documentclass[onecolumn]{aastex63}
\usepackage{lineno}
\received{XXX}
\revised{XXX}
\accepted{XXX}
\submitjournal{ApJ}

\shorttitle{Nucleosynthesis in High Magnetic Fields}
\shortauthors{Famiano et al.}

\begin{document}
%\linenumbers
\title{Nuclear Reaction Screening, Weak Interactions, and r-Process Nucleosynthesis in High Magnetic Fields}

\correspondingauthor{Michael Famiano}
\email{michael.famiano@wmich.edu}

\author[0000-0003-2305-9091]{Michael Famiano}
\affiliation{%
	Physics Department, Western Michigan University, Kalamazoo, MI 49008-5252 USA
}
\affiliation{National Astronomical Observatory of Japan, Mitaka, Tokyo 181-8588 Japan}%Lines break automatically or can be forced with \\
\author[0000-0002-2999-0111]{A. Baha Balantekin}%
\email{baha@physics.wisc.edu}
\affiliation{National Astronomical Observatory of Japan, Mitaka, Tokyo 181-8588 Japan}
\affiliation{Department of Physics, University of Wisconsin, Madison, WI 53706 USA
}%
\author[0000-0002-8619-359X]{T. Kajino}
\email{kajino@nao.ac.jp}
\affiliation{National Astronomical Observatory of Japan, Mitaka, Tokyo 181-8588 Japan}
\affiliation{Graduate School of Science, The University of Tokyo, 7-3-1 Hongo, Bunkyo-ku, Tokyo 113-0033, Japan}
\affiliation{School of Physics, and International Research Center for Big-Bang Cosmology and Element Genesis, Beihang University, 37 Xueyuan Rd., Haidian-district, Beijing 100083 China}
\author[0000-0003-3083-6565]{M. Kusakabe}
\email{kusakabe@buaa.edu.cn}
\affiliation{School of Physics, and International Research Center for Big-Bang Cosmology and Element Genesis, Beihang University, 37 Xueyuan Rd., Haidian-district, Beijing 100083 China}
\author[0000-0003-2595-1657]{K. Mori}
\email{kanji.mori@grad.nao.ac.jp}
\affiliation{National Astronomical Observatory of Japan, Mitaka, Tokyo 181-8588 Japan}
\affiliation{Graduate School of Science, The University of Tokyo, 7-3-1 Hongo, Bunkyo-ku, Tokyo 113-0033, Japan}

\author[0000-0002-8965-1859]{Y. Luo}
\email{ydong.luo@grad.nao.ac.jp}
\affiliation{National Astronomical Observatory of Japan, Mitaka, Tokyo 181-8588 Japan}
\affiliation{Graduate School of Science, The University of Tokyo, 7-3-1 Hongo, Bunkyo-ku, Tokyo 113-0033, Japan}

\begin{abstract}
Coulomb screening and 
weak interactions in a hot, magnetized plasma are investigated.  Coulomb screening is evaluated in a relativistic
thermal plasma in which electrons and positrons are in equilibrium.  In addition to temperature effects, effects on weak screening from
a strong external magnetic field are evaluated.  In high fields, the
electron transverse momentum components are quantized into Landau levels.  
The characteristic plasma screening length at high temperatures and
at high magnetic fields is explored.  In addition to changes to
the screening length, changes in weak interaction rates are estimated. It is found that high fields can result in increased $\beta$-decay
rates as the electron and positron spectra are dominated by Landau levels.  Finally, the effects 
studied here are evaluated in a simple r-process model.  It is found that 
relativistic Coulomb screening has a small effect on the final abundance 
distribution.  While changes in weak interaction rates in strong magnetic fields
can have an effect on the r-process evolution and abundance distribution, the field 
strength required to have a significant effect may be larger than what is currently 
thought to be typical of the r-process environment in collapsar jets or neutron star 
mergers.  If r-process sites exist in fields $\gtrsim 10^{14}$ G effects from 
fields on weak decays could be significant.
\end{abstract}


%\maketitle

\section{Introduction}
Nearly all modern nuclear astrophysics studies rely on knowledge 
of thermonuclear reaction rates between two  
or more reacting particles.  
The rate at which nuclei in a hot plasma interact is 
governed by the reaction cross section and the velocities
of the reacting nuclei in their center of mass.
In  general, the reaction rates for an environment at a 
certain temperature are taken as the average rates, which
are deduced by integrating over the reaction
cross section (as a function of energy) weighted by the 
Maxwell-Boltzmann
energy distribution of the reactants in the plasma involved, 
known as the thermonuclear reaction rate (TRR), $\left\langle\sigma v\right\rangle$ \citep{illiadis,boyd08}.  
For 
resonances in the cross section at specific energies, the
evaluation is similar, but the cross section also has
a term defining the resonance.

In a hot plasma,
the background electrons create a 
``screening'' effect between
two reacting 
charged particles
\citep{wu17,liu16,spitaleri16,yakovlev14,kravchuk14,potekhin13,quarati07,
	shaviv00,adelberger98,shalybkov87,wang83,
	www82,itoh77,jancovici77,graboske73,dewitt73,salpeter69,salpeter54}. 
Coulomb screening reduces
their Coulomb barrier because the effective
charge between two particles is reduced.  The commonly-used ``extended'' \citep{jancovici77,itoh77} screening and recent evaluations of screening
from relativistic effects have been explored \citep{famiano16,luo20}.
In evaluating the screening effect, even a small shift in the potential energy can result in significant changes in
the classical turning points of the WKB approximation, resulting in an increase in the reaction rate.  It should be noted that other positively charged nuclei in 
a plasma also 
increase the reaction rate as positive and negative charges are redistributed in the presence of a ``point-like'' nuclear potential. Though this 
adjustment to thermonuclear rates has been known for a long time \citep{salpeter54}, 
effects from relativistic, magnetized plasmas have not been fully addressed.  

Closely tied to the equilibrium abundances of electrons and positrons is pair
production, which occurs at high-enough temperatures in which the tail of the 
Fermi distribution exceeds the pair-production threshold.  Pair production has been studied in stellar cores of very massive stars \citep{kozyreva17, woosley17, spera17, takahashi18} and as a neutrino 
cooling mechanism \citep{itoh96}. 
Also, though electron capture reactions have been previously studied \citep{itoh02,liu07}, the simultaneous effects of external 
magnetic fields and relativistic pair production on reaction rate screening (fusion and
electron capture) in magnetized plasmas
have not been fully considered.
For temperatures and magnetic fields that are high enough, 
electrons and positrons can exist
in non-negligible equilibrium abundances.
In a magnetized plasma, the electron and 
positron energy distributions are  altered by the external field.

In a hot plasma, 
the
background charges include the surrounding electrons, positrons, and other nuclei.   Classically, for a non-relativistic charge-neutral medium the electrostatic potential $\phi$ of a charge $Ze$ in the
presence of a background charge density can be computed via
the Poisson-Boltzmann equation:
\begin{equation}
\label{PB}
\nabla^2\phi(r) = -4\pi Ze\delta(\mathbf{r}^3) -4\pi\sum_{z\ge-1} ze n_z\exp\left[-\frac{ze\phi(r)}{T}\right],
\end{equation}
where the last term is a sum over all charges in the medium with charge $ze$ and number density $n_z$, including non-relativistic electrons ($z=-1$).
This description is almost universally used in astrophysical calculations involving nuclear reactions. 
Here, the electron degeneracy must be calculated or estimated explicitly to
accurately determine the energy and density distribution. (Natural units are used: $k=\hbar=c=1$.)

 However, for hot, magnetized plasmas electrons and positrons must be expressed in equilibrium using Fermi-Dirac statistics.  
 The lepton number density in the presence of an external
field is modified by the
presence of Landau levels and changes from the zero-field form \citep{grasso01,kawasaki12}:
\begin{eqnarray}
\label{num_dens_terms}
    n(B=0,T) & =& \frac{1}{\pi^2}\int\limits_0^\infty
    \frac{p^2dp}{\exp\left[\frac{E-\mu}{T}\right]+1}
    -\frac{1}{\pi^2}\int\limits_0^\infty
    \frac{p^2dp}{\exp\left[\frac{E+\mu}{T}\right]+1},
    \\\nonumber
    n(B\ne 0,T) & =&
    \frac{eB}{2\pi^2}\sum\limits_{n=0}^\infty(2-\delta_{n0})
    \left[
    {
    \int
    \limits_0^\infty 
    \frac{dp_z}
    {\exp\left[\frac{\sqrt{E^2+2neB}-\mu}{T}\right]+1}
    -\int
    \limits_0^\infty 
    \frac{dp_z}
    {\exp\left[\frac{\sqrt{E^2+2neB}+\mu}{T}\right]+1}
    }
    \right].
\end{eqnarray}
In the above Equation, $E=\sqrt{p_z^2+m^2}$, where the z direction is parallel to the
magnetic field.
The term $\delta_{n0}$
accommodates the degeneracy for the higher Landau levels, and the index $n$ 
takes into account the Landau level as well as the z-component of electron 
spin.  As
$B\rightarrow 0$, the summation in the second relationship in Equation \ref{num_dens_terms} becomes an integral, and the zero-field 
number density results.

 The Poisson-Boltzmann
equation must then be replaced with the equivalent equation assuming Fermi statistics with
a magnetic field, $B$, and chemical potential, $\mu$:
\begin{eqnarray}
	\label{PF}
	\tiny
	\nabla^2\phi(r) &=& -4\pi Ze\delta^3(\mathbf{r}) -4\pi\sum_{z>0} ze n_z\exp{\left[-\frac{ze\phi_r}{T}\right]}
	\\\nonumber
	+&\frac{eB}{\pi}&\sum\limits_{n=0}^{\infty} g_n\int\limits_0^\infty{dp \left[\frac{1}{\exp(\sqrt{E^2+2neB}-\mu- e\phi_r)/T+1}-\frac{1}{\exp(\sqrt{E^2+2neB}+\mu+e\phi_r)/T+1}\right]}
	\\\nonumber
	+&4\pi&\sum\limits_{z>0}zen_z
	\\\nonumber
	-&\frac{eB}{\pi}&\sum\limits_{n=0}^{\infty} g_n\int\limits_0^\infty{dp \left[\frac{1}{\exp(\sqrt{E^2+2neB}-\mu)/T+1}-\frac{1}{\exp(\sqrt{E^2+2neB}+\mu)/T+1}\right]}
\end{eqnarray}
where the sum in the third term accounts for the quantized transverse momentum of electrons and positrons in a high magnetic field, and 
$g_n=2 - \delta_{n0}$ accounts for Landau level degeneracy. The relativistic effects come from
the high thermal energy, $T\sim m_e$, the Landau level spacing for field strengths with $\sqrt{eB}\sim m_e$, or both \citep{kawasaki12, grasso01}.  
The last two terms in Equation \ref{PF} account for
the redistribution of charge on
the uniform charge background.  
For a charge-neutral plasma, the sum of
these last two terms is zero before the charge $Ze$ is introduced.
Here, electrons are assumed to be relativistic while ions
are still treated classically; the non-relativistic
nuclei are treated with
Boltzmann statistics.

At lower temperatures and higher densities, the electron degeneracy is higher, and a first-order
solution to the Poisson equation is invalid.  The 
chemical potential must be accounted for in the relativistic treatment of Equation \ref{PF} and computed
using the electron-positron number density assuming charge neutrality. The screening is very strong, and $E_C/kT\gg 1$.  
The thermal energy is
less important, and the potential is modified by the difference in Coulomb energy
before and after the reaction -- the so-called ion sphere model \citep{clayton,salpeter69,salpeter54}.  

Perhaps ``intermediate screening,'' where $E_C/kT\sim 1$, is the most
complicated.
In this regime, screening enhancement has been computed in one of two ways.
One method is to solve the Poisson-Boltzmann equation numerically \citep{graboske73}.
In this case, numerical fits or tables might be used for astrophysical codes.
For many computational applications, an empirical interpolation between strong and
weak screening is computed \citep{www82,salpeter69}.

The ``screening enhancement factor'' (SEF) $f$, relates the screened rate
to the unscreened rate
by $\left\langle\sigma v\right\rangle_{scr}=f\left\langle\sigma v\right\rangle_{uns}$.
The value of $f$ can be deduced from the WKB approximation in the thermonuclear reaction rates as
$f=e^H$ \citep{graboske73,jancovici77,salpeter54,salpeter69,www82}, where $H$ is a unitless value derived from the
specific type of screening employed \citep{das16,kravchuk14,itoh77,alastuey78,dewitt73,quarati07}.
As mentioned above, the intermediate exponent $H_I$ is often determined using strong and weak screening values, 
$H_I = H_SH_W/\sqrt{H_S^2+H_W^2}$.   This method is used
commonly in astrophysics codes incorporating
nuclear reaction networks \citep{mesa11, mesa15, mesa18, libnucnet}.
\begin{figure}[b]
\gridline{
\fig{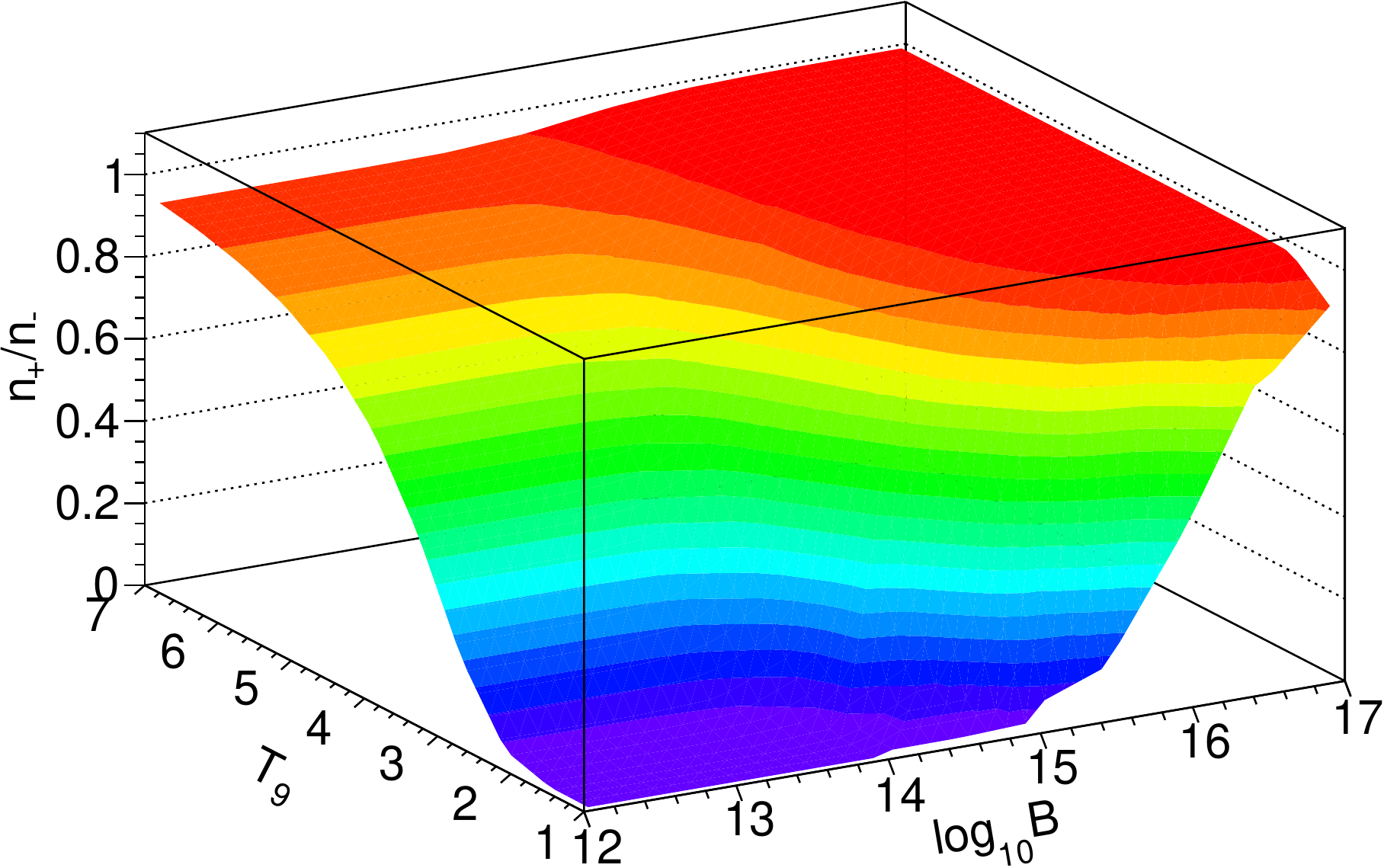}{0.49\textwidth}{(a)}
\fig{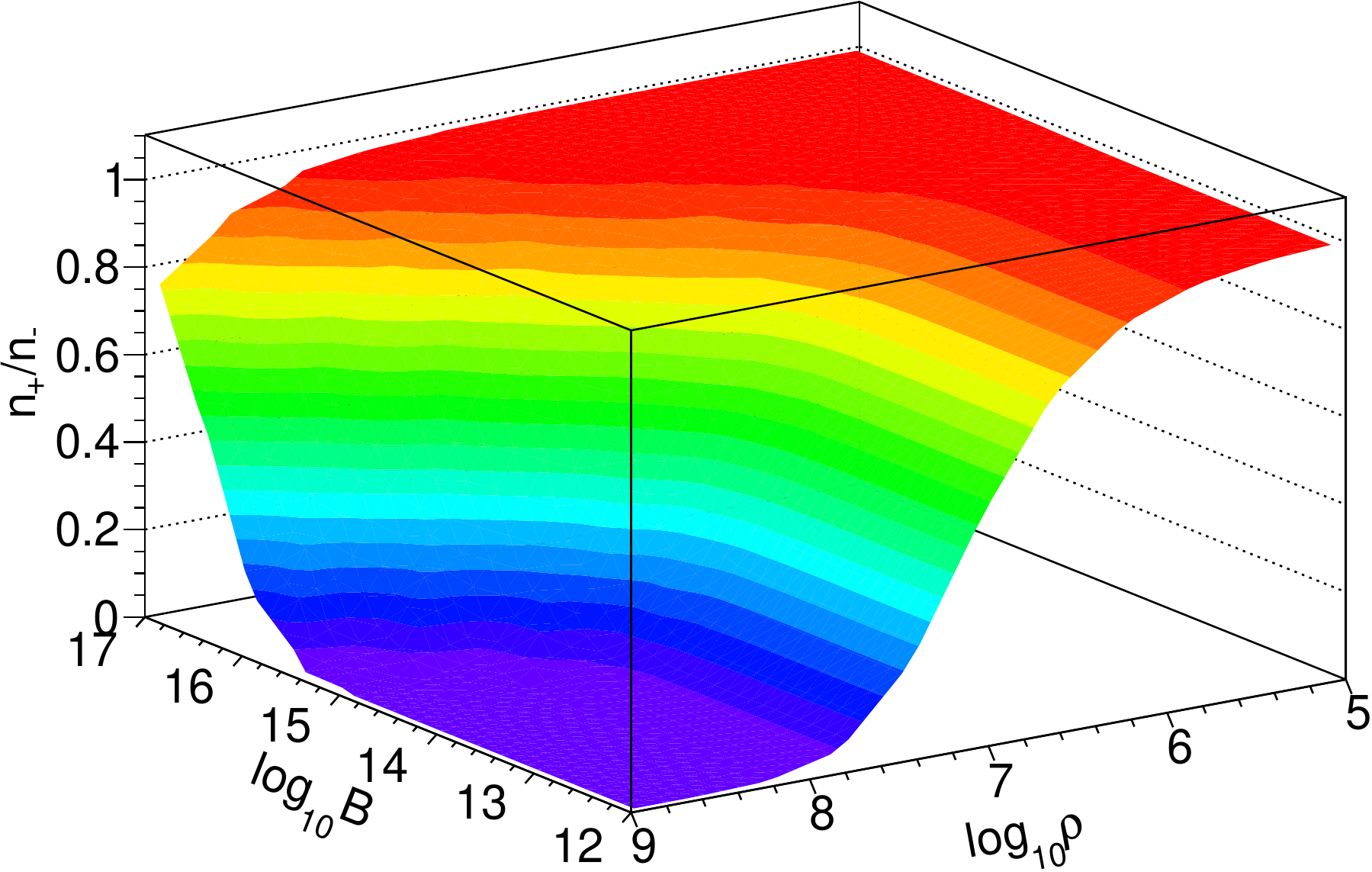}{0.49\textwidth}{(b)}
}
	\caption{\label{screen_modes}
	(a)
	    Positron-electron ratio as a function of temperature and magnetic field in a neutral plasma at $\rho Y_e = 5\times10^5$ g/cm$^3$. 
 (b) Positron-electron ratio as a function of density and magnetic field in a neutral plasma at temperature $T_9=7$ and $Y_e = 0.5$.
	    The number densities are computed up to 2000 Landau levels.}
\end{figure}

An example of the importance of including thermal and magnetic field effects 
is shown in Figure \ref{screen_modes}.
Shown in the figure is the ratio of positron to electron number density as a 
function of temperature and magnetic field (where $T_9$ is the 
temperature in billions of K) at a density and electron fraction $\rho Y_e=5\times10^{5}$ g/cm$^{3}$ taking into account the electron 
chemical potential at high density.  Relativistic effects
become increasingly important 
in this region as the positron number density becomes a significant fraction of 
the electron number density.  The increased overall number of charges of any sign contribute to the screening effect, and this will be explored in this paper.

The goal of this paper is to evaluate the effects of
reaction rate screening in relativistic electron-positron plasmas found in hot, magnetized stellar environments.  Results from this work will be 
applied to an example nucleosynthesis process in a magnetohydrodynamic (MHD)
collapsar jet.
In this paper, the effects of weak screening corrections in a magnetized,
relativistic plasma will be evaluated.  A useful approximation which can be
used effectively in computational applications is developed.  The effects
of screening in a sample astrophysical site are evaluated.  In addition,
the effects of strong magnetic fields on the weak interactions in a
plasma are explored.
\section{Weak Screening Limit}
\subsection{First Order Expansions: Debye-H\"uckel and Thomas-Fermi}
In the high temperature, low density ``weak screening'' limit, the Coulomb energy between two reacting nuclei is lower than the thermal energy, $E_C/kT\ll 1$,
as is the electron chemical potential. The electrons are mostly 
non-degenerate, and Equations \ref{PB} and \ref{PF} can be expanded to first order in 
potential, 
$\mathcal{O}(\phi)$, known as the Debye-H\"uckel approximation.
A corresponding Debye length, $\lambda_D$ can be derived, resulting in
a Yukawa-type potential, $\phi(r)\propto (e^{-r/\lambda_D})/r$ as opposed to the usual $1/r$ unscreened Coulomb relationship. For  lower
temperatures and higher densities resulting in higher electron degeneracy, 
the Thomas-Fermi screening length is more appropriate.  This is defined by the first order approximation:
\begin{equation}
\frac{1}{\lambda_{TF}^2} \equiv 4\pi e^2\frac{\partial n}{\partial\mu}.
\end{equation}
This is derived from the density of states at the Fermi surface \citep{ichimaru93}, but it is also equivalent to the first-order expansion in potential as the chemical
potential is used as a mathematical surrogate for the potential with the same
results. 
This relationship can also be deduced from the solution
of the Schwinger-Dyson equation for
the photon propagator \citep{kapusta06}. 
The contribution to the screening length from the surrounding nuclei must also be included, and this can be significant in some cases.
\begin{figure}
	\includegraphics[width=\linewidth,]{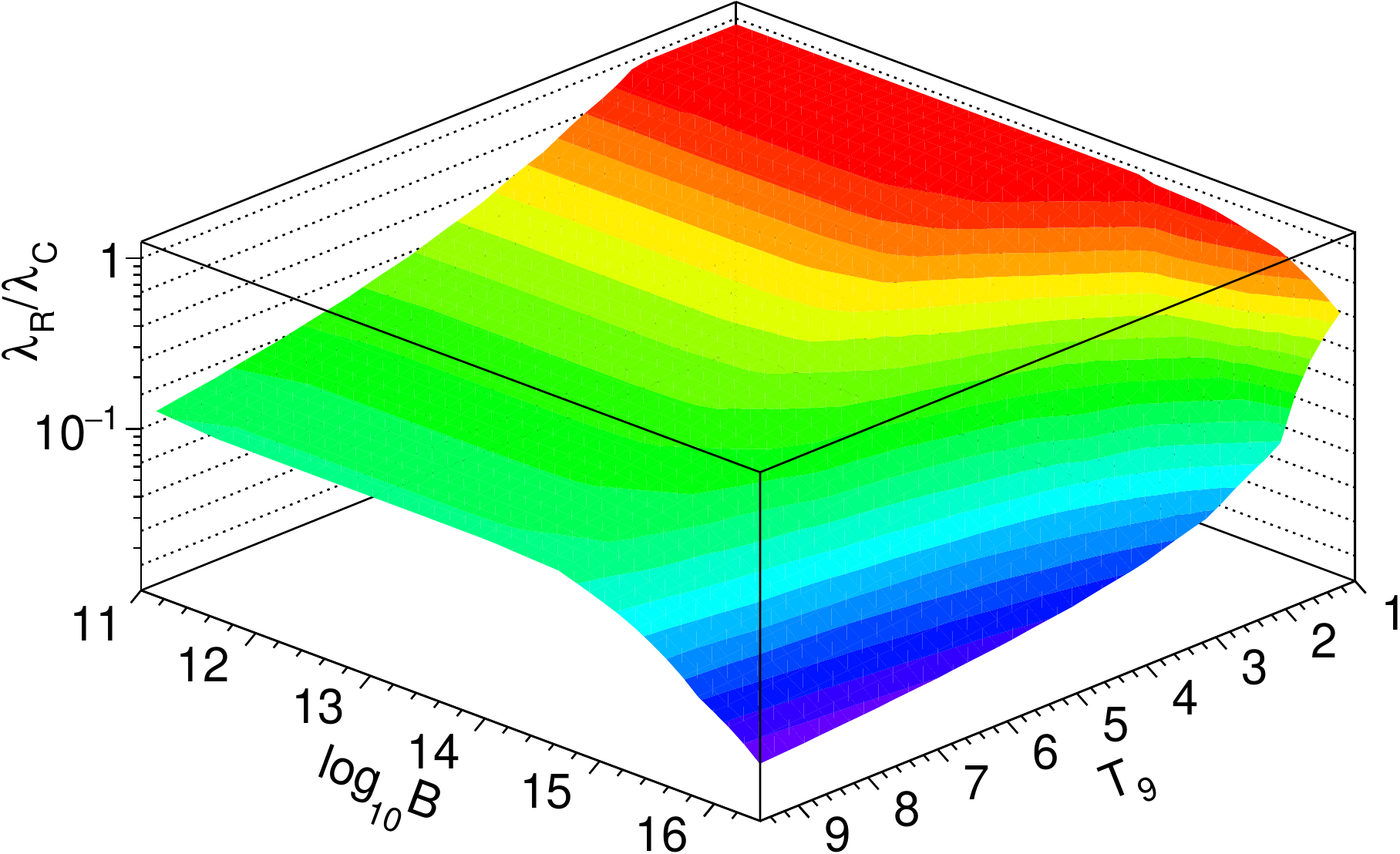}
	\caption{\label{debye_fig}The ratio of classical to relativistic electron screening lengths for a neutral plasma as a function of temperature and magnetic field (G) at a constant electron density, $\rho$Y$_e=5\times10^5$ g/cm$^3$. }
\end{figure} 

The chemical potential can be determined using Equation \ref{num_dens_terms} for a plasma of density
$\rho$, electron fraction $Y_e$, and net electron density $n_- - n_+$. For most astrophysical applications, a
static plasma is assumed with a net charge density of zero.

The ratio of the relativistic Thomas-Fermi electron-positron screening length, $\lambda_{TF}$, to the classical Debye length, $\lambda_D$, is shown in Figure \ref{debye_fig} as a function of temperature and magnetic field at $\rho Y_e=5\times10^{6}$ g/cm$^{3}$. In this figure, only the electron-positron screening length ratio is shown to emphasize the difference that high temperature and magnetic fields can induce in a plasma.  In astrophysical calculations, the screening length
from other nuclei must also be included, $1/\lambda^2 = 1/\lambda_{ion}^2+1/\lambda_{-,+}^2$.  There is a significant difference
between the classical and relativistic screening lengths at high temperature and field. Because the screened rates depend exponentially
on the screening lengths at low density/high-temperature, even small changes in the screening length can be significant.  The relativistic electron screening length can be quite small at high-enough temperature or $B$ field. 
\begin{figure}
\gridline{
\fig{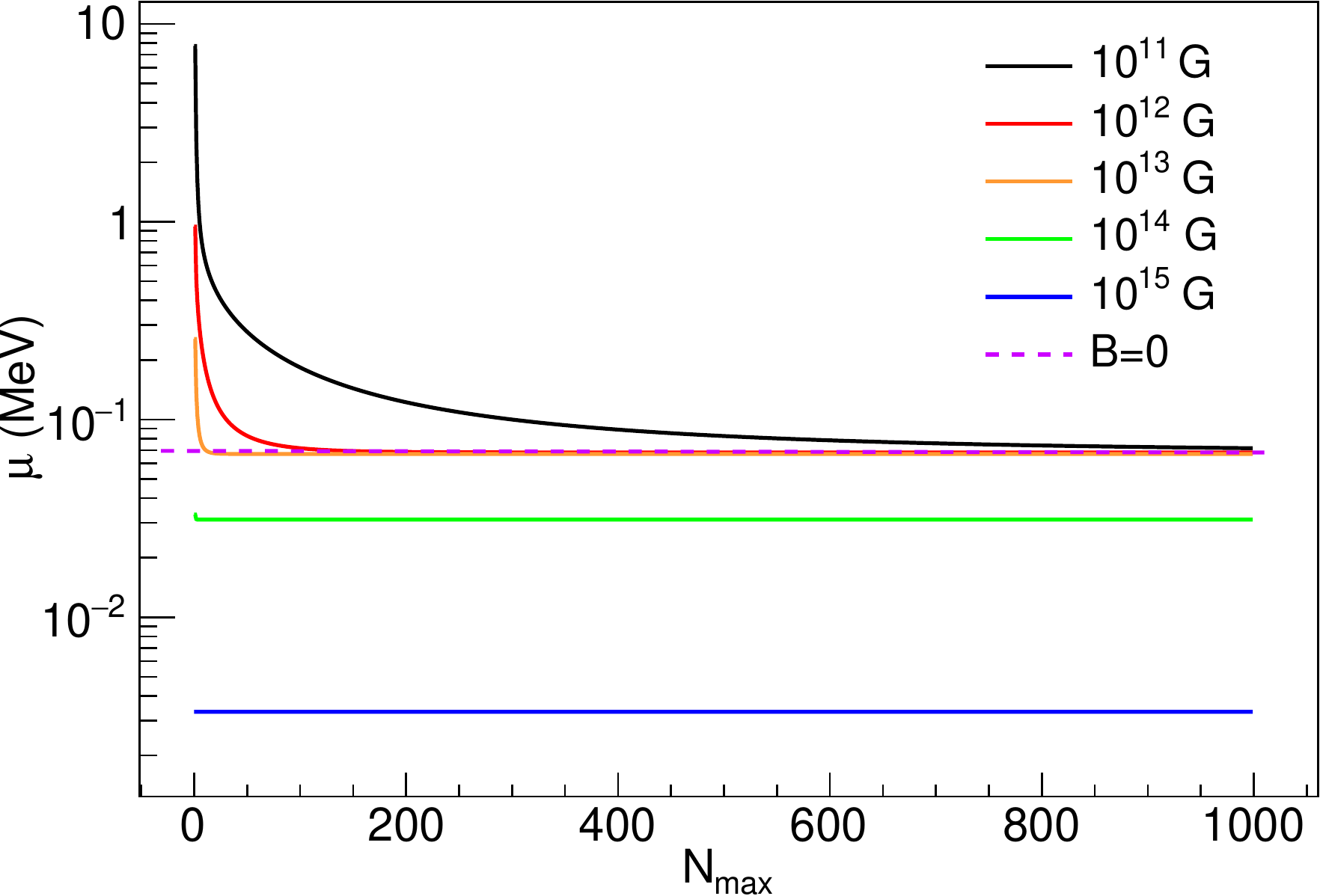}{0.49\textwidth}{(a)}
\fig{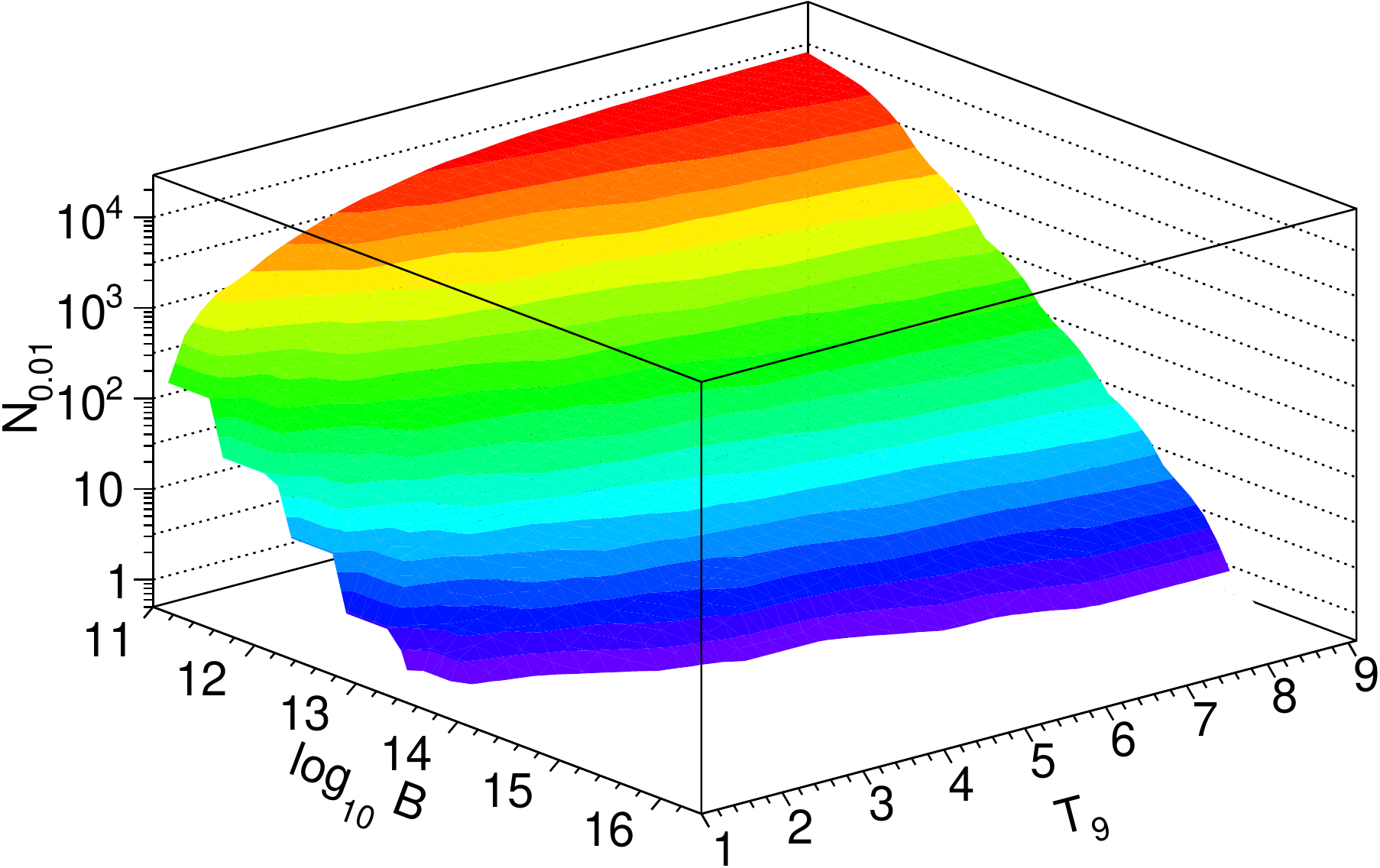}{0.49\textwidth}{(b)}
}
	\caption{\label{landau_sens} (a)
	Electron chemical potential as a function of the maximum
	number of Landau levels included in summation over Landau levels
	at $T_9=2$, $\rho$Y$_e = 5\times10^4$ g cm$^{-3}$ for various magnetic
	fields.  The dashed line is the chemical potential for an ideal 
	Fermi gas.
	(b) Number of Landau levels necessary to approach an equilibrium 
	electron number density with a maximum uncertainty of 1\%, $N_{0.01}$ at
	a $\rho$Y$_e=5\times10^4$ g cm$^{-3}$ as a function of magnetic field (G) and $T_9$.}
\end{figure}

It is noted that at higher density or lower temperature, intermediate
	screening depends more heavily on the electron chemical potential. The increased electron chemical potential 
	results in the electron-positron number densities 
	which 
	approach classical (non-relativistic) values.  That is $n_- \rightarrow \rho N_A Y_e$ and $n_+ \rightarrow 0$.
Because of this, first-order weak screening is replaced by an ion-sphere screening model or a type of geometric mean between the ion-sphere and weak screening model \citep{www82,salpeter69}. 

In determining the equilibrium electron-positron number density and the screening lengths, 
computational models may truncate the number of Landau levels that are counted
in the evaluation, or the sum may be replaced by an integral in a low-field approximation \citep{kawasaki12}.  For high fields, one can
determine the number of Landau levels necessary to sum over to obtain
a certain accuracy in the computation.  This is illustrated in Figure \ref{landau_sens}.  In Figure \ref{landau_sens}a, the 
computed electron chemical potential is shown as a function of the maximum
Landau level, $N_{max}$ included in the sum in Equation \ref{num_dens_terms} for 
$T_9=2$, and $\rho Y_e=5\times 10^4$ g cm$^{-3}$.  As the number of Landau levels summed over increases, the chemical potential converges to its equilibrium value.  For a field
of 10$^{15}$ G, the convergence is immediate, and the approximation where only the lowest Landau level is considered is valid.  For
10$^{14}$ G, the convergence occurs rapidly at $N_{max}=1$, and the difference between
this approximation and the $N_{max}=1$ summation is small.  At lower fields, 
a summation over more Landau levels is necessary in order to achieve a reasonable
accuracy.  

It is also interesting to note that at higher fields, the electron chemical potential is
reduced as the level density is adjusted by the presence of Landau levels.  At the highest
fields, the electron transverse momentum is discrete and increases with field.  The energy necessary to fill higher Landau levels is large compared to the thermal energy of
the plasma, $kT$, and electrons are forced into the lowest-energy levels.  However, if the field is low-enough such that $\sqrt{eB}\ll kT$, the chemical potential approaches
that of an ideal Fermi gas, and the plasma can be treated as such.  In this
case, the field can be ignored.

Similarly, in Figure \ref{landau_sens}b, 
truncating the sum over Landau levels at a specific number is explored by
examining the number of Landau levels necessary to achieve a desired uncertainty.  Shown on the right side of this figure is
the number of Landau levels necessary, $N_{0.01}$, such that the relative difference between the sum over $N_{0.01}$ Landau levels and the equilibrium number density is less than 1\%:
\begin{equation}
1 - \frac{\sum\limits_{i=0}^{N_{0.01}-1} h_i}{\sum\limits_{i=0}^{N_{large}} h_i}<0.01,
\end{equation}
where $h_i$ are individual terms in the number density in Equation \ref{num_dens_terms}.
That is, the relative difference between the number density if only 
$N_{0.01}$ Landau levels are used and if a sufficiently large 
number of Landau levels is used is less than 0.01. For this 
figure, the density times the electron fraction is $\rho Y_e=5\times10^{4}$ g cm$^{-3}$.  
For fields that are
high-enough, $B\gtrsim 10^{13}$ G, each successive term in the
sum drops
by roughly an order of magnitude, $h_{i+1}/h_i\sim 0.1$.  Here, a value of $N_{large}$
of 10$^{4}$ is assumed. From the left side of the figure, it is seen that even at low
fields, sums up to terms less than 10$^4$ Landau levels are 
sufficient to characterize the plasma, indicating a choice of $N_{large}=10^4$ to
be sufficient.  Even at low fields, the last $\sim3000$
Landau levels in the sum contribute less than 1\% to the total
electron-positron number density.
 For a lower field, it is necessary to include several hundred (or more) Landau levels in the sum for an
accurate calculation.  For the 
very high field, however, one can achieve a high accuracy by including only the lowest Landau level, known at the ``lowest Landau level approximation'' or the LLL approximation.  A discussion of the accuracy and utility of the LLL approximation in
evaluating the TF length will be given later.

\iffalse
At a more fundamental level, the electron/positron chemical potential decreases with increasing magnetic field. The degree to which this
can be approximated either through a truncation of the sum over
Landau levels or by using the expansion of Equation \ref{numdens} 
is shown in Figure \ref{mu_approx}.  In this figure, the numerically calculated chemical
potential as a function of the maximum summed Landau level is shown
for a homogeneous plasma with temperature, $T_9=3$, density,
$\rho$=10$^7$ g/cm$^{3}$, and electron fraction $Y_e=0.5$.   In this figure, chemical potentials for various magnetic fields are plotted.
Since the expansion of Equation \ref{numdens} is independent of
any summation, the lines are horizontal.  
\fi

As can be seen from Figures \ref{screen_modes} and \ref{landau_sens}a, the effect of the 
magnetic field becomes negligible roughly below 10$^{13}$ G.  The
electron-positron population is determined almost exclusively by
the system temperature and density.  In this region, the 
thermally calculated chemical potential without magnetic fields is almost identical
to that computed if magnetic field effects are accounted for and the positron
number density approaches zero as stated previously.  For
the temperature and density used for Figure \ref{landau_sens}a, 
the electron chemical potential if no field were present would be 0.76 MeV. 
Above 10$^{13}$ G, the chemical potential decreases with field. 
\subsection{High-Field Approximation: Euler-MacLaurin Formula in Momentum}
\label{low_p_high_b}
At high fields and high temperatures, the chemical potential is low, and the electron-positron Fermi distribution is constrained to relatively low momentum.  In this case, we consider an Euler-MacLaurin expansion in momentum 
using the Euler-MacLaurin formula. The 
net electron number density can be written as:
\begin{equation}
    n_e=eB\frac{ T}{2\pi^2}\sum\limits_{\tilde{p},n=0}^\infty
    g_n
    \left(\frac{2-\delta_{p0}}{2}\right)
    \frac{\sinh\tilde{\mu}}{\cosh\sqrt{\tilde{p}^2+\tilde{m}^2+n\gamma}+\cosh\tilde{\mu}},
\end{equation}
where $\gamma\equiv 2eB/T^2$ and terms with a tilde are divided by $T$, $\tilde{x}\equiv x/T$.  
These terms are unitless.  It is noteworthy that,
for the Euler-MacLaurin formula, the higher-order
derivatives are zero, meaning that the sum above is complete.
In the case of a strong magnetic
field, the LLL approximation yields:
\begin{equation}
    n_e=eB\frac{T}{2\pi^2}
    \sum\limits_{\tilde{p}=0}^\infty \left(\frac{2-\delta_{p0}}{2}\right)
    \frac{\sinh\tilde{\mu}}{\cosh\sqrt{\tilde{p}^2+\tilde{m}^2}+\cosh\tilde{\mu}},
\end{equation}
resulting in a linear dependence on the external 
magnetic field.

The Thomas-Fermi screening
length in a strong magnetic field is derived as:
\begin{eqnarray}
\label{EM1}
    \frac{1}{\lambda_{TF}^2} & = &4\pi e^2\frac{\partial n}{\partial\mu}\\\nonumber
    ~& =& eB\frac{e^2} {\pi}\frac{\partial}{\partial\tilde{\mu}}
    \sum\limits_{n=0}^{\infty}g_{n}
    \int\limits_0^{\infty}\frac{d\tilde{p}}
    {
        \exp{
                \left[
                    \sqrt{
                            \tilde{p}^2+\tilde{m}^2+n\gamma
                        }\mp\tilde{\mu}
                \right]
            }
        +1
    }\\\nonumber
    ~& =& eB\frac{e^2}{\pi}
    \sum\limits_{n=0}^{\infty}g_{n}
    \int\limits_0^{\infty}\frac{\partial}{\partial\tilde{\mu}}
    \frac{d\tilde{p}}
    {
        \exp{
                \left[
                    \sqrt{
                            \tilde{p}^2+\tilde{m}^2+n\gamma
                        }\mp\tilde{\mu}
                \right]
            }
        +1
    }\\\nonumber
    ~& =& eB\frac{e^2}{\pi}
    \sum\limits_{n=0}^{\infty}g_{n}
    \int\limits_0^{\infty}
    \frac{d\tilde{p}}
    {
        1+\cosh{\left(\sqrt{\tilde{p}^2+\tilde{m}^2+n\gamma}\mp\tilde{\mu}\right)}
    },
\end{eqnarray}
where the $\mp$ corresponds to the electron/positron number density, and the sum over
both electron and positron densities is implied.

The Euler-MacLaurin fomula, expanded in momentum, yields an easily-computed
form for the integral term above:
\begin{eqnarray}
\label{EM_p_b}
    \frac{1}{\lambda_{TF}^2}
    &=&eB\frac{e^2} {\pi}\sum_{\tilde{p},n=0}^{\infty}g_n
\left(\frac{2-\delta_{p0}}{2}\right)
        \frac{1+\cosh{\tilde{\mu}}\cosh\sqrt{\tilde{p}^2+\tilde{m}^2+n\gamma}}
        {
        \left(
        \cosh{\tilde{\mu}}+\cosh{\sqrt{\tilde{p}^2+\tilde{m}^2+n\gamma}}
        \right)^2}
\end{eqnarray}
where the sum over $n$ is a sum over Landau Levels and the sum over $\tilde{p} = p/T$ results from the 
Euler-MacLaurin formula for Equation \ref{EM1}.

One can approximate a sum over several Landau levels and only up to a maximum
value of $\tilde{p}$ in the above equation:
\begin{equation}                \frac{1}{\lambda_{TF}^2}\propto\sum_{n=0}^{\infty}g_n
    \left[
        \sum_{\tilde{p}=0}^{\infty} ...
        \right]
        \rightarrow
    \sum_{n=0}^{n_{max}}g_n
    \left[
        \sum_{\tilde{p}=0}^{\tilde{p}_{max}} ...
        \right] + R_{\tilde{p}},
\end{equation}
where $n_{max}$ is the highest Landau level included in the sum, and $\tilde{p}_{max}$ is the highest term included
in the Euler-MacLaurin formula. The remainder induced by
truncating the sum is $R_{\tilde{p}}$.

\begin{figure}
    \centering
    \gridline{
     \fig{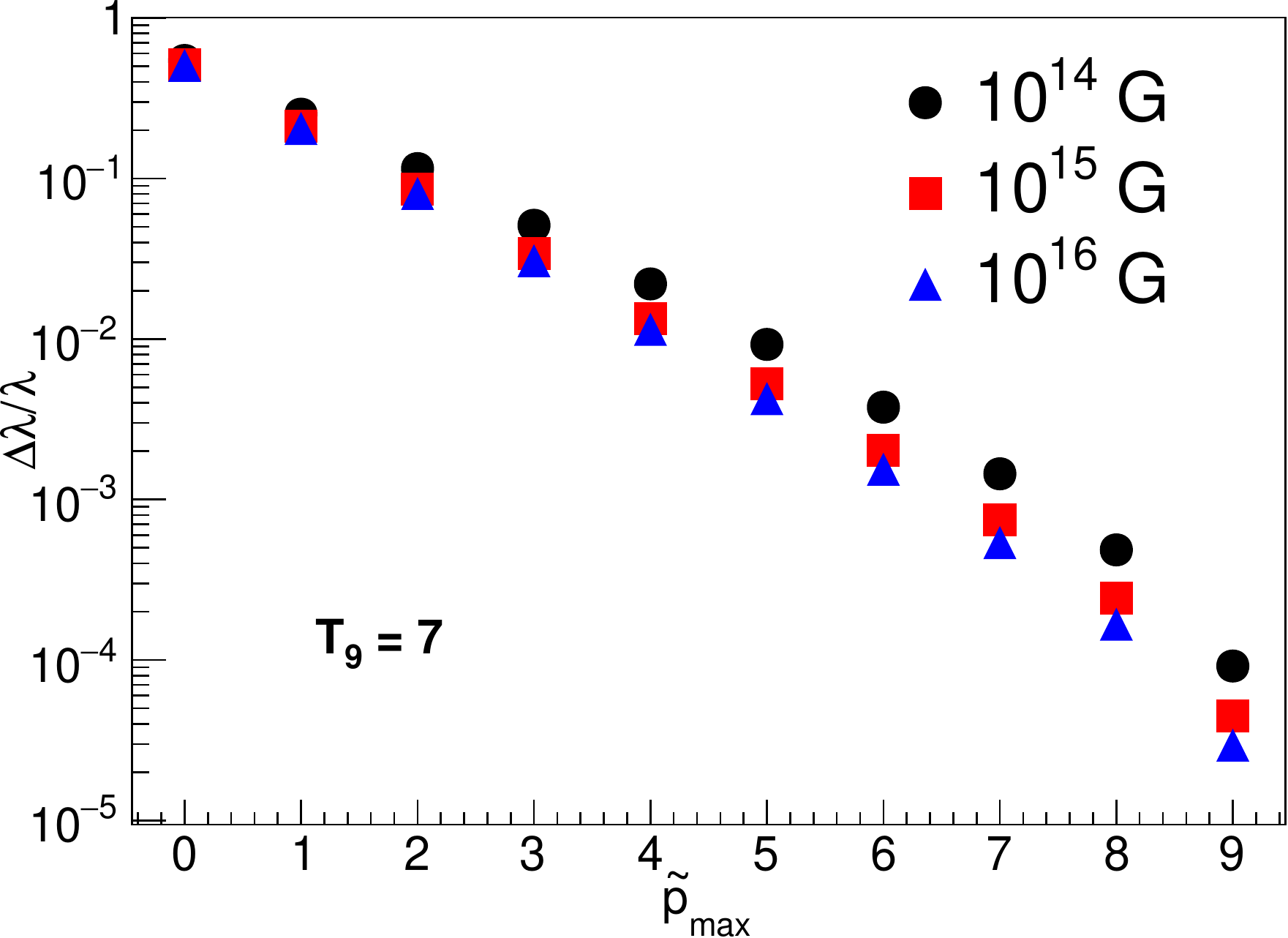}{0.49\textwidth}{(a)}
     \fig{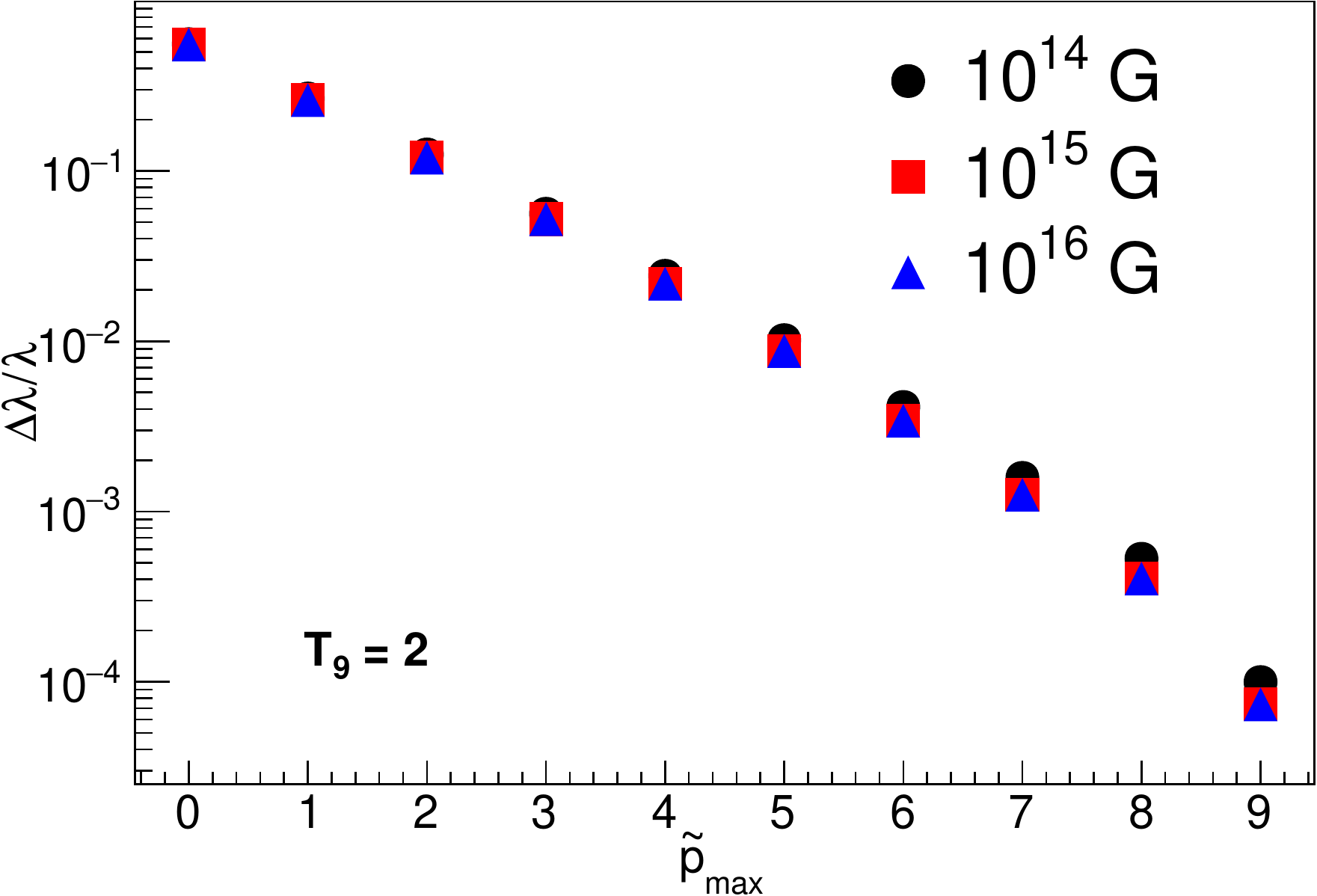}{0.49\textwidth}{(b)}
    }
    \caption{
        Relative error in Euler-MacLaurin formula compared
        to exact numerical computation for the integration in Equation \ref{EM1} as a function of maximum $\tilde{p}$ in the sum.
        Computations are for $\rho Y_e = 5\times 10^{4}$ g cm$^{3}$ 
        at temperatures (a) $T_9= 7$ and (b) $T_9=2$.  The maximum
        Landau level calculated in each case is $N_{max} = 2000$.
        }
    \label{error_gr}
\end{figure}

At high magnetic field, the electron chemical potential is much smaller than the Landau level spacing. In this case, the sum over $\tilde{p}$ converges rapidly, and the summation can be
truncated to a few terms.  For the purposes of computation, the limitation of the sum
may be determined to truncate at $\tilde{p}_{max}$ where the difference in successive terms is smaller than some uncertainty, $\varepsilon$:
\begin{equation}
    \frac{f_{\tilde{p}_{max}}-f_{(\tilde{p}_{max}-1)}}
    {f_{\tilde{p}_{max}}}<\varepsilon
\end{equation}
As an example, the relative error in $\lambda_{TF}$, $\Delta\lambda/\lambda = 1 -\lambda_{McL}/\lambda_{exact}$ (where the Thomas-Fermi length deduced
from the truncated sum is $\lambda_{McL}$ and that deduced from
Equation \ref{EM1} is $\lambda_{exact}$) induced by truncating the Euler-MacLaurin sum to a 
maximum index of $\tilde{p}_{max}$ is shown in Figure \ref{error_gr} for 
temperatures $T_9$=7 and 2, 
at $\rho Y_e=5\times$10$^4$ g cm$^{-3}$, and three
values of the external magnetic field.  
Even for a low value of $\tilde{p}_{max}=5$, the uncertainty is less than
1\%. 

\begin{figure}
\gridline{
\fig{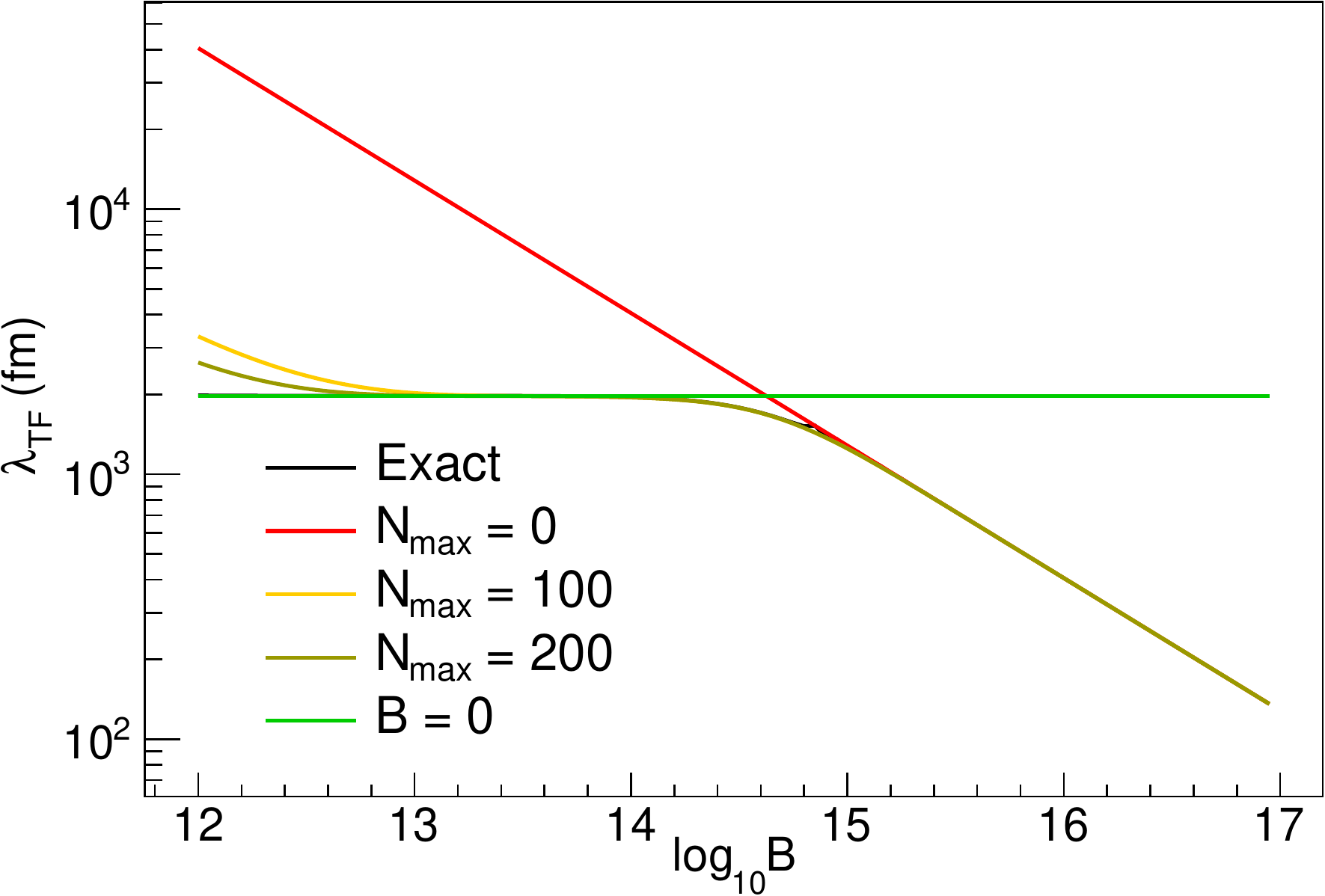}{0.49\textwidth}{(a)}
\fig{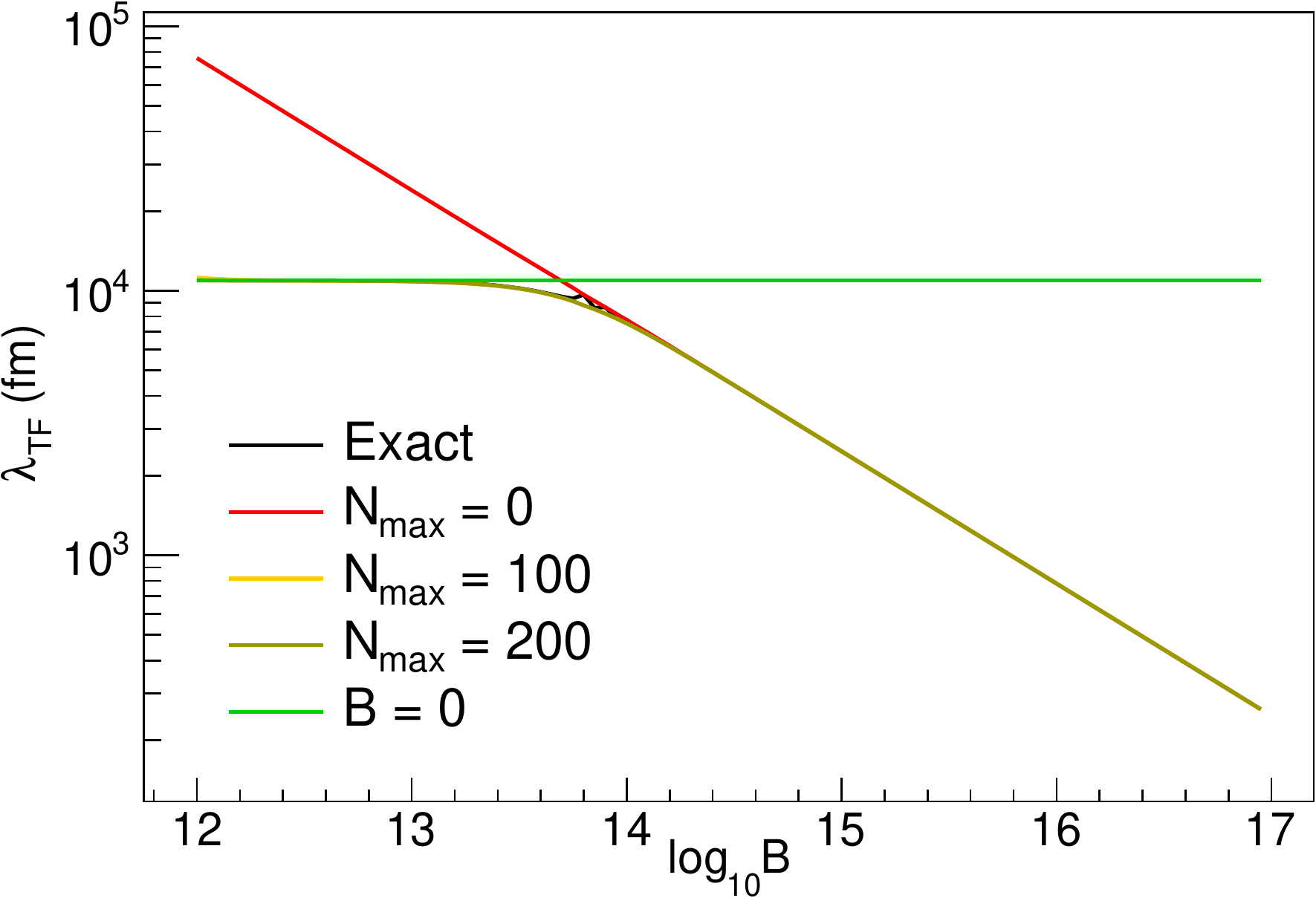}{0.49\textwidth}{(b)}
}
    \caption{Electron Thomas-Fermi screening length using the 
    approximation described in Section \ref{low_p_high_b} for sums up to various maximum
    Landau levels, as indicated in the figure.
    In both figures, $\rho Y_e = 5\times10^4$ g cm$^{-3}$.  The temperature is (a) $T_9=7$ and (b) $T_9=2$. 
    For the figure (b), the lines corresponding to $n=100$ and $n=200$ lie on top of each other.
    In this figure, only the lowest 12 terms
    in the Euler-MacLaurin sum are computed, $\tilde{p}_{max} = 12$.
    }
    \label{low_p_high_b_compare}
\end{figure}

The validity of this approximation in determining the screening length at temperatures $T_9 = 2$ and 7 at $\rho Y_e = 5\times$10$^4$ g cm$^{-3}$ is shown in Figure 
\ref{low_p_high_b_compare}.  In this figure,
the approximation given in Equation \ref{EM_p_b}
is used to determine the TF screening lengths.
For each line in the figure, only the lowest
12 terms in the sum over $\tilde{p}$ are used.  That is
$\tilde{p}_{max} = 12$.  The maximum number of Landau levels
summed over is indicated for the various 
results in the figure.

One sees that the lowest Landau level (LLL, $N_{max}$ = 0) approximation performs quite well at high fields 
(log(B)$\gtrsim$14).  At lower fields, more 
Landau levels must be included in the sum.

\begin{figure}
\gridline{
\fig{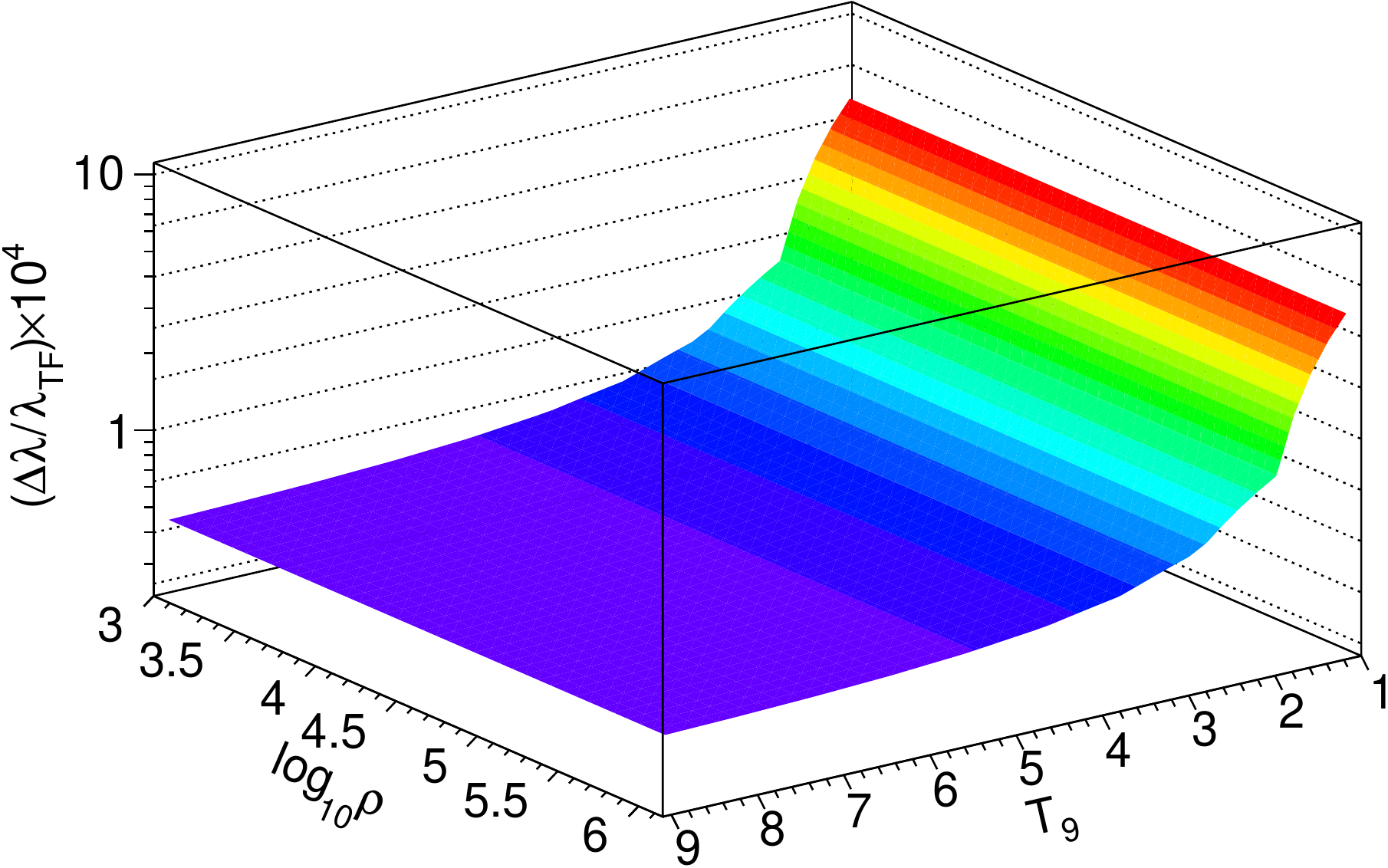}{0.49\textwidth}{(a)}
\fig{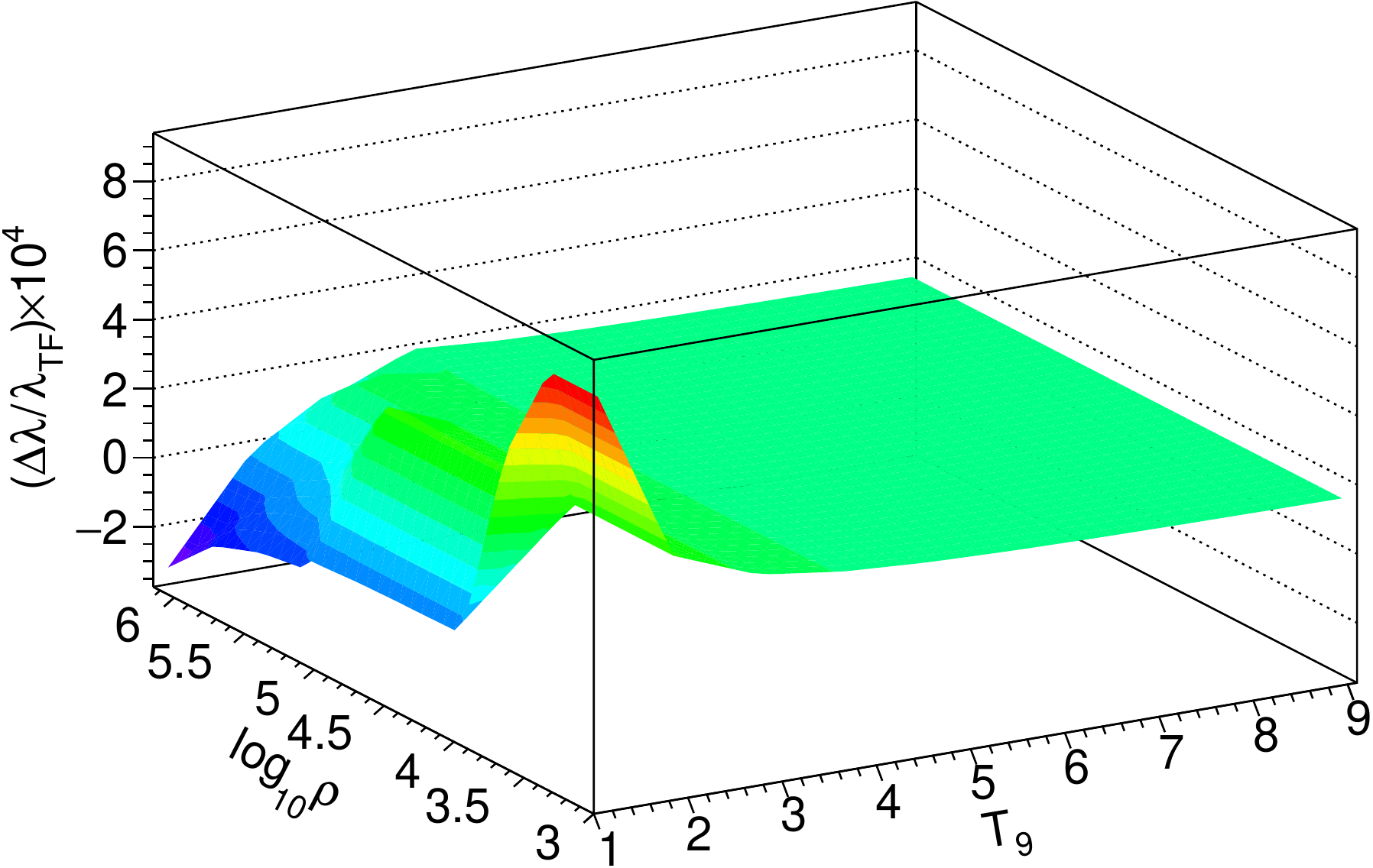}{0.49\textwidth}{(b)}
}
    \caption{Relative difference in TF screening length Euler-MacLaurin approximation to the exact computation 
    as a function of T and $\rho$ for (a)$ B = 10^{16}$ G and 
    (b)$ B = 10^{13}$ G for
    various temperatures and densities.  Here, only the LLL approximation is used for the 10$^{16}$ G case and the lowest 2000 Landau levels are used
    for the 10$^{13}$ G case.  In both panels, $Y_e=0.5$.
    }
    \label{delta_lambda}
\end{figure}

For completeness, the dependence of this approximation on temperature and 
density is shown
in Figure
\ref{delta_lambda}, which shows the relative error
in the TF length computed with Equation \ref{EM_p_b} compared 
to that 
computed with Equation \ref{EM1}.  It is seen that -- in the weak screening
regime -- there is almost no dependence on density and a small dependence on 
temperature.  Even at low fields, the (Figure \ref{delta_lambda}b), the 
error is relatively small.  At lower temperatures, the error is somewhat
larger.  However, this area would very likely correspond to
non-relativistic or intermediate screening.

For a lower field of $B = 10^{13}$ G, the approximation of Equation  \ref{EM_p_b}
is shown in Figure \ref{delta_lambda}b, including the lowest 2000 Landau 
levels ($N_{max} = 2000$) and $\tilde{p}_{max} = 12$.  The TF screening length is still
fairly well approximated over a wide range of temperatures and densities even at lower
B field if a sufficient number of Landau levels are included in the sum.  For most
temperatures and densities, the screening length is within about 10\% of the actual Thomas-Fermi length.  However, it is also noted that if the field
is low enough, $\lambda_{TF}$ for $B=0$ is an excellent approximation, and
the effect of the field can be ignored.  
\section{Weak Interactions}
In addition to the inclusion of magnetized plasma effects on screening of the Coulomb potential and modifications to the electron-positron chemical potential, effects on weak interaction rates have also been examined.  Weak interactions can be altered by changes to the electron Fermi-Dirac distribution function and the electron energy spectrum in weak decays \citep{luo20,grasso01,fassio69}.  In addition, the shifts to the electron-positron chemical potentials in the thermal plasma are also modified.
The shift in chemical potentials can change the Fermi-Dirac functions, altering the available states for capture and decay as well as the Pauli blocking factors.  This can 
influence all of the weak interactions.  Also, the electrons
and positrons involved in weak interactions
are constrained to Landau levels, creating nearly-discrete
energy spectra, especially at high fields.  

In the presence of 
magnetic fields, the phase space ($d^3p$) of the interactions is changed by the presence of Landau
Levels.  The density of states is (in natural units):
\begin{equation}
    dn \propto \frac{d^3p}{(2\pi)^3} = \sum\limits_{n=0}^{\infty}\left(2 - \delta_{n0}\right)\frac{eB}{2\pi^2}dp_z.
\end{equation}
The corresponding shift in the lepton energy spectra can have dramatic effects
on the weak interaction rates in a magnetized plasma.  With the inclusion of
density distributions modified by the existence of Landau levels,
the approximate weak interaction rates can be rewritten 
with the momentum component parallel to the magnetic 
field vector  and the discrete transverse momentum components \citep{luo20,grasso01,fassio69}:
\begin{eqnarray}
\label{mag_beta_minus}
    \Gamma_{\beta^-} & = &\kappa \frac{eB}{2} \sum\limits_{n=0}^{N_{max}}\left(2-\delta_{n0}\right)\int\limits_{\omega_\beta}^{Q}\frac{E(Q-E)^2}{\sqrt{E^2 - m_e^2-2neB}}\left(1-f_{FD}(E,\mu_e)\right)\left(1-f_{FD}(Q-E,-\mu_\nu)\right)dE,
    \\
    \label{mag_beta_plus}
    \Gamma_{\beta^+} & = &\kappa \frac{eB}{2} \sum\limits_{n=0}^{N_{max}}\left(2-\delta_{n0}\right)\int\limits_{\omega_\beta}^{-Q}\frac{E(-Q-E)^2}{\sqrt{E^2 - m_e^2-2neB}}\left(1-f_{FD}(E,-\mu_e)\right)\left(1-f_{FD}(-Q-E,-\mu_\nu)\right)dE,
    \\
    \label{mag_beta_EC}
    \Gamma_{EC} & = &\kappa \frac{eB}{2} \sum\limits_{n=0}^{N_{max}}\left(2-\delta_{n0}\right)\int\limits_{\omega_{EC}}^\infty\frac{E(E-Q)^2}{\sqrt{E^2 - m_e^2-2neB}}f_{FD}(E,\mu_e)\left(1-f_{FD}(E-Q,\mu_\nu)\right)dE,
    \\
    \label{mag_beta_PC}
    \Gamma_{PC} & = &\kappa \frac{eB}{2} \sum\limits_{n=0}^{N_{max}}\left(2-\delta_{n0}\right)\int\limits_{\omega_{PC}}^\infty\frac{E(E+Q)^2}{\sqrt{E^2 - m_e^2-2neB}}f_{FD}(E,-\mu_e)\left(1-f_{FD}(E+Q,-\mu_\nu)\right)dE,
\end{eqnarray}
where the following quantities are defined \citep{arcones10, hardy09}:
\begin{eqnarray}
    \omega_{EC/PC} &\equiv & \mbox{max}\left[\pm Q,m_e\right],
    \\\nonumber
    \omega_{\beta} &\equiv & \sqrt{m_e^2+2neB},
    \\\nonumber
    N_{max}&\le &\frac{Q^2 - m_e^2}{2eB},
    \\\nonumber
    \kappa & \equiv & \frac{B\ln 2}{K m_e^5},
    \\\nonumber
    B & \equiv & 1+3g_A^2 
    =
    \left\lbrace
    \begin{array}{ll}
      5.76, & \text{nucleons}, \\
      4.6, & \text{nuclei},
    \end{array}
    \right.
    \\\nonumber
    K &\equiv & \frac{2\pi^3\hbar^7\ln 2}{G_V^2 m_e^5} = 6144 \mbox{ s}.
\end{eqnarray}
Here the transition $Q$ value is the difference in nuclear masses.

In Equations \ref{mag_beta_minus} -- \ref{mag_beta_PC}, the 
Fermi Dirac distributions are cast to accommodate the electron
energy of individual Landau levels.  The energy distribution of
an electron in the nth Landau level is:
\begin{equation}
    f_{FD}(E,\mu_e) = \frac{1}{\exp{\left[\frac{\sqrt{E^2 + 2neB}-\mu_e}{T}\right]}+1}.
\end{equation}
For positrons, the chemical potential is negative.  

\begin{figure}
\gridline{
\fig{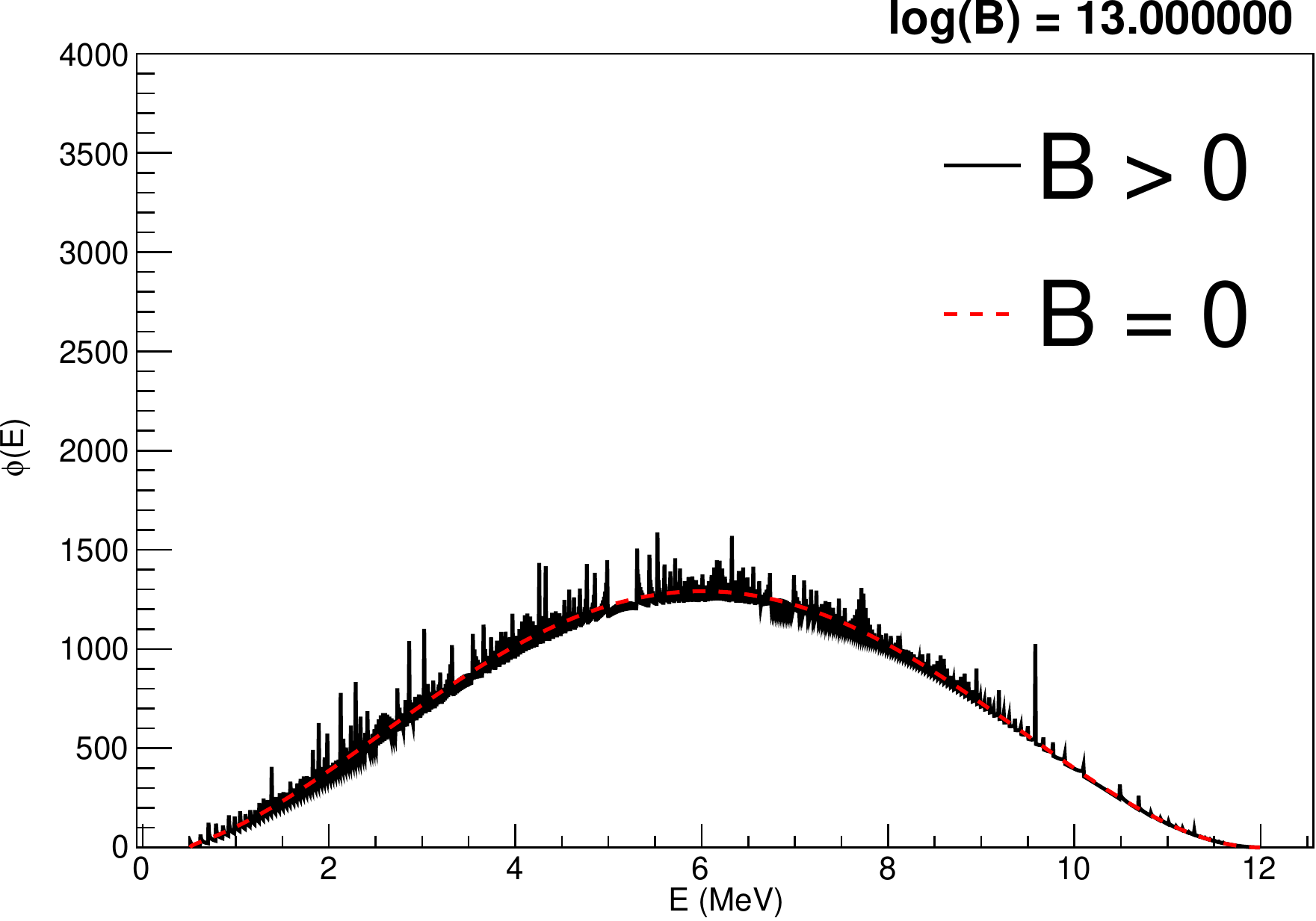}{0.32\linewidth}{}
\fig{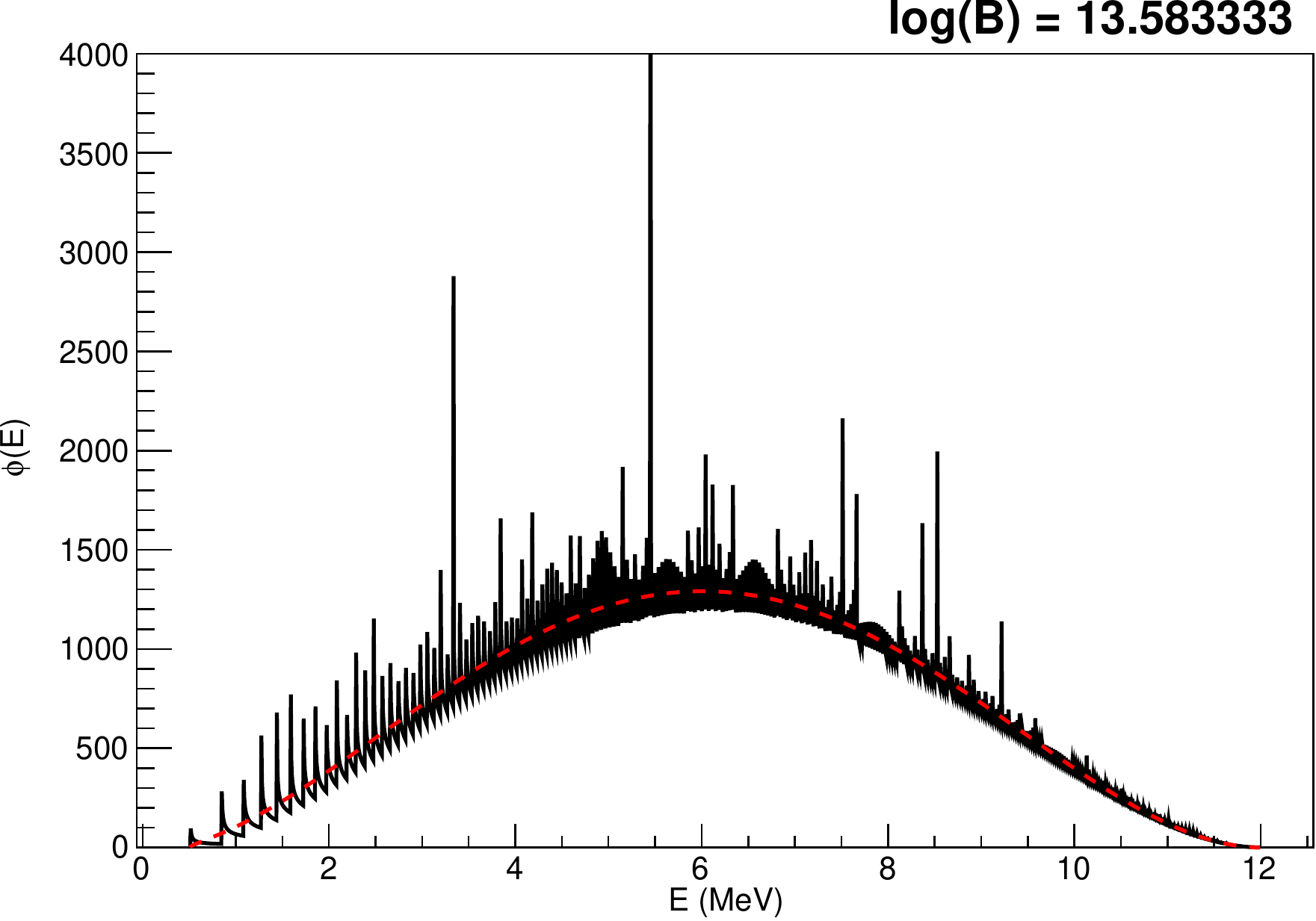}{0.32\linewidth}{}
\fig{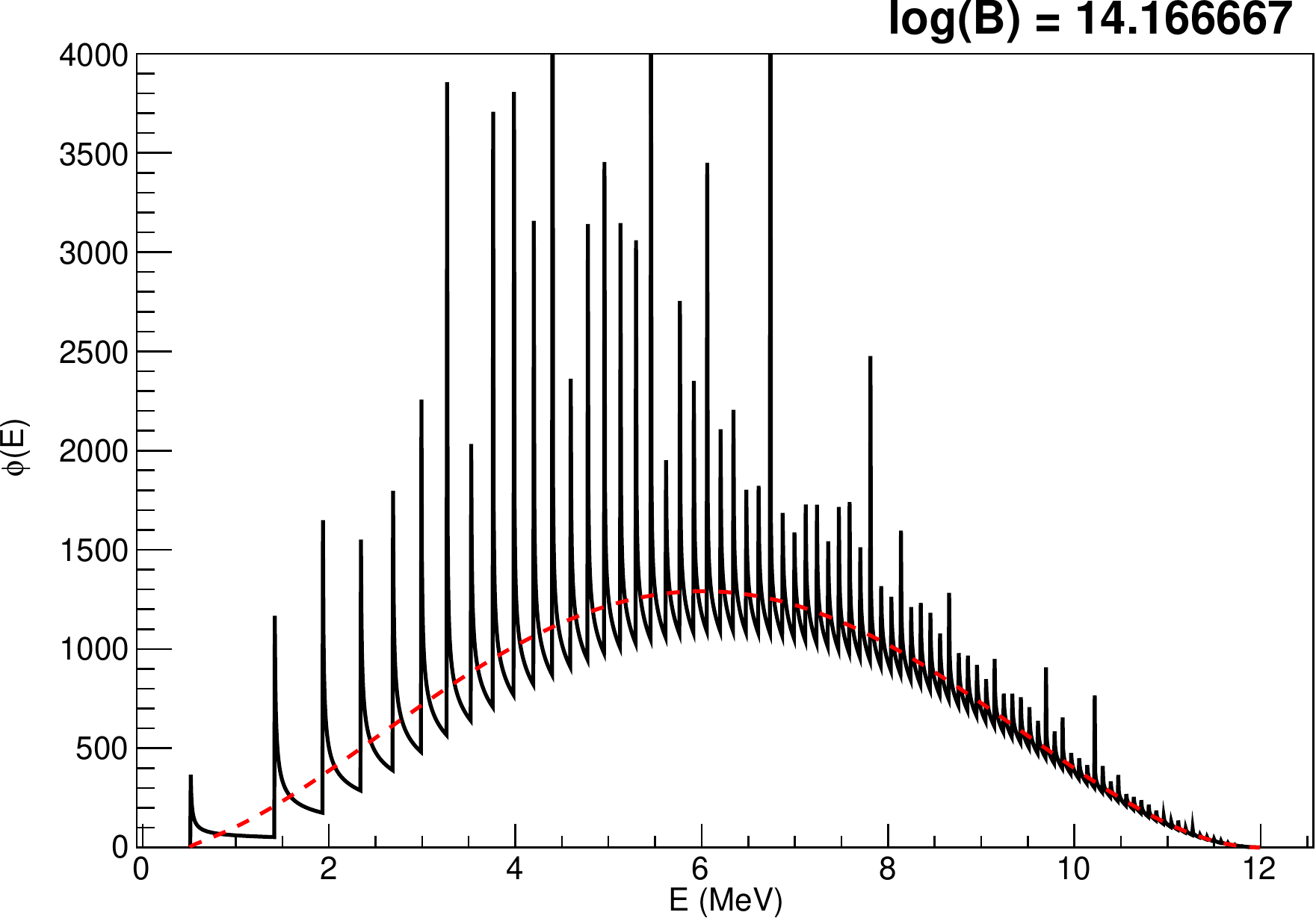}{0.32\linewidth}{}
}
\gridline{
\fig{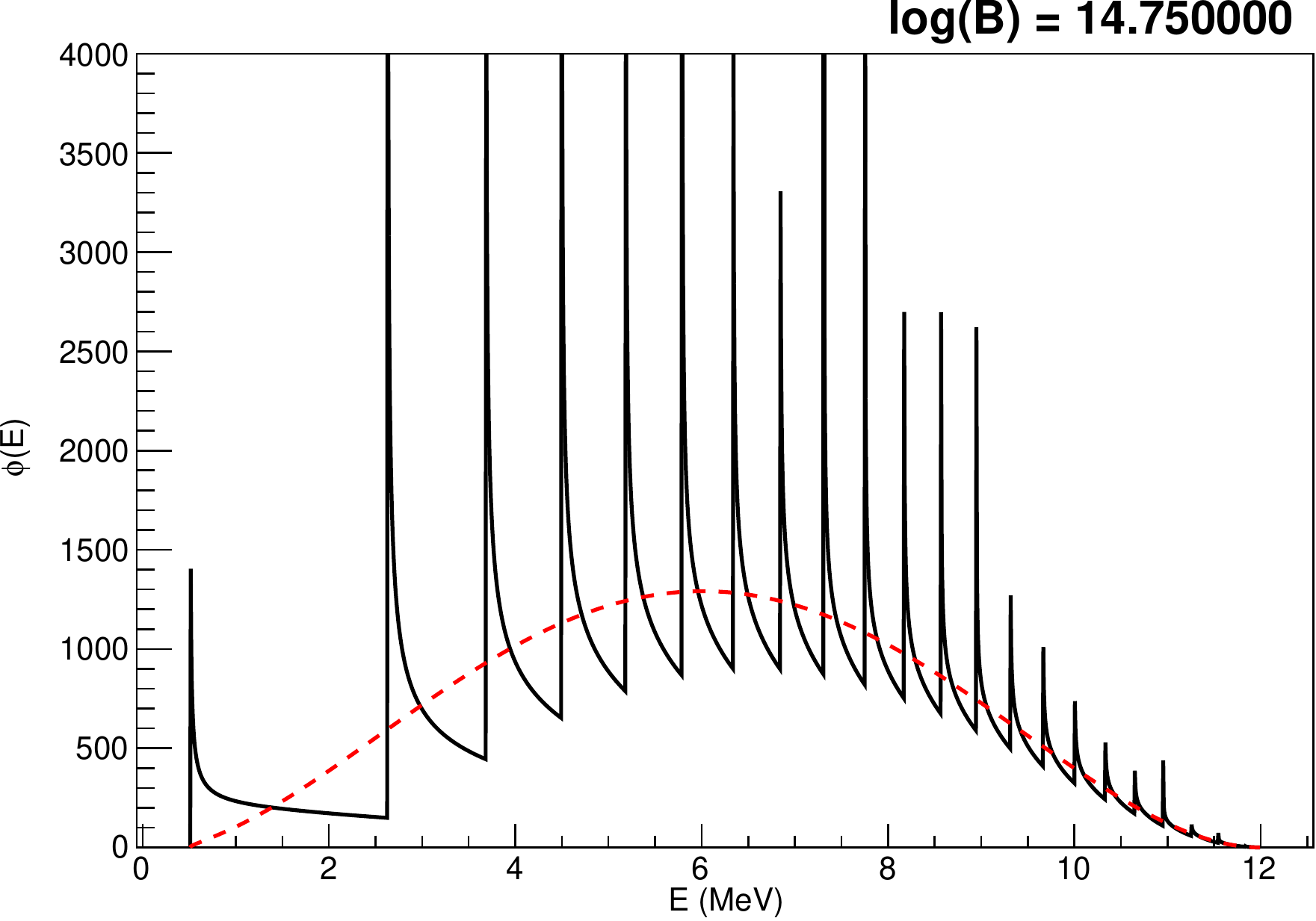}{0.32\linewidth}{}
\fig{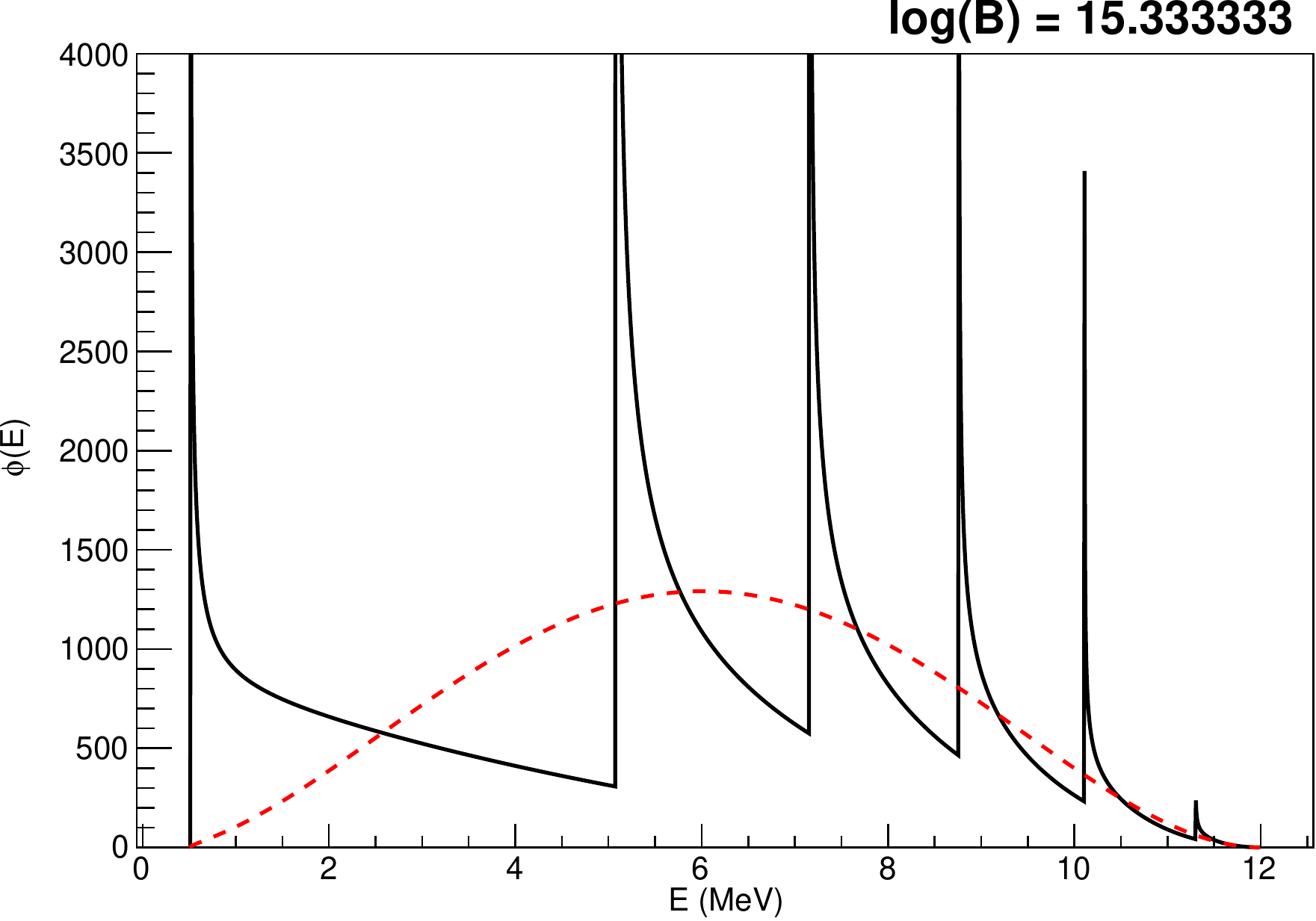}{0.32\linewidth}{}
\fig{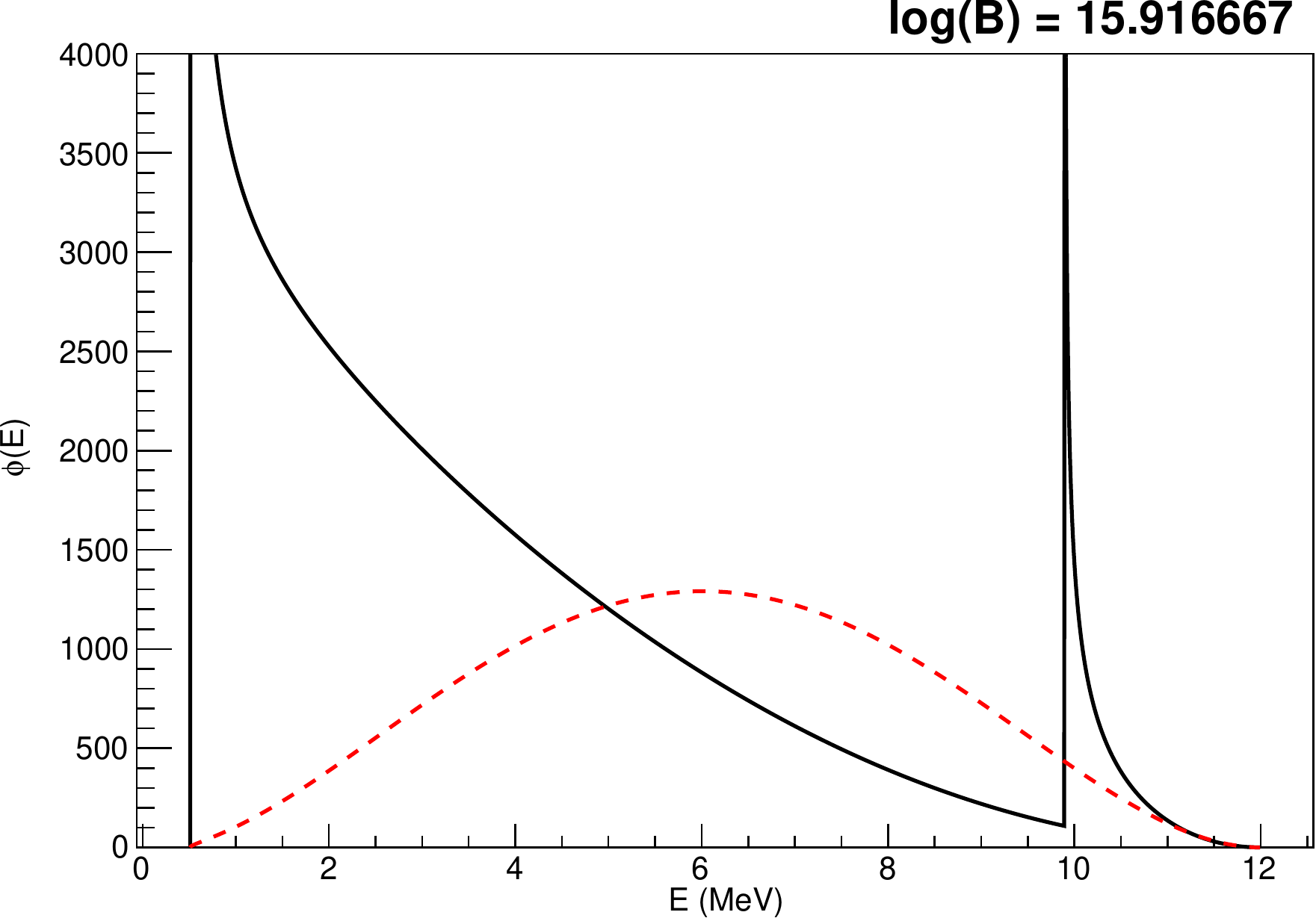}{0.32\linewidth}{}
}
    \caption{Evolution of the $\beta^-$-decay spectrum with magnetic field for
    six different fields indicated in each panel.  The red, dashed line indicates the spectrum for $B=0$, and
    the black line indicates the spectrum for the magnetic field indicated in each figure.  For this
    series of figures, the decay Q value is 12 MeV, and the values of $T_9$ and $\rho Y_e$ are 2 and 500 g cm$^{-3}$ 
    respectively.  The magnetic field units are G.}
    \label{beta_spec_evolution}
\end{figure}
\begin{figure}
\gridline{
\fig{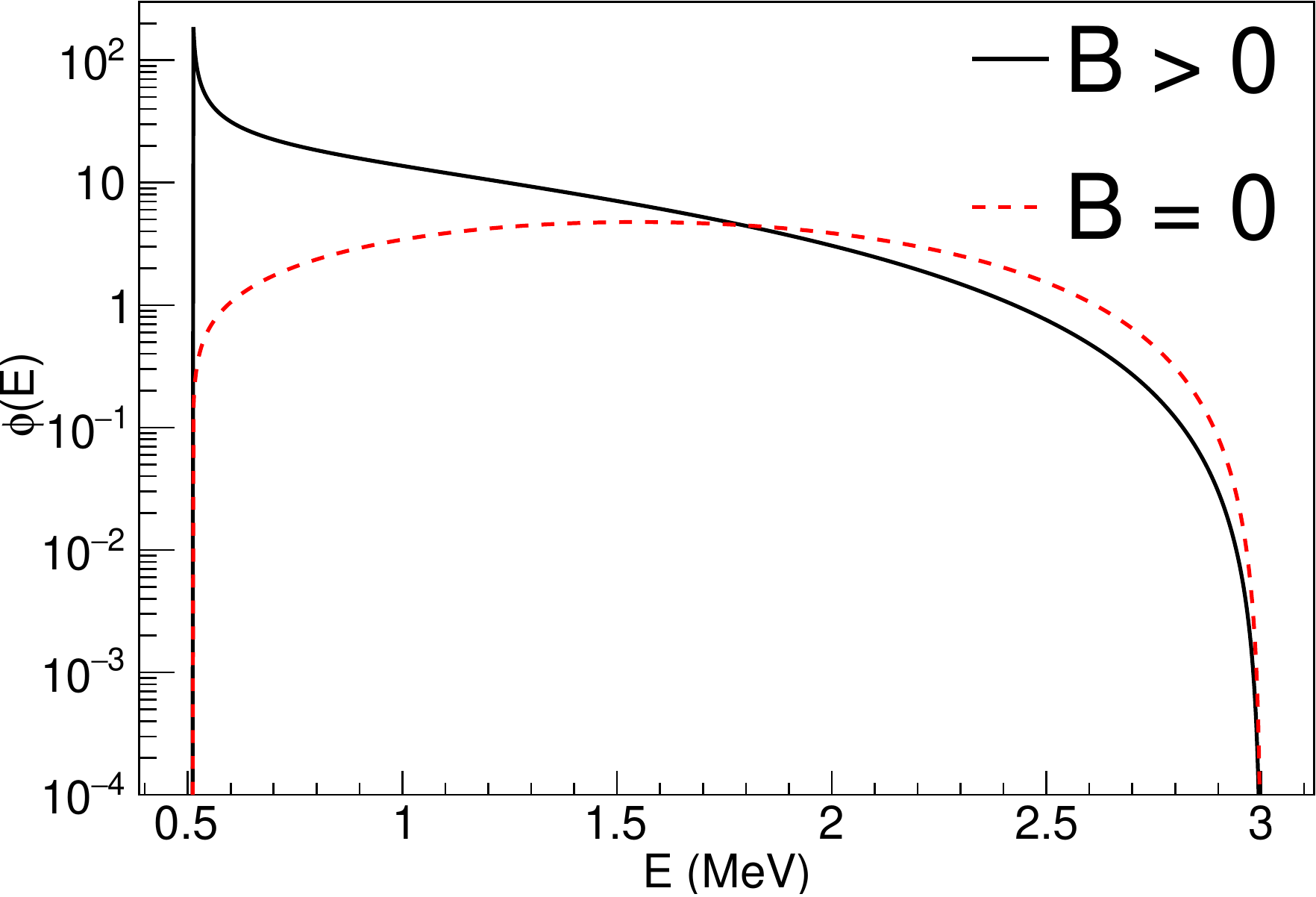}{0.49\textwidth}{(a)}
\fig{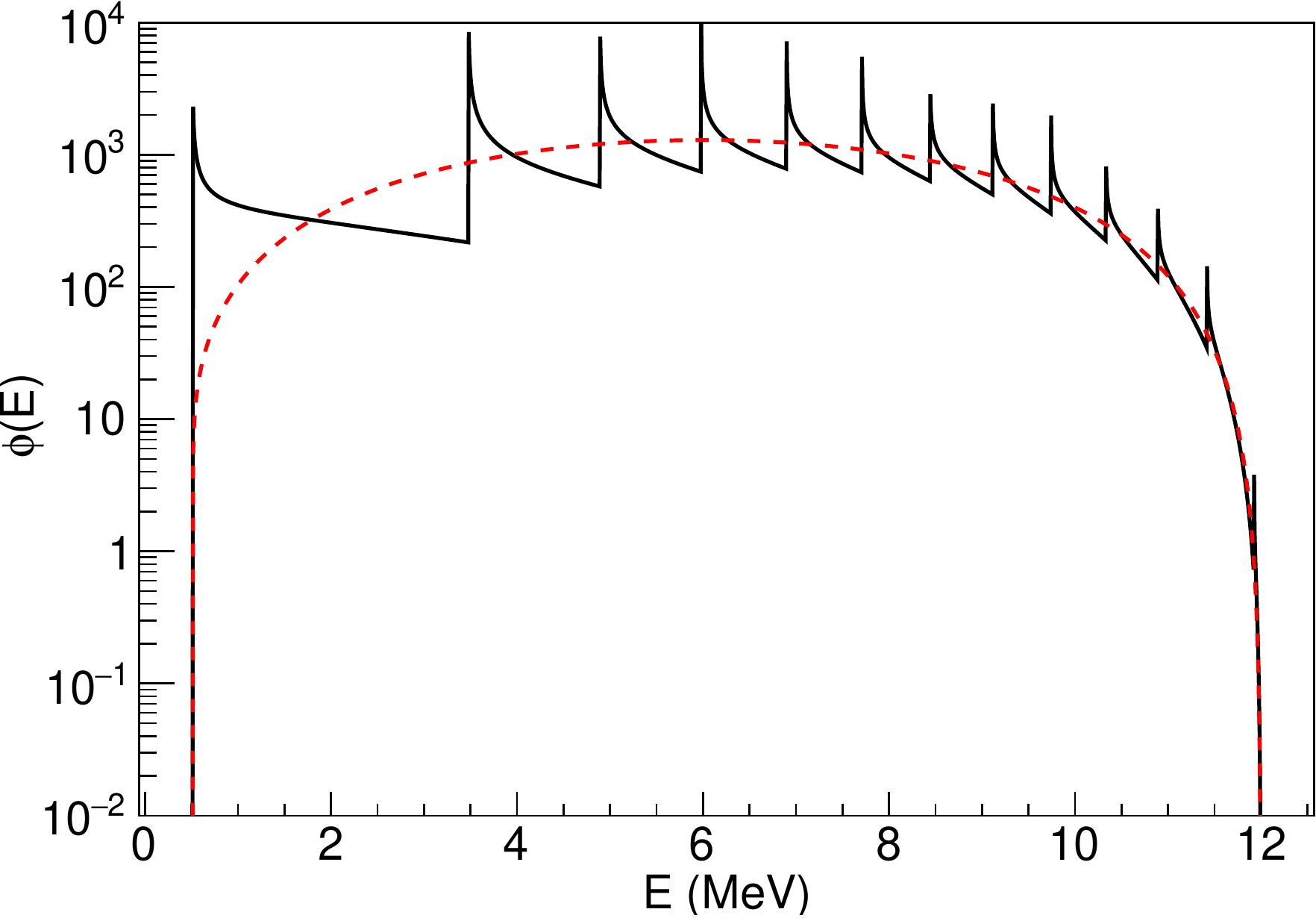}{0.49\textwidth}{(b)}
}
\gridline{
\fig{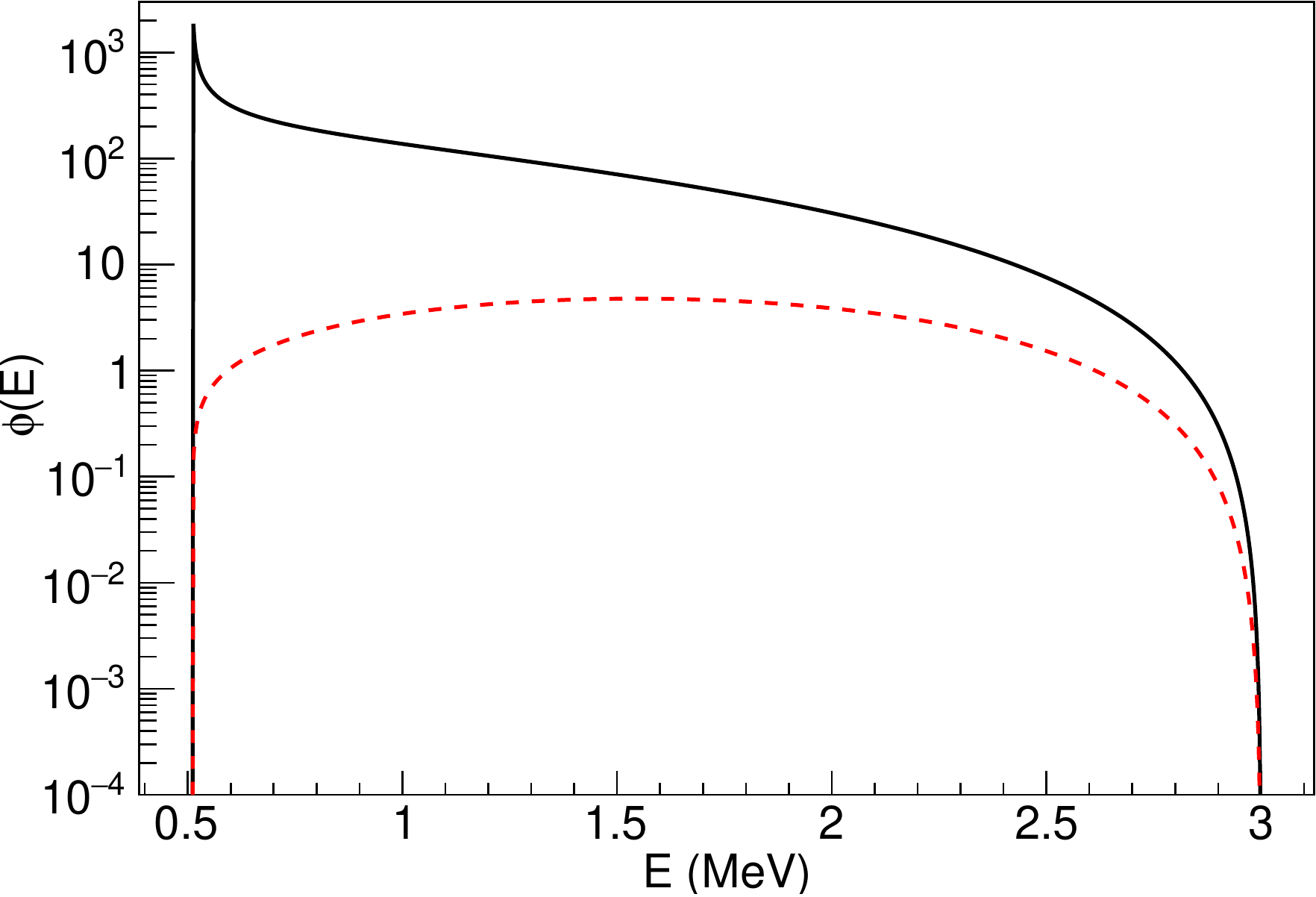}{0.49\textwidth}{(c)}
\fig{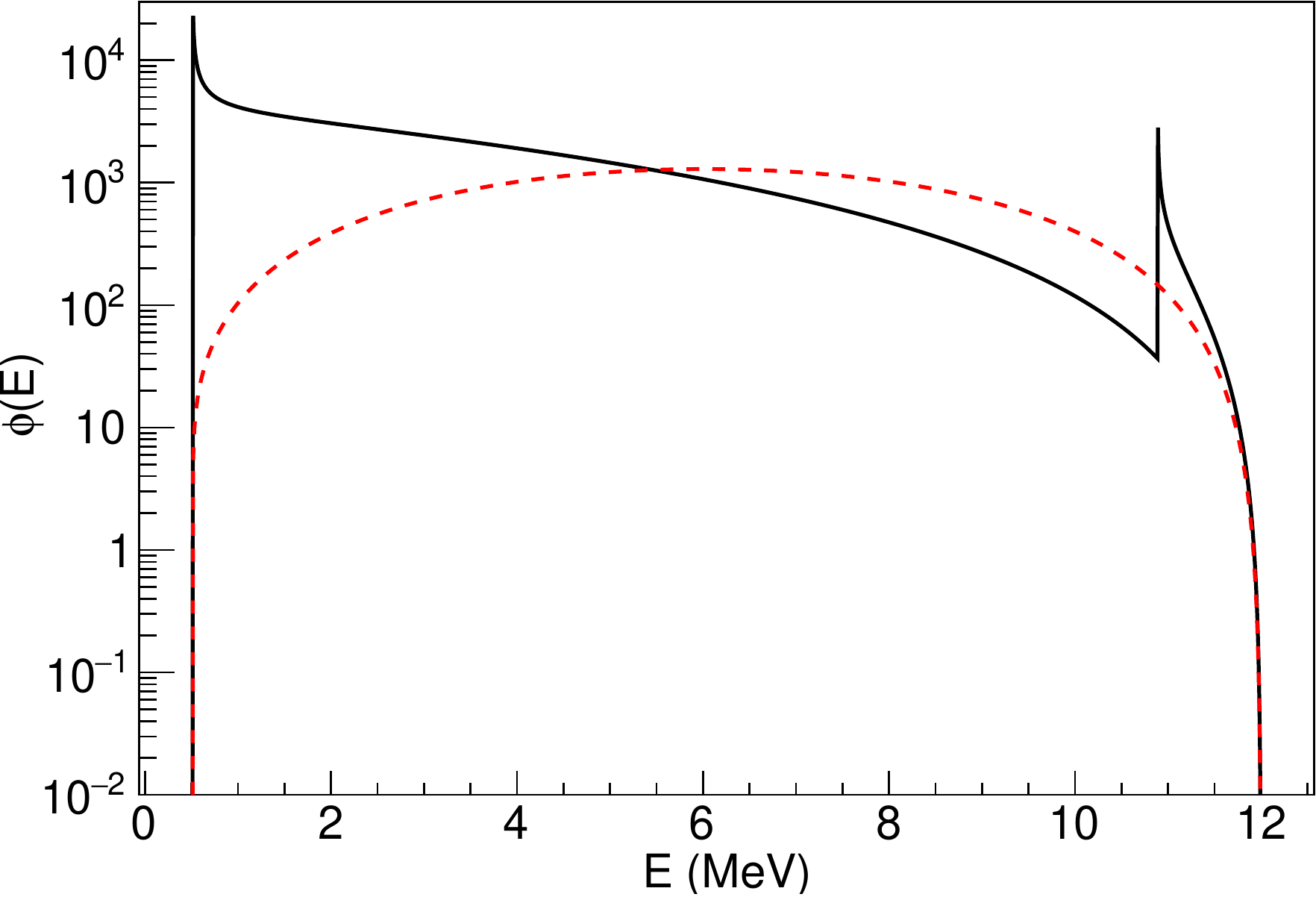}{0.49\textwidth}{(d)}
}
    \caption{Electron $\beta$-decay spectra for $B = 10^{15}$ G (a,b)
    and 10$^{16}$ G (c,d) for low $Q$ values (a,c) and high $Q$ values (b,d).  The
    spectra are calculated at $T_9=2$ and $\rho Y_e$=500.  The red, dashed lines
    correspond to the spectra for $B=0$.}
    \label{beta_spec}
\end{figure}
Unlike the case of an ideal Fermi gas, the electron-positron energy 
spectrum in weak interactions is not thermal, and the LLL approximation is 
not necessarily applicable.  For example, the evolution with magnetic field of the $\beta^-$
spectrum for a nucleus with a 
decay Q value of 12 MeV at $T_9=2$ and $\rho Y_e=500$ g cm$^{-3}$ is shown in
Figure \ref{beta_spec_evolution}.  This spectrum
is the integrand of Equation \ref{mag_beta_minus}.  In the case of a non-zero
field, the $\beta$ spectrum is a sum of
individual spectra for each Landau level with the maximum Landau level energy
less than the decay $Q$ value, $\sqrt{2neB + m_e^2}\le Q$.

For a lower field, the Landau level spacing is much less than the $Q$ value of 
the decay $\sqrt{eB}\ll Q$. An electron can
be emitted into any of a large number of Landau levels with level energies less
than the electron energy.  The Landau level spacing is
quite small in this case.  For decays to many possible Landau levels, the 
integrated spectrum
is closer in value to the zero-field spectrum.  In other words,
as $eB\rightarrow0$, the integrated non-zero-field spectrum
approaches the zero-field spectrum.  The sum in Equation \ref{mag_beta_minus}
becomes an integral, and the Landau level spacing
$eB = \Delta p^2\rightarrow d^2p$.  The sum over all Landau levels approaches 
the zero-field spectrum

As the magnetic field increases, such that $\sqrt{2neB}\sim Q$, fewer Landau 
levels contribute to the total spectrum.  For a very few levels, the  
zero-field and non-zero-field spectra can be dramatically different, and the 
decay rates
can be magnified for higher fields.

This could be potentially important for an r process that proceeds in a high 
magnetic field, such as in a collapsar jet or NS merger, for example.  Because the r process 
encompasses nuclei with a wide range of $\beta^-$ decay Q values, the effects 
of
an external magnetic field can be significant. 
This is shown in Figure
\ref{beta_spec}, which shows the electron energy spectrum in $\beta^-$ decay 
for several cases of Q-value and magnetic fields.  This spectrum is also the 
integrand of Equation \ref{mag_beta_minus}.  Spectra are computed
for $\beta^-$ decays at $T_9$=2 and $\rho Y_e$ = 500 g cm$^{-3}$.
\begin{figure}
\gridline{
\fig{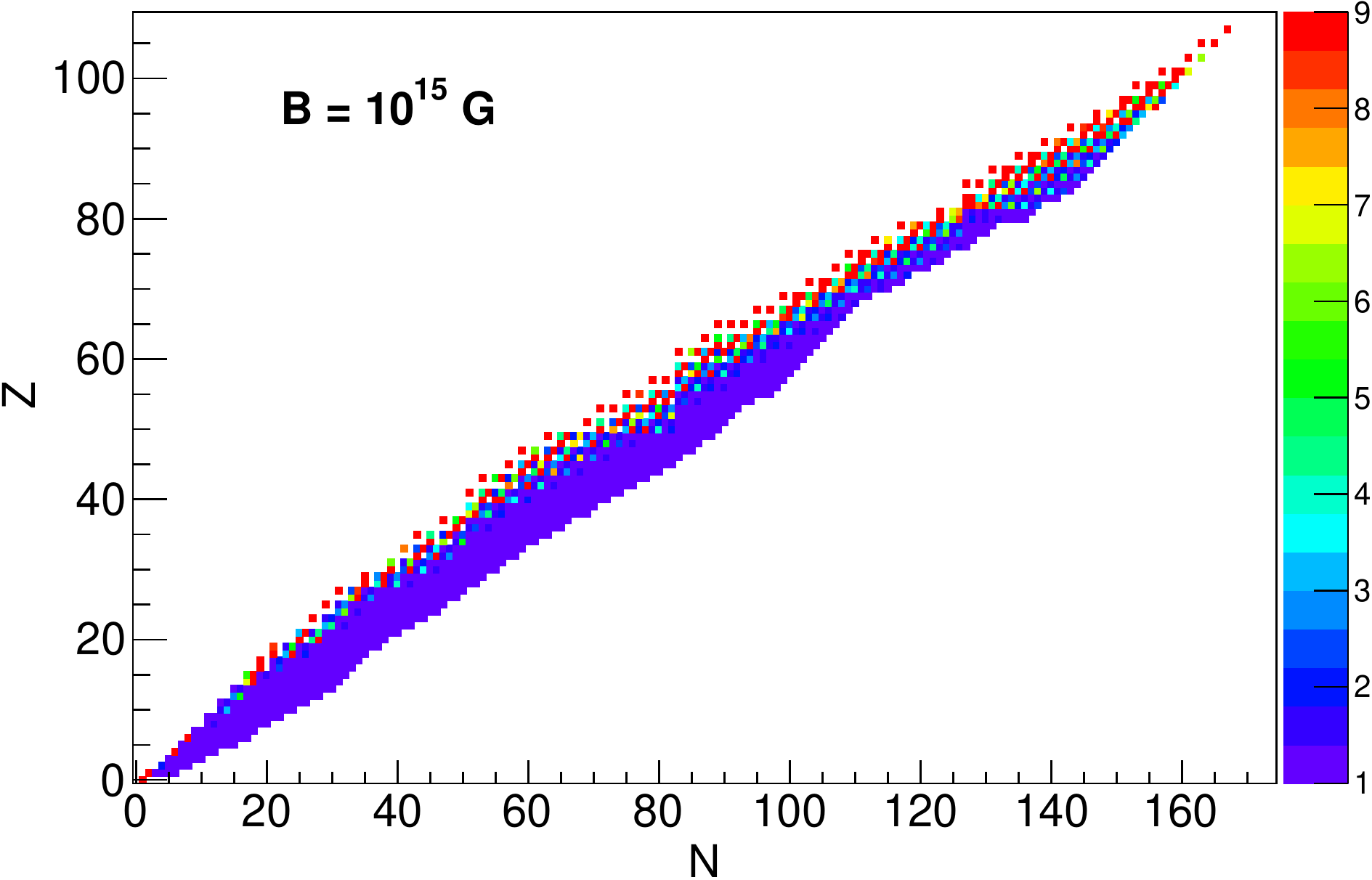}{0.49\textwidth}{(a)}
\fig{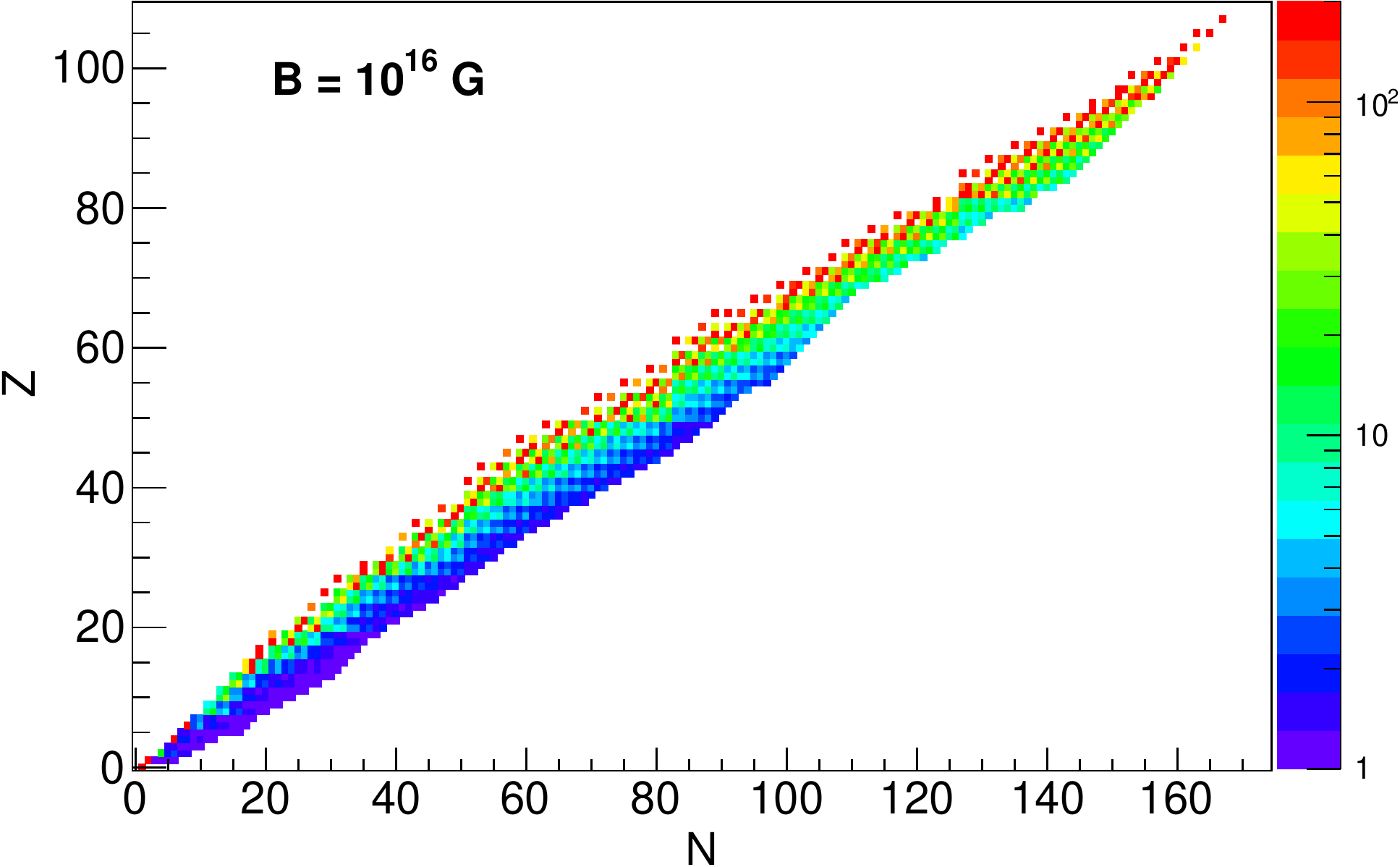}{0.49\textwidth}{(b)}
}
\gridline{
\fig{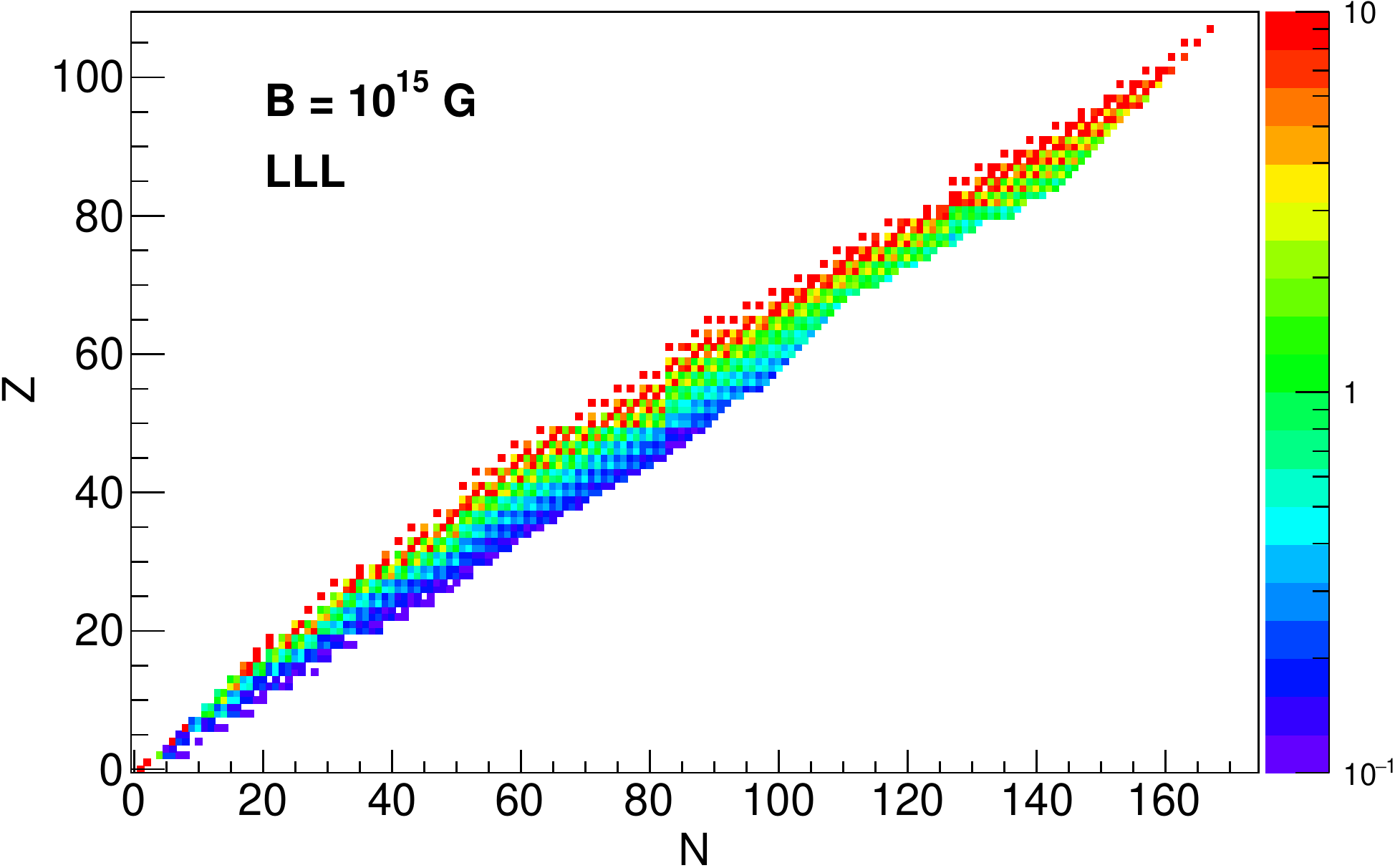}{0.49\textwidth}{(c)}
\fig{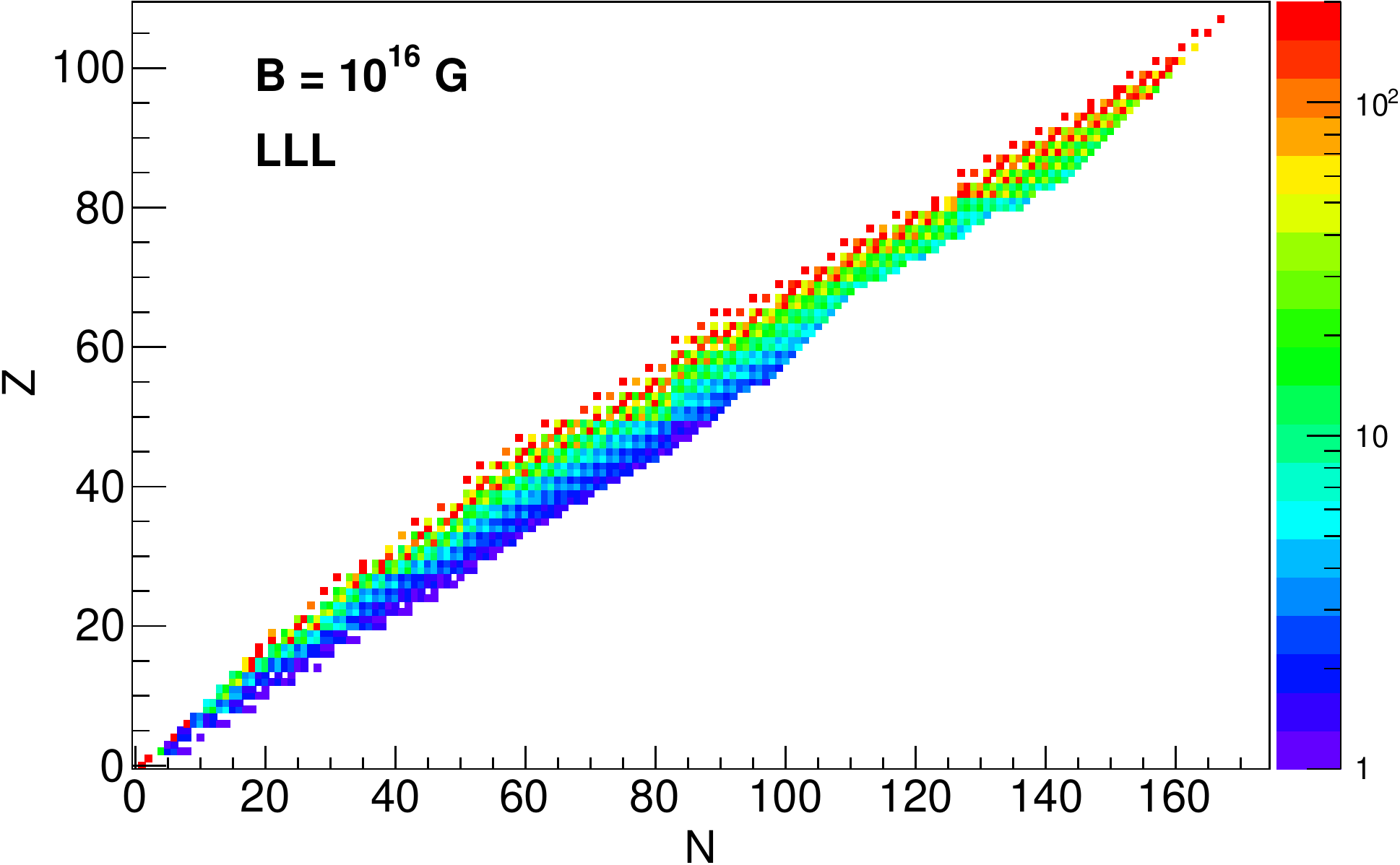}{0.49\textwidth}{(d)}
}
    \caption{Ratio of $\beta^-$ decay rates for decays of nuclei unstable
    against $\beta^-$ decay in a non-zero
    field to those in a zero field, $\Gamma(B\ne 0)/\Gamma(B=0)$ for magnetic
    fields $B=10^{15}$ G (a,c)
    and $B=10^{16}$ G (b,d). The top row corresponds to 
    ratios for which all relevant Landau levels are included in the
    decay calculation, while the bottom row is for calculations for which 
    only the lowest
    Landau level is included in the calculations. In all figures, $T_9 = 2$,
    $\rho Y_e = 500$ g cm$^{-3}$.  Note the difference in scales in each figure.}
    \label{rate_ratios}
\end{figure}

In this figure, four cases are shown for each combination of two $Q$ values of 3 
MeV and 12 MeV and two cases of magnetic field of 10$^{15}$ G and 10$^{16}$ G. 
For the low $Q$ value of 3 MeV, the electrons can only be emitted into the lowest
Landau level for both fields, and the sum in Equation 12 consists of only the $n=0$ term;
$N_{max} = 0$.  However, at a higher $Q$ value of 12 MeV, the electron
can be emitted into any of a number of Landau levels.  For example,
an electron emitted with an energy of 6 MeV could fall into the $N=0$, 1, or
2 Landau level.  The integration is thus a sum over all Landau levels up to
the maximum possible Landau level within the $\beta$ spectrum; $N_{max}=11$ 
in this case.  For a field of 10$^{15}$ G, the Landau level spacing 
$\sqrt{eB}=2.43$ MeV, which is less than the decay $Q$ value, so multiple 
Landau levels contribute to the $\beta$ spectrum.

For a higher field of 10$^{16}$ G, with a Landau level spacing of 7.69 MeV, 
even at high $\beta^-$ decay $Q$ values, only a few (or one) Landau levels 
can be occupied by the emitted electrons.  Further, as indicated
in Figure \ref{beta_spec}, for decay spectra that occupy very 
few Landau levels,
the integrated spectrum, which is proportional to the total decay rate, can
be significantly higher than the zero-field spectrum.  

The relationship between the Landau level spacing and the $\beta$-decay $Q$ value
is important in considering the astrophysical r process.  Because
the r process proceeds along a path of potentially very neutron-rich nuclei,
the $\beta^-$ decay $Q$ values can be quite large, $\sim$ 10 MeV.  Thus,
for an r process in a high-field environment, the
decay rates could be quite sensitive to the field.  However, because the $Q$ values are large,
one cannot necessarily assume that the decay rates can be computed with
just the LLL approximation.

The influence of high magnetic fields on $\beta^-$ decay is shown in Figure 
\ref{rate_ratios} for two assumptions of the magnetic field and two assumptions of Landau levels (whether the LLL approximation
is used or not)
at a temperature $T_9 = 2$ and $\rho Y_e = 500$.  Here, the ratios of decay 
rates in a
non-zero field to those in  a zero field $\Gamma(B\ne0)/\Gamma(B=0)$ 
are plotted for each $\beta^-$ unstable nucleus with $Q$ values
taken from the AME2016 evaluation \citep{ame2016}.  

Several findings are noted in this figure.  First, for nuclei closer to 
stability, the $Q$ values are much lower, and the rate ratio is higher. This
is because electrons are emitted in only a few (or one) Landau levels.
These nuclei would correspond to the schematic cases of 
Figures \ref{beta_spec}a and c.

For the higher field of 
10$^{16}$ G, the figures for the LLL assumption and the assumption for all 
relevant Landau levels are very similar, indicating that the LLL is the primary
contributor to the electron spectra in $\beta^-$ decay for all nuclei
at this field strength. At this field, the difference between the zero-field 
and non-zero-field computations is
significant, and the increase in rates is much higher.
However, for a field of 10$^{15}$ G, inclusion of
only the LLL underestimates the total rate.  Including all relevant Landau 
levels
in the rate computation is necessary.

For more neutron-rich nuclei, more Landau levels are filled by the emitted
electron, and the $\beta$ spectrum more closely matches the zero-field
spectrum.   Thus, the ratio approaches unity.  This would correspond
to the case represented schematically in Figure \ref{beta_spec}b.

For a higher field, 
the ratio is close to unity only for the most neutron-rich nuclei, where
the $Q$ values are high enough fill multiple Landau levels in
the decay.  For the
nuclei closer to stability, the $Q$ values are low enough that only a single
Landau level is filled by the ejected electron, resulting in a 
decay spectrum that is significantly different than the zero-field case.
For the $B=10^{16}$ G case for nuclei close to
stability, the larger rates would correspond to the decay
spectrum represented schematically in Figure \ref{beta_spec}c.
\section{Effects of External Magnetic Fields in r-Process Nucleosynthesis}
As an example, r-process nucleosynthesis in a collapsar jet trajectory is examined.  
It is 
thought that the 
magnetic fields associated with collapsar 
jets and neutron star mergers (NSMs) could be as 
high as 10$^{16}$ G \citep{nakamura15, kiuchi15, kiuchi14,takiwaki09}.  Such strong
fields 
are formed by amplifying initially weak fields associated with the accretion 
region.  While these fields may be near
the surface of the objects, these will be considered
as a possible upper limit in nucleosynthesis associated with
collapsars and NSMs.  Within the actual jet region in this model,
fields have been computed to be $\sim 10^{12 - 14}$ G \citep{harikae}.  
Other 
evaluations of magnetic fields in collapsars or neutron star mergers
have resulted in similar fields near the surface or the accretion
disk, with some estimates up to and exceeding 10$^{17}$ G \citep{price,ruiz}.
While the field in the actual nucleosynthesis site may vary significantly,
a few field cases are examined here to show the field magnitudes necessary
to result in significant differences in the final r-process abundance
distribution. Some of the fields investigated in
the r-process nucleosynthesis studied here may very well exceed 
realistic values or those in nature and are thus illustrative in 
conveying field-strength effects in nucleosynthesis processes.
Temperature effects, on the other hand are computed for the actual computed
environmental temperature of the r-process site.
Here, the
effects of Coulomb screening in the early stages of the r process as well as
the
effects from the enhancement of weak interaction rates by the external field
are examined. 

Several nucleosynthesis scenarios are investigated to evaluate the effects 
on 
r-process nucleosynthesis.  These scenarios are listed in Table 
\ref{screen_mag_models}, where the notation X(F)$_{\mbox{log}B}$ is used; the label `X' refers to a specific screening and weak interaction treatment at a field $B$, and `F' indicates the inclusion of fission cycling or not.  
For example, model A$_{14}$ is model A at a magnetic field of 10$^{14}$ G without fission cycling while model
AF$_{14}$ is the same model with fission cycling included.
The various models summarized are:
\begin{itemize}
    \item No Coulomb screening and no magnetic field effects. (Models A$_{\mbox{log}B}$ and AF$_{\mbox{log}B}$)
    \item Default classical screening in which weak screening is determined by electrons in a Maxwell distribution \citep{jancovici77,itoh79}. 
     (Models B$_{\mbox{log}B}$ and BF$_{\mbox{log}B}$.)
    \item Relativistic screening in which the weak screening TF length is
    determined from electrons in an ideal Fermi gas \citep{famiano16}.
     (Models C$_{\mbox{log}B}$ and CF$_{\mbox{log}B}$.)
    \item Relativistic screening including effects on the TF length from an 
    external magnetic field on the Fermi gas \citep{luo20}.
     (Models D$_{\mbox{log}B}$ and DF$_{\mbox{log}B}$.)
    \item Relativistic including effects on the TF length plus magnetic field effects on weak 
    interaction rates assuming
    the LLL approximation.  (Models E$_{\mbox{log}B}$ and EF$_{\mbox{log}B}$.)
    \item Relativistic including effects on the TF length plus effects on weak 
    interaction rates including
    all contributing Landau levels to the $\beta^-$ decays.  (Models F$_{\mbox{log}B}$ and FF$_{\mbox{log}B}$.)
\end{itemize}
In Table \ref{screen_mag_models}, the models indicated by $B=0$
are those for which the magnetic field effects are not included
in the evaluation of screening or weak interactions.  Model E includes
effects of the magnetic field on weak interactions, but only the LLL
approximation is used.  Model F includes weak interaction effects for all
relevant Landau levels in $\beta^-$ decays. 
\begin{table}
    \centering
        \caption{Models used to evaluate the effects of screening
        from temperature and magnetic fields as well as effects from
        magnetic fields on weak interactions.  For each model, the subscript is the magnetic field strength.  }
    \begin{tabular}{|c|c|c|}
    \hline
         \textbf{Model} &  \textbf{Screening} & \textbf{Weak Interactions}\\
         \hline
         A(F)$_{\mbox{log}B}$ &  None & $B = 0$\\
         B(F)$_{\mbox{log}B}$ &  Classical & $B = 0$\\
         C(F)$_{\mbox{log}B}$ &  Relativistic ($B=0$) & $B = 0$\\
         D(F)$_{\mbox{log}B}$ &  Relativistic (B$\ne$0) & $B = 0$\\
         E(F)$_{\mbox{log}B}$ &  Relativistic ($B\ne0$) & $B \ne 0$, LLL only\\
         F(F)$_{\mbox{log}B}$ &  Relativistic ($B\ne0$) & $B \ne 0$, All LL\\
         \hline
    \end{tabular}

    \label{screen_mag_models}
\end{table}

\begin{figure}[b]
\gridline{
\fig{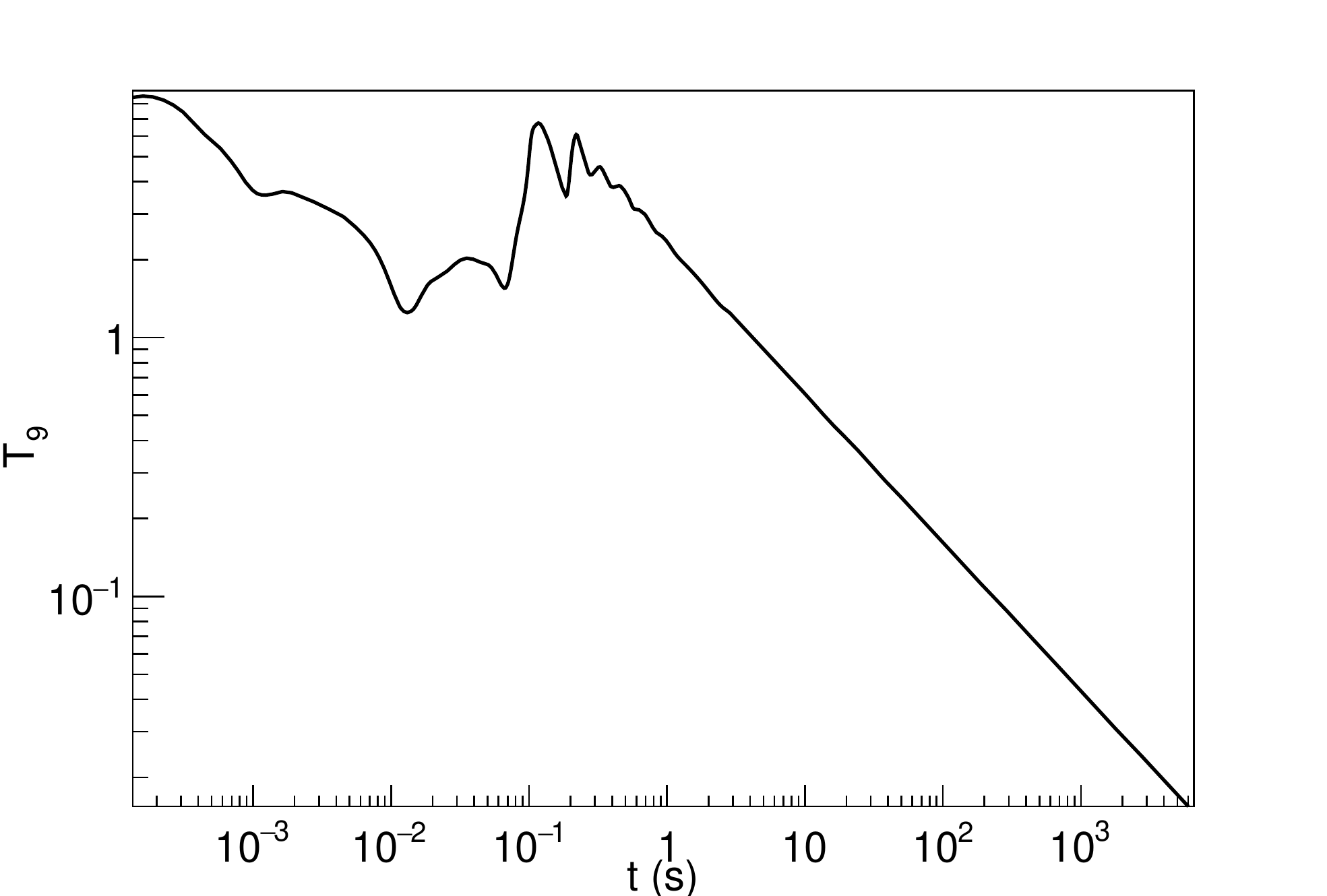}{0.49\textwidth}{(a)}
\fig{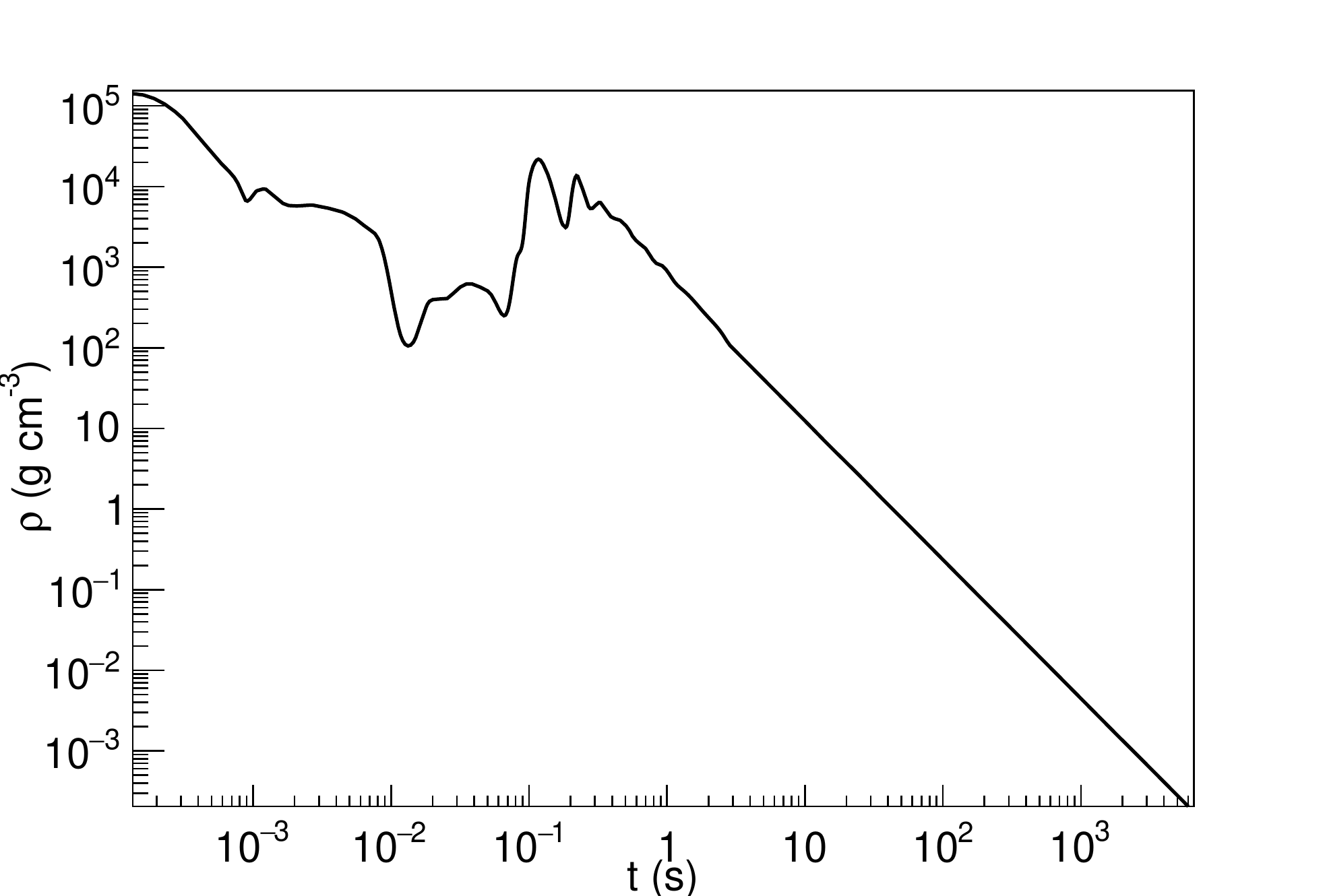}{0.49\textwidth}{(b)}
}
    \caption{Trajectory used for the MHD r-process nucleosynthesis
    calculation. (a) Temperatures, $T_9$. (b) Density.}
    \label{MHD_traj}
\end{figure}

In order to evaluate the effects of magnetic fields on screening and
weak interactions in a possibly highly magnetized plasma in
the r process, a single trajectory from the MHD jet model  of 
\citet{nakamura15} was used.  This trajectory is shown in Figure
\ref{MHD_traj}.  Several values of a static, external magnetic field were 
evaluated. 
Because the field may not be well understood in
many sites, this evaluation is taken to be qualitative only as
a demonstration of the magnitude of the effects of strong
external fields in nucleosynthetic sites.  Nucleosynthesis in
static fields, $14\le \mbox{log}(B) \le 16$, was evaluated.  

For the r-process calculation, the initial composition
was assumed to be protons and neutrons with $Y_e = 0.05$
as given in \cite{nakamura15}.
The nuclear reaction network code \texttt{NucnetTools} \citep{libnucnet} was 
modified to include thermodynamic effects and screening effects at high 
temperature and magnetic fields.  
The reaction network was a
full network which was truncated at $Z = 98$.  

The weak interaction rates
were computed using the relationships in Equations \ref{mag_beta_minus} -- 
\ref{mag_beta_PC}.  These rates
are ground-state transitions only.  However,
the purpose of this initial evaluation is not an evaluation of accurate
weak interaction rates, but a description of the effects 
of strong magnetic fields on nucleosynthetic processes.  If transitions
\textit{to} excited states are included, the rates are expected to be even
more sensitive to external fields because of the smaller transition
Q value relative to the Landau level spacing (Figure \ref{beta_spec}), while transitions \textit{from} excited states may be less sensitive as the Q values are larger, though one must also account for changes in transition order when including excited states.

The nucleosynthesis was computed to 6000 s. In order to do this, an 
extrapolation of the \cite{nakamura12} trajectories to low T and low $\rho$ was made 
because the published trajectories stop at 2.8s.
At low-enough temperatures and densities, neutron captures decline, and 
only  
$\beta$-decays and subsequent smoothing
ensues.  The temperature and density extrapolation was done 
 assuming an adiabatic expansion for
$t>2.8$ s, $\log(T_9)\propto \log(\rho) \propto \log(t)$. This extrapolation 
allows the temperature and density to drop significantly to follow 
the processing 
further
in time while examining effects from late-time fission cycling.  Clearly, 
there is still 
some nucleosynthesis during this phase, and this is used to evaluate 
long-term 
effects of the nucleosynthesis.  

To include screening effects, relativistic weak screening was used for 
$T_9>0.3$.  For lower temperatures, the classical
Debye-H\"uckel screening was used in models C - F (see Figure \ref{debye_fig}).  In model B, 
classical Debye-H\"uckel screening was used for all temperatures.  
For the strong magnetic field, the 
Thomas-Fermi length of 
Equation \ref{EM_p_b} was used. For weaker fields, the difference 
between the screening lengths
for the relativistic case at $B=0$ and at $B\ne 0$ is negligible as shown 
in 
Figure \ref{low_p_high_b_compare}. 
Thus, to improve the speed of the network calculations, the LLL
approximation 
was assumed with the 
expansion of Equation \ref{EM_p_b}.  In order to determine whether to use
the 
LLL approximation or the 
thermal screening length (with $B=0$), the inverse screening 
length, $k\propto 
1/\lambda$, was computed in each case, and the 
maximum value was used:
\begin{equation}
    k \rightarrow \mbox{max}\left[k(B=0),k(B\ne 0)\right]
\end{equation}
The resultant corresponding screening length is then determined by Equation 
\ref{EM_p_b} at high fields 
and the relativistic length computed in prior work \citep{famiano16} at lower 
fields.  Certainly, there is a 
small transition region between the low-field and the high-field values shown 
in Figure \ref{low_p_high_b_compare}
where the screening length is overestimated slightly.   In this region, the 
screening length could be overestimated by as much at $\sim$15\%, with a 
resultant shift in the overall reaction rates of about 15\%.  This can be 
corrected by relaxing the LLL approximation and including as few as 10 
Landau levels in the
length calculation.  However, it is ignored in this evaluation because the 
correction is small compared to the
change in screening length from the magnetic field.  The r-process is not 
expected to be dominated by screening as it is primarily a neutron capture 
process, and the time spent in this transition region for the r-process is 
expected to be brief compared to the entire r-process.  Future, more accurate 
evaluations may include this small correction. 

\begin{figure}
    \centering
    \includegraphics[width=\textwidth]{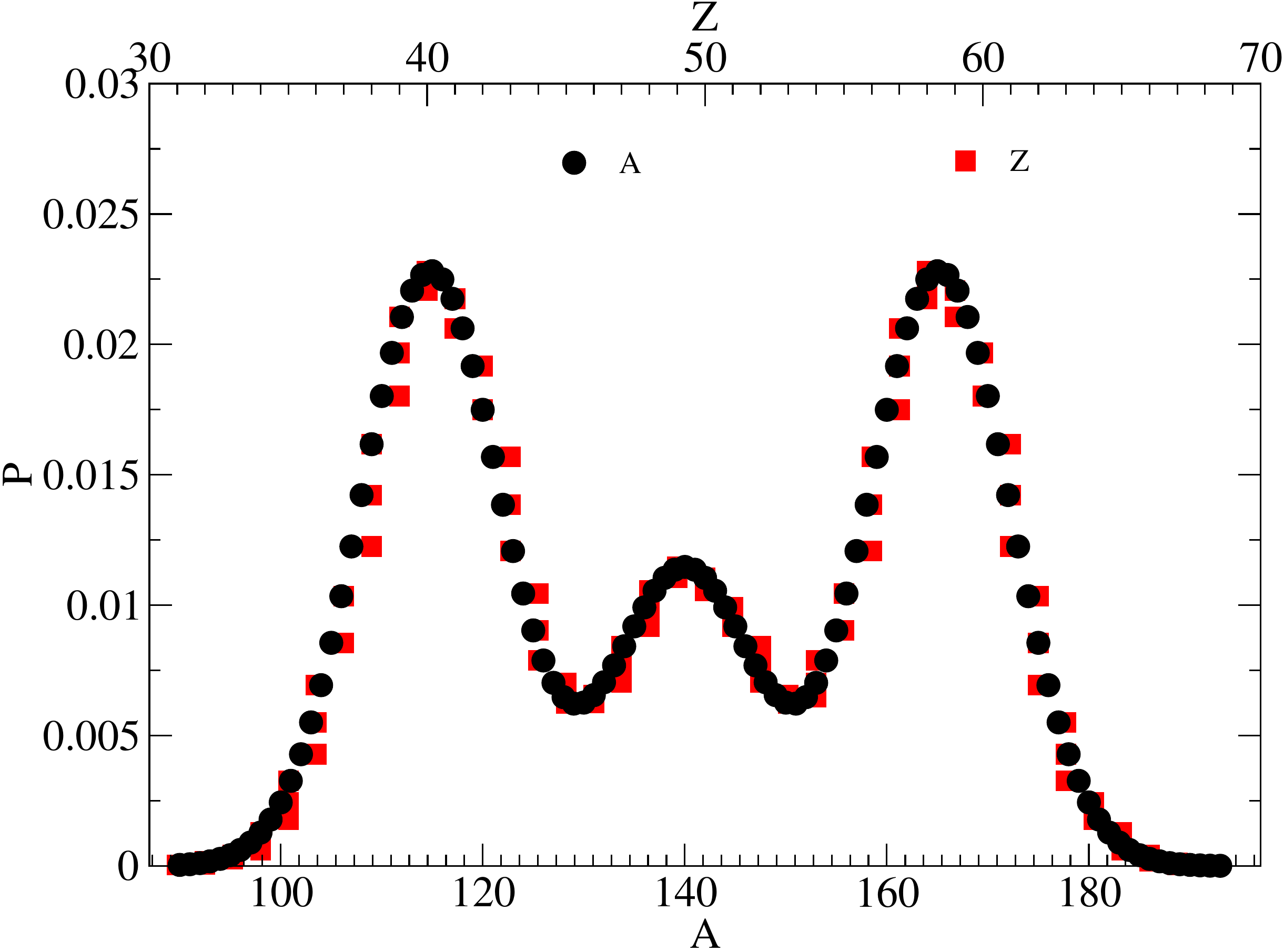}
    \caption{Fission probability distribution in 
    product mass (black circles) and $Z$ (red squares) for
    $^{282}$Cf showing the trimodal structure, which results from a combination
    of symmetric and asymmetric
    fission.}
    \label{fission_dist}
\end{figure}

Effects from fission cycling were included in a rudimentary 
fashion following 
the prescription of
Shibagaki et al. \citep{shibagaki16}. In this model,
fission was implemented for the Cf isotopic chain, $^{270 - 295}$Cf.
Fission rates were assumed to be 100 s$^{-1}$ for all nuclei in this isotopic 
chain.  The fission parameters in the \cite{shibagaki16} model
are shown in Table \ref{fission_parms}.  With this parameter
set, the fission distribution for $^{282}$Cf is 
shown in Figure \ref{fission_dist}.  Clearly, this fission
model is overly simplistic and does not represent the 
full details of the nuclear structure necessary for a proper
determination of fission.  However, as we will discuss later, it is
necessary to include fission in a collapsar/NSM r process, and this model
provides an appropriate level of detail to capture the overall
effects of intense magnetic fields on $\beta$ decays in this site.
Fission of
the Cf isotopic chain here is meant to replace 
neutron-induced fission, $\beta$-delayed fission, and 
spontaneous fission of all fissile 
nuclei produced in the r-process path.  As such, the fission 
product 
distribution can be contrasted with that developed using
more accurate models.  For example, the
evaluation of fission using the GEF 2016 and FREYA models 
\citep{vassh2019} predicts similar neutron emission
in fission, though the fission product distribution for
the Cf nuclei is generally symmetric for spontaneous fission 
with asymmetric components for neutron-induced fission.  
Fission induced by $\beta$-decay of the Cf chain 
has been predicted to be predominantly symmetric for $N>180$
with asymmetric components at lower mass \citep{vassh2019,kodama75}.

\begin{table}
    \caption{Fission parameters used in this evaluation.  Fission model 
    taken from \citet{shibagaki16}}
    \label{fission_parms}
    \centering
    \begin{tabular}{|c|c|c|}
    \hline
         \textbf{Parameter}& \textbf{Description} & \textbf{Value} \\
         \hline
         \hline
        $W_{i}$ & Intermediate fragment probability & 0.2\\\hline
        $W_{H/L}$ & Heavy/Light fragment probability & 0.4 \\\hline
        $N_{loss}$ & Average neutrons/fission & 2 \\\hline
        $\sigma$ & Width of fragment distribution & 7 \\\hline
        $\alpha$ & Relative difference of centroids of fragment distributions & 0.18 \\\hline
    \end{tabular}
\end{table}
\subsection{r-Process Abundance Distributions}
\begin{figure}
\gridline{
\fig{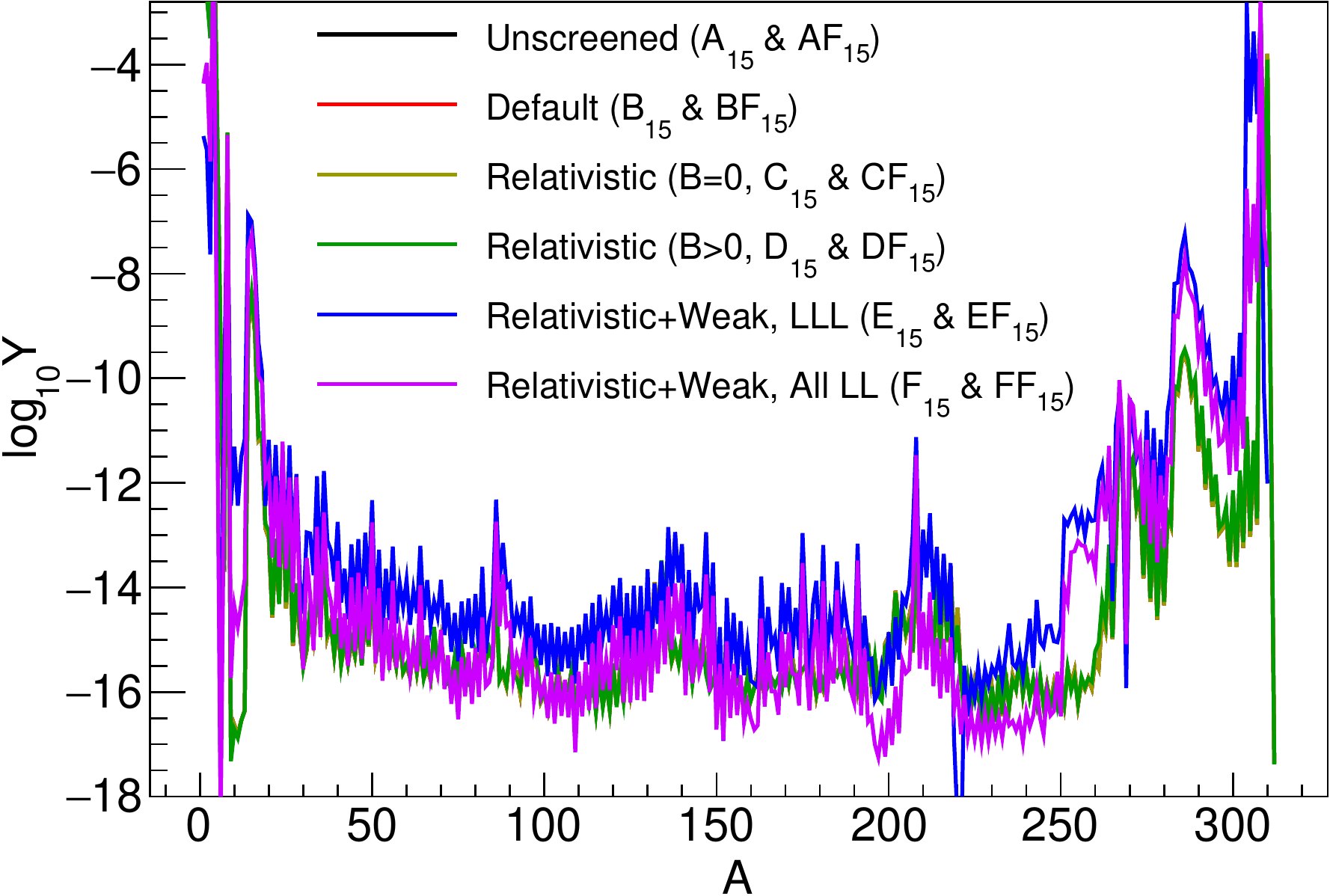}{0.48\textwidth}{(a)}
\fig{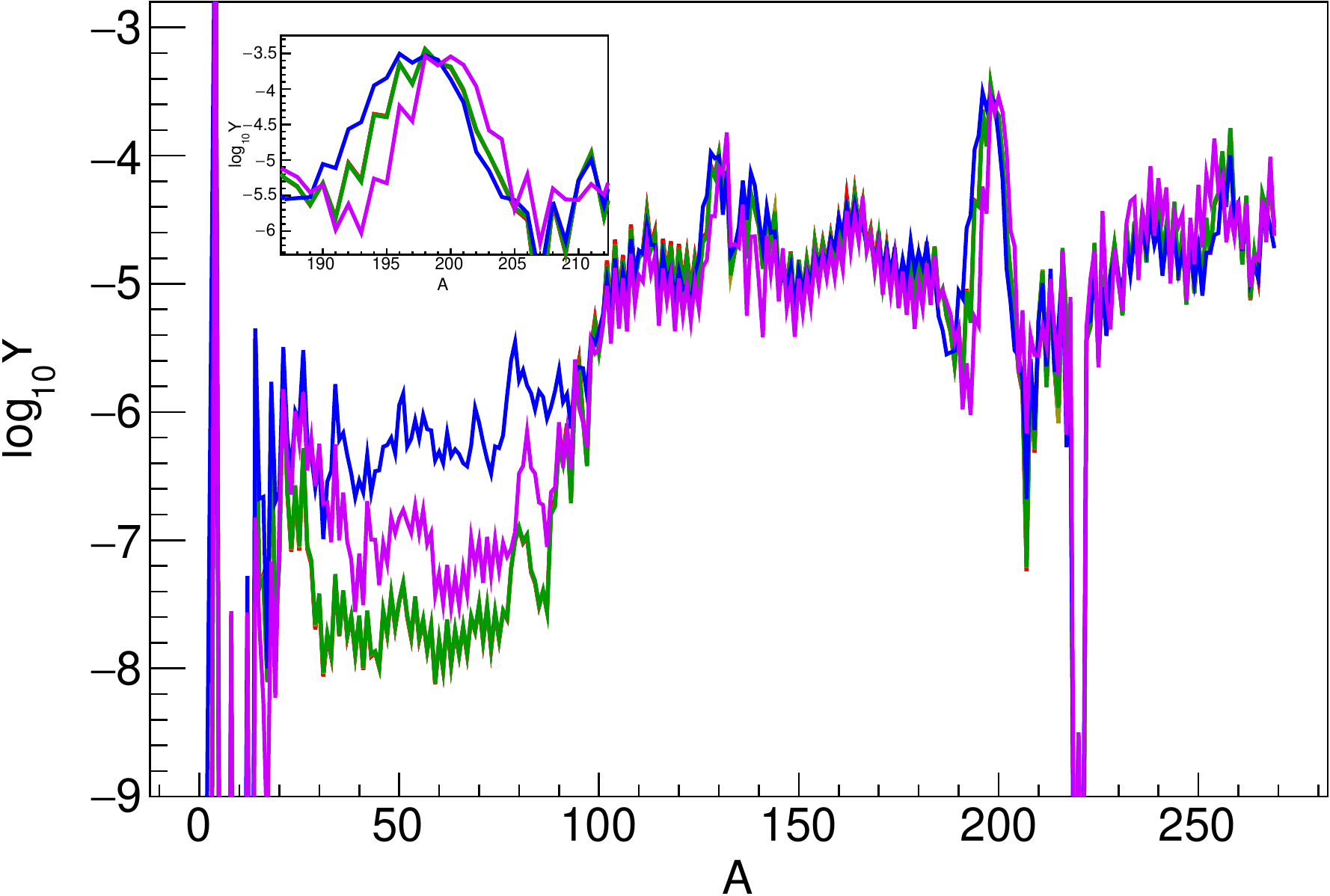}{0.48\textwidth}{(b)}
}
\gridline{
\fig{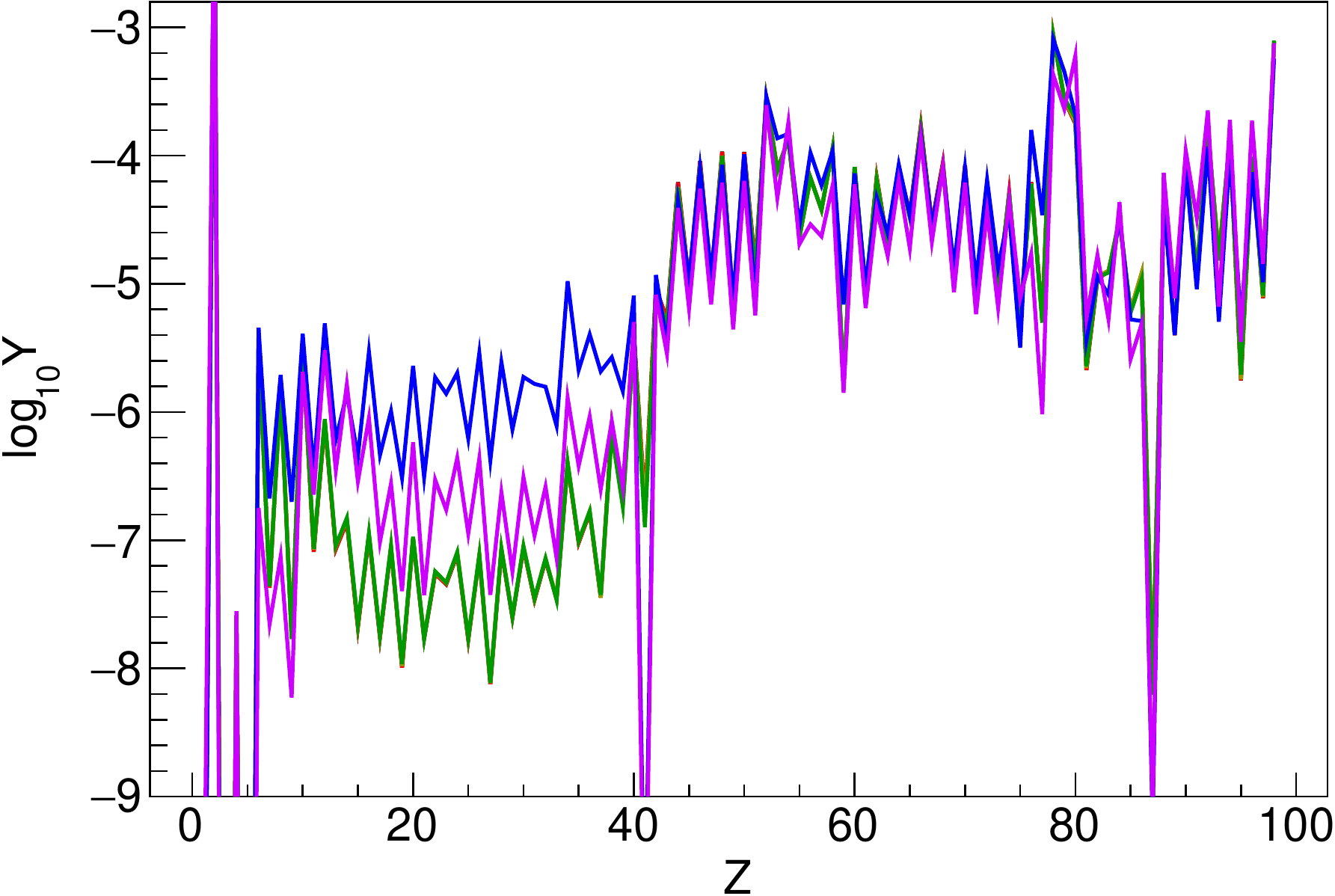}{0.48\textwidth}{(c)}
\fig{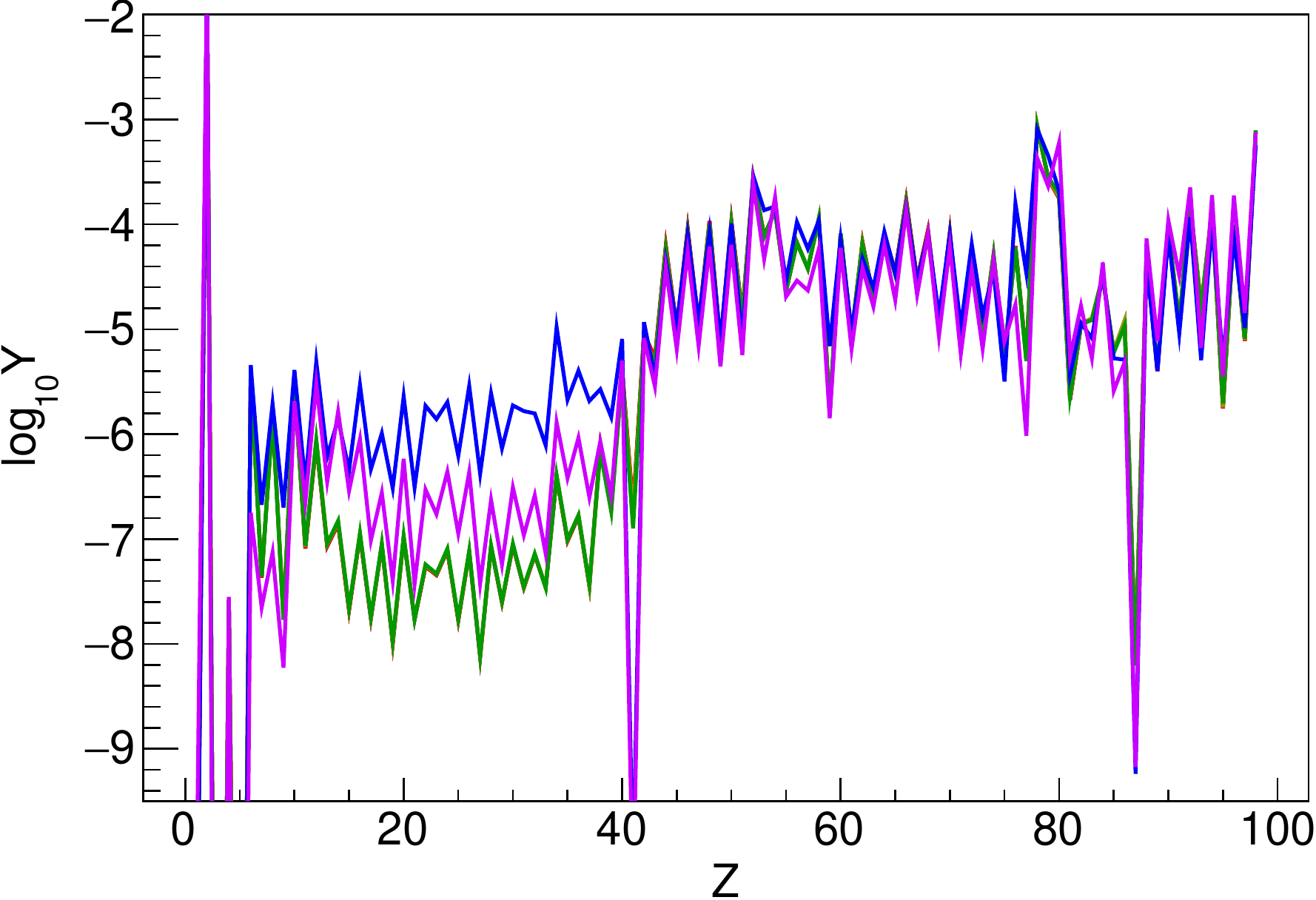}{0.48\textwidth}{(d)}
}
    \caption{Abundances at $t=6000$ s for MHD models for the
    adiabatic trajectory in Figure \ref{MHD_traj} with an external field of 10$^{15}$ G. 
    Plots (a) and (c) show nucleosynthesis results without fission, and plots (b) and (d)
    shows nucleosynthesis results with fission.
 For the models with fission, the points for
    the default screening model nearly coincide with those for the 
    unscreened model, and the points for the relativistic screening
    model for $B=0$ nearly coincide with those for the relativistic 
    screening model with $B=10^{15}$ G.}
    \label{MHD_abun}
\end{figure}

\begin{figure}
\gridline{
\fig{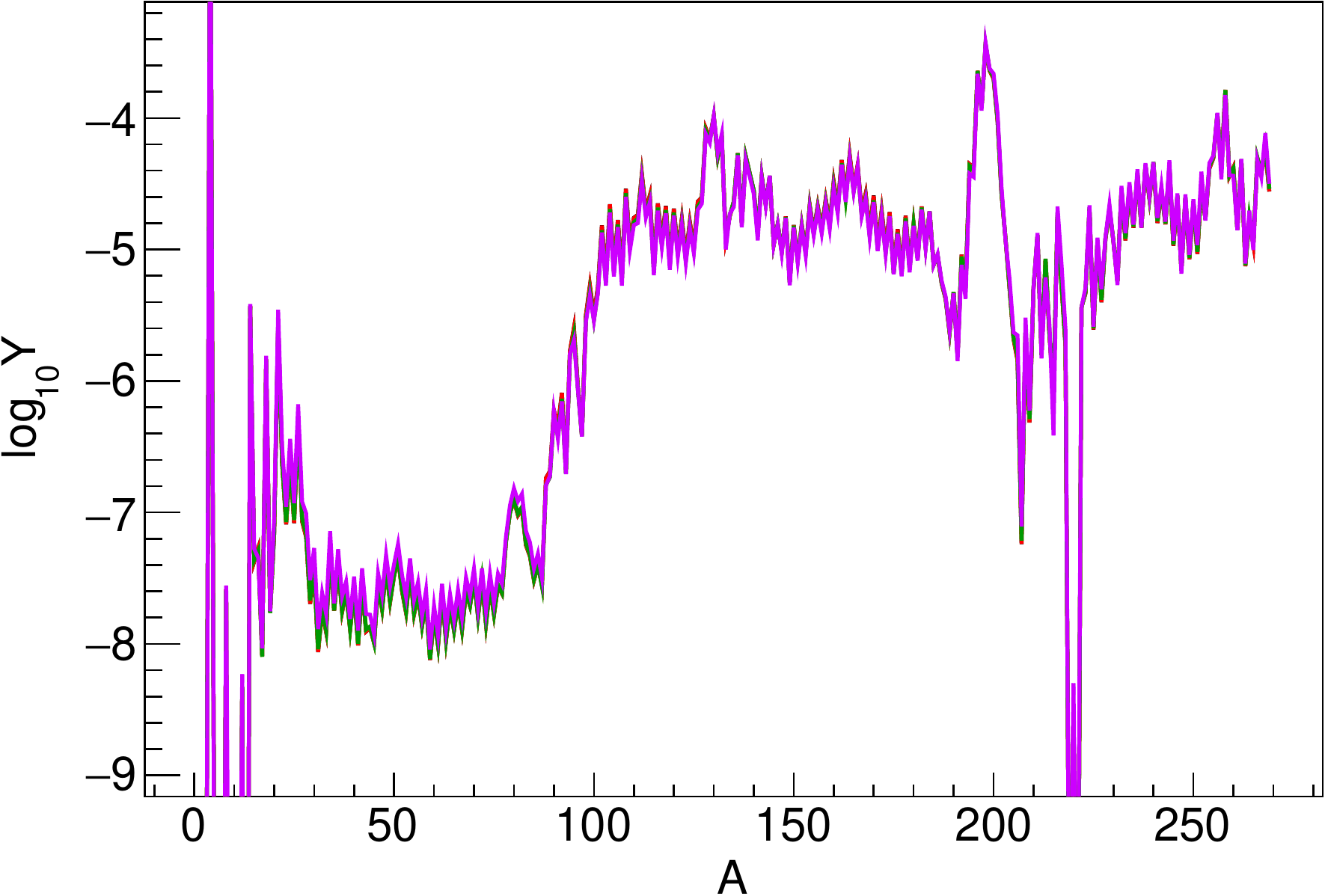}{0.49\textwidth}{(a)}
\fig{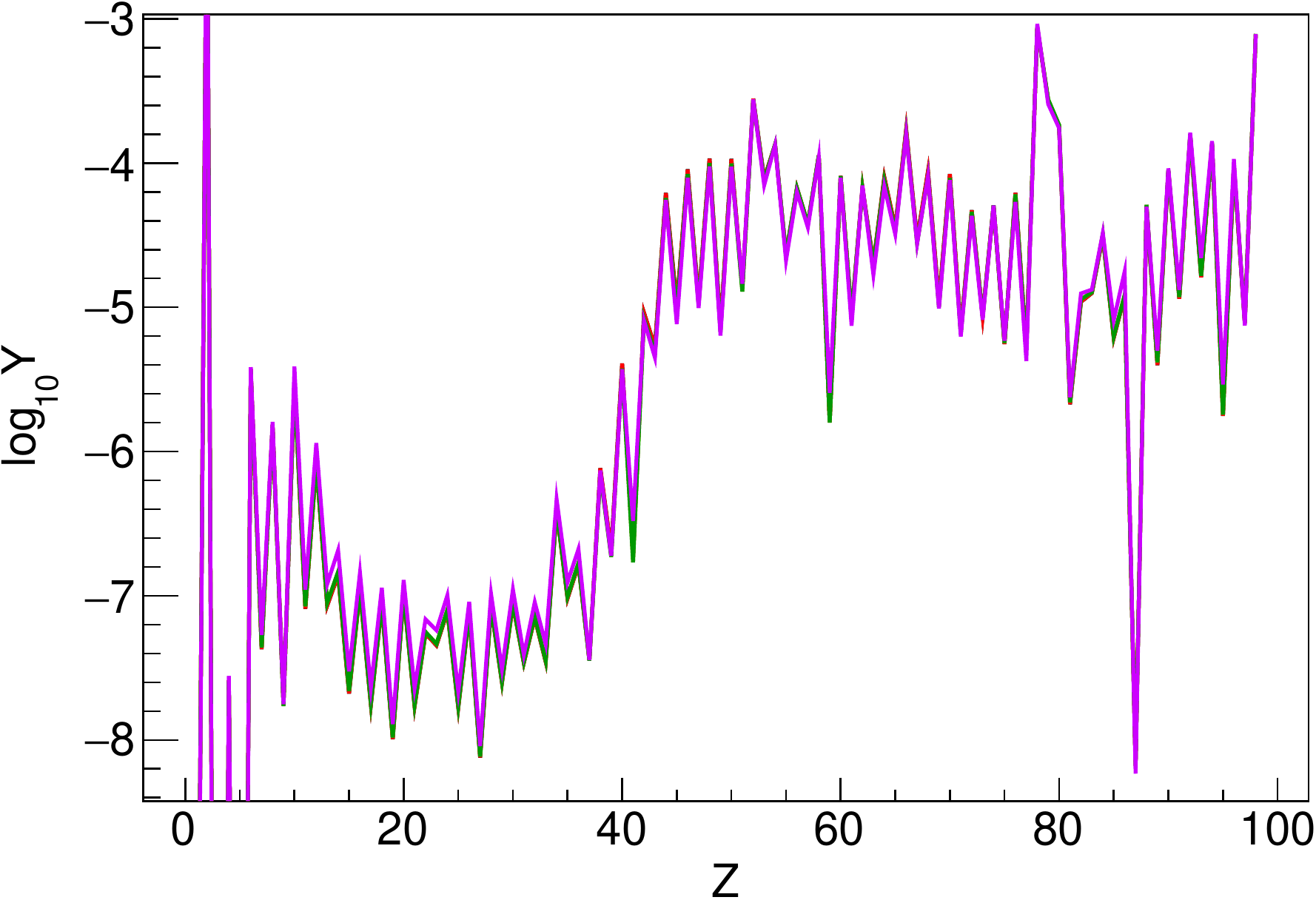}{0.49\textwidth}{(b)}
}
    \caption{Abundances at $t=6000$ s for MHD models for the
    adiabatic trajectory in Figure \ref{MHD_traj} for a field of 10$^{14}$ G 
    including fission.  
    The colors are the same as those in Figure \ref{MHD_abun}.
    The
    LLL approximation is not shown.  }
    \label{MHD_abun_10}
\end{figure}

\begin{figure}
\gridline{
\fig{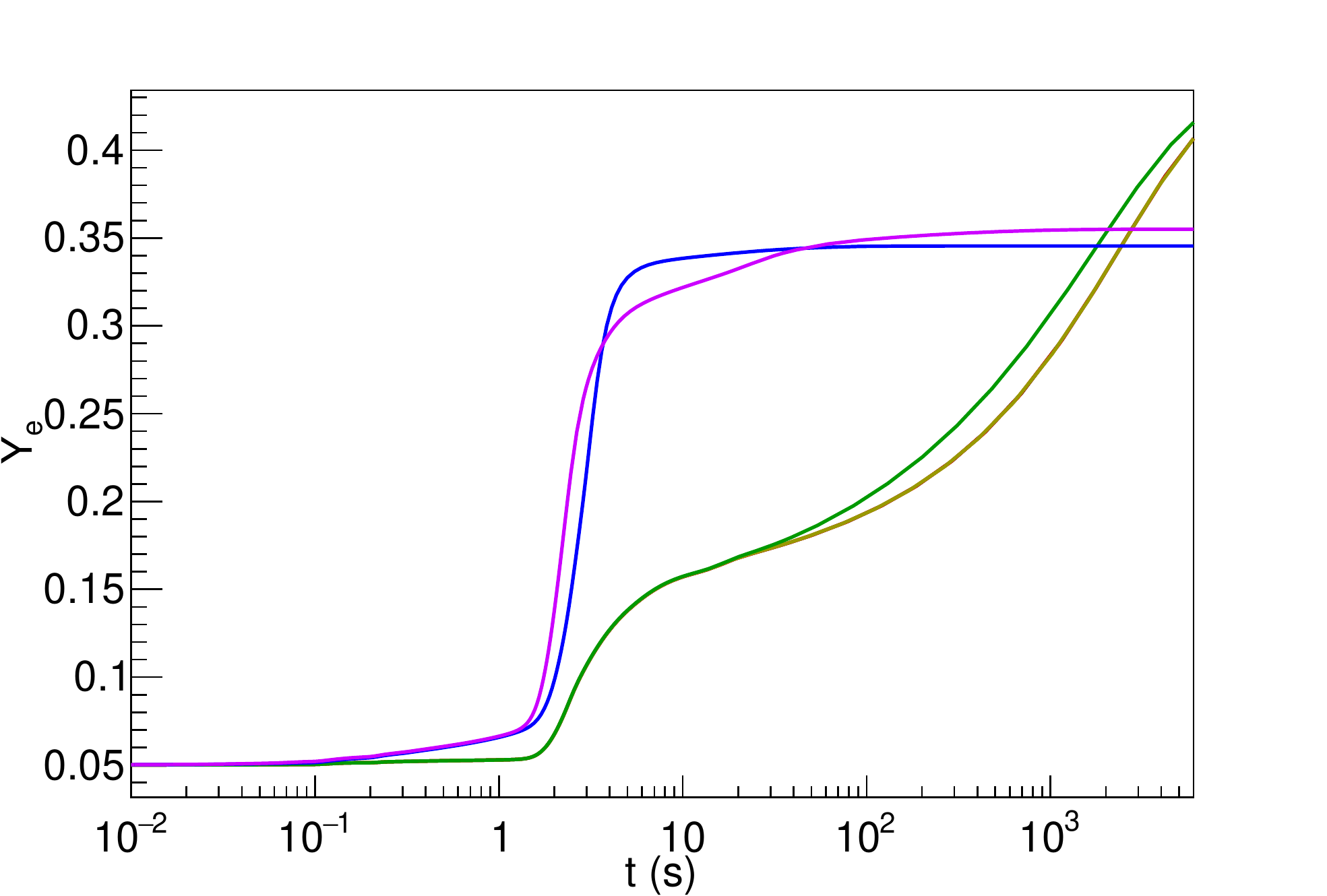}{0.49\textwidth}{(a)}
\fig{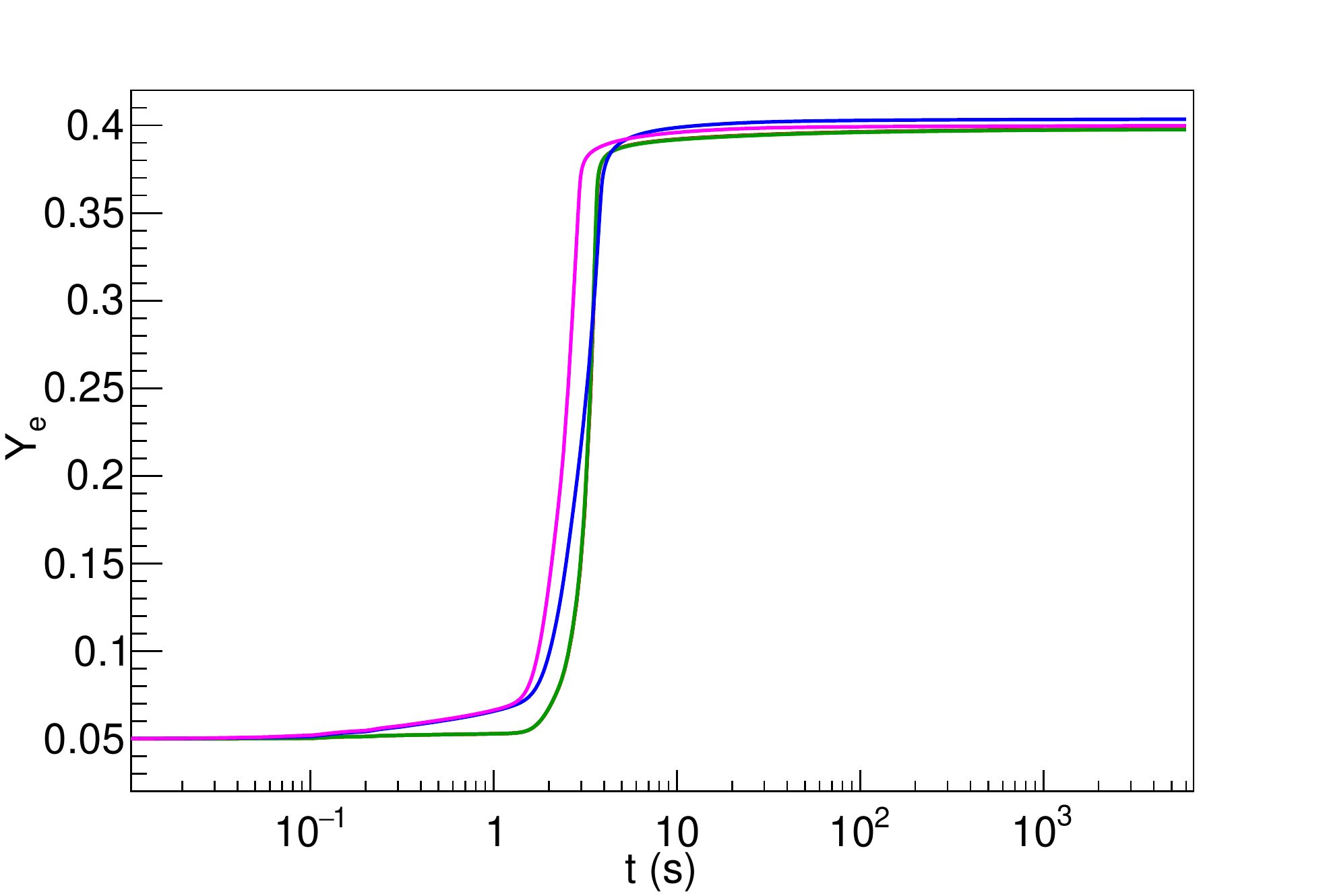}{0.49\textwidth}{(b)}
}
\gridline{
%\fig{legend-crop.pdf}{0.5\textwidth}{}
}
    \caption{Electron fractions as a function of time for 
    trajectories without fission (a) and with fission (b). In both 
    figures the lines for no screening, default screening, and relativistic ($B=10^{15}$ G) screening coincide.  In the right figure, screening
    with enhanced weak interactions deviates from the other models. The colors are the same as those shown in Figure \ref{MHD_abun}.}
    \label{MHD_ye}
\end{figure}

The final abundance distributions for all six models studied with and without fission
are shown in Figure \ref{MHD_abun} for a field of 10$^{15}$ G.  
Figure \ref{MHD_abun_10} shows the final
abundance distributions for models including fission at a field 
of 10$^{14}$ G.  (All models except E$_{14}$ and EF$_{14}$ shown in Figure \ref{MHD_abun_10}.)
The electron fraction $Y_e$ is plotted for
all models in Figure \ref{MHD_ye}.

In all cases, Coulomb screening of nuclear reactions has a minimal effect on the overall reaction network. This is not surprising as the primary fusion
reaction is neutron capture, which is immune to screening.  While the inclusion
of magnetic fields creates a slight enhancement in the overall abundance for the 
heavier nuclei due to the enhancement of charged-particle reactions 
early in the r process (e.g., proton and alpha captures), this enhancement is
minimal.  Likewise, the effects from default screening and 
relativistic screening are negligible in this treatment.

However, the
inclusion of enhanced weak rates does have an effect on the overall
resultant reactions.  For a full treatment, including accurate
computations of the weak rates with contributions from all relevant 
Landau levels, the overall $\beta^-$ rates are higher, resulting
in a more rapid progression to the heaviest nuclei.  As can be seen in the
case for no fission in Figure \ref{MHD_abun}, the rapid $\beta$-decay rates
results in a large abundance of nuclei near the endpoint of the reaction 
network ($Z=98$).  The nucleosynthesis progresses to the Cf
isotopic chain, where the abundance builds up.  At this point, the only possible
reactions are (n,$\gamma$), (n,$\alpha$), neutron-induced fission, and photospallation reactions as a result of truncating the network at Z=98.  This results in additional neutron
production and minimal production of $\alpha$ particles.  Of course, this is
an unrealistic scenario because of the artificial termination point in 
the nucleosynthesis, but it does convey the increased nucleosynthesis
speed from the high magnetic field in a very neutron-rich environment.

The LLL approximation for $\beta^-$ decay rates is also shown in this 
figure. In this case, the Landau level spacing is generally less than the 
decay Q value, except for a few low mass nuclei with $Z\lesssim20$.  This results
in overall slower $\beta$ decay rates, resulting in a slower progress to the heavy mass nuclei and a larger relative abundance at the low mass nuclei. 

The right side of Figure \ref{MHD_abun} shows the final abundance
distributions if fission cycling is included in the network calculation.
As expected, there is very little difference between the 
abundance distributions if nuclear screening is included in the 
reaction network.  However, the inclusion of 
$\beta^-$ decay enhancement results in a an enhancement of
the low-mass nuclei, $(Z,A)\lesssim (40,100)$.  For the heavier mass
nuclei, fission products dominate the abundance distribution.  
As fission becomes dominant, heavier-mass nuclei are enhanced
in abundance relative to that of the low-mass nuclei, and one 
notices a relative increase in abundance for $Z\gtrsim40$ for 
all models.

However, there is also an enhancement of the abundances of the low-mass
nuclei with field-enhanced decay rates relative to the abundances of
nuclei without them.  This is likely a result of the more rapid progression of 
the r-process to the fissile nuclei.  There are two effects that can
be considered in this case.  First, from Figure \ref{rate_ratios}, 
it can be seen that the enhancement of the $\beta^-$ decay rates
is less for lower mass nuclei than for the higher mass nuclei.  While this 
enhancement is small, it results in a somewhat slower progression of
the r-process through these lower-mass progenitors \textit{relative to} the progression through higher mass nuclei.  Thus, a slight
buildup of abundance relative to the high-mass nuclei can result.  This is 
particularly noticeable if only the LLL is taken into account.  The rate
differences are more pronounced, and the the enhancement of low-mass 
nuclear abundance is larger.

To a lesser extent, the neutrons produced in fission can also slightly enhance
the production of lower-mass nuclei.  It is assumed
that two neutrons are produced in each fission in this model.  
Because of the very large initial neutron abundance, the progression to
fission is not surprising in this scenario.  However, for the the case
in which decays are enhanced by the magnetic field, the progress to fissile nuclei 
is more rapid.  
Thus, more 
fission neutrons are produced in the r-process. These can be used
as fuel for subsequent processing.  Of course, 
neutrons produced in fission are captured by \textit{all} 
progenitor nuclei, and not just the low-mass nuclei.  
The slightly less-enhanced decay rates of the low-mass nuclei, on the other
hand, result in
an abundance that is likely even more enhanced than in the absence
of fission.

From Figure \ref{MHD_abun}, one also notes that there is a slight 
shift to higher mass in the final abundance distribution for
the field-enhanced case.  This is because the more rapid
decay rates result in a slight shift of the r-process path closer
to stability than in the case with zero field.  This shift is 
prominent at the abundance peaks.  
For an r-process path 
that is closer to stability, the path intersects the 
magic numbers at a higher mass, resulting in the slight shift
by a few mass units.
This is shown in the inset for the A$\sim$195 abundance peak
in Figure \ref{MHD_abun}b.

Given the prominent contribution to the final abundance distribution by fission, 
it it thus emphasized that -- in the collapsar model here -- fission
cycling is an integral part of r-process calculation.

An evaluation at a field of 10$^{14}$ G is shown in Figure \ref{MHD_abun_10}. 
In this figure, the abundances at $t=6000$ s for a calculation including fission 
are shown, and the LLL approximation has been
removed for clarity.  As expected, for the lower field, the decay rates are
closer to the zero field decay rates, and the overall shift in the abundance
distribution is smaller, though a small increase in abundance is noted for $A<100$.
 This trend
is consistent with the non-zero field trends observed but to a lesser extent.

The electron fraction as a function of time, $Y_e$, is shown for all
six models with and without fission cycling in Figure \ref{MHD_ye} at a field of
10$^{15}$ G.
For each case, it is observed that screening has a minimal
effect on the evolution of the electron fraction.  During the early stages of the r process, the high-temperature environment is in nuclear statistical
equilibrium (NSE).  As the environment cools and expands, reactions dominate with a small time window during
which charged-particle reactions (e.g., ($\alpha$,$\gamma$), 
($\alpha$,n), etc.) may occur.  These would
be affected by Coulomb screening.

Without
fission, the dominant contribution to $Y_e$ is from the Cf isotopic 
chain.  In the case of field-enhanced decay rates, because the
progression to the Cf chain is more rapid, an equilibrium $Y_e$ 
occurs very rapidly, with a more rapid progression if all Landau
levels are included in the decays, as expected.  It is also
noted that a complete inclusion of all Landau levels results in
a slightly higher equilibrium $Y_e$ as the r-process path is closer to 
stability.  For the other calculations, the $Y_e$ is lower as
the r-process path is more neutron-rich as explained previously.

Figure \ref{MHD_ye}b shows the evolution of $Y_e$ in the more
realistic case including fission cycling in the calculation. 
Here, as the r-process becomes dominated by fission products, the 
equilibrium $Y_e$ is similar in all cases.  However, it can be seen
that inclusion of the field-enhanced rates results in an earlier
rise in the electron fraction owing to a more rapid r process combined
with a more rapid decay to stability.
\begin{figure}
\gridline{
\fig{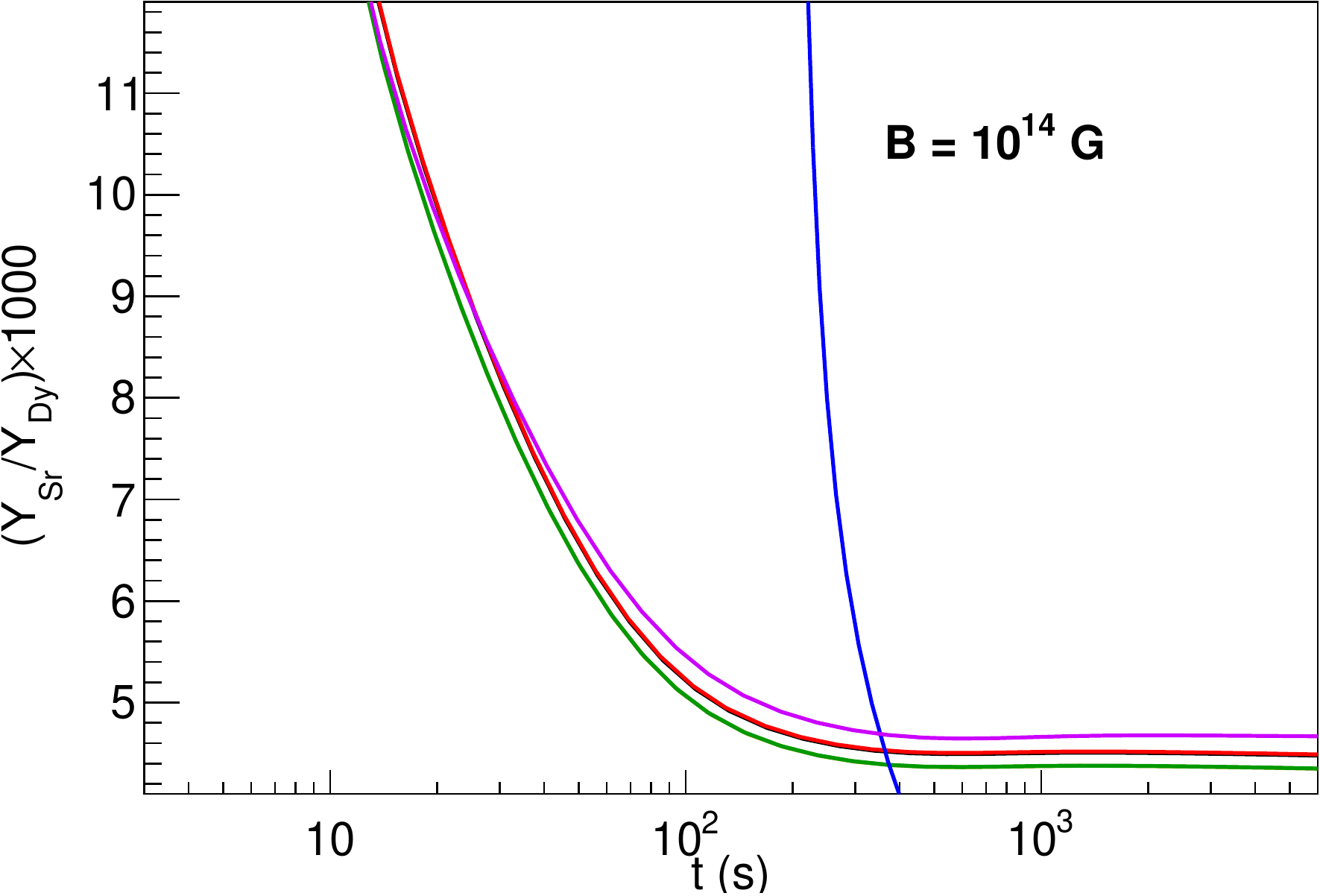}{0.49\textwidth}{(a)}
\fig{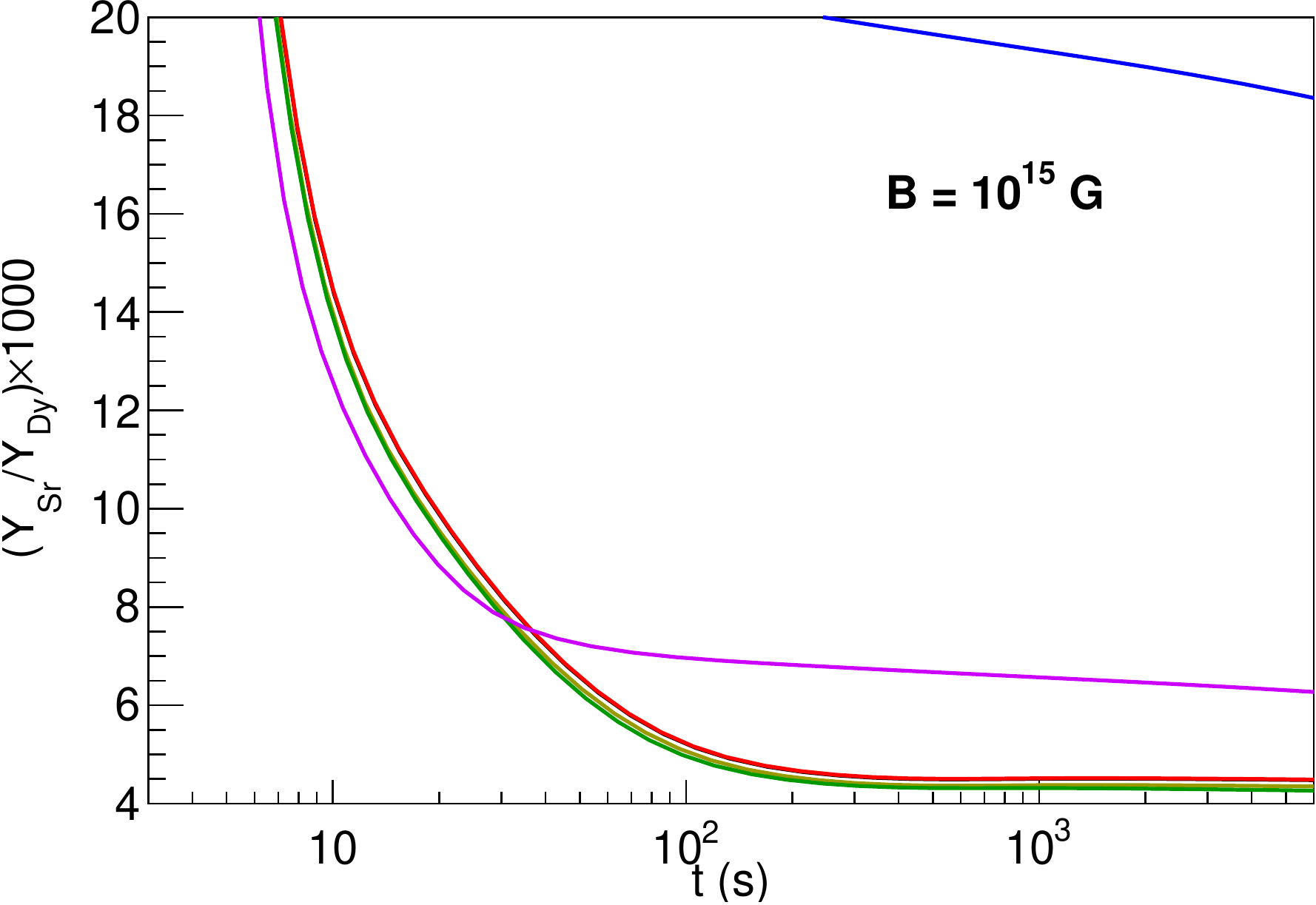}{0.49\textwidth}{(b)}
}
\gridline{
\fig{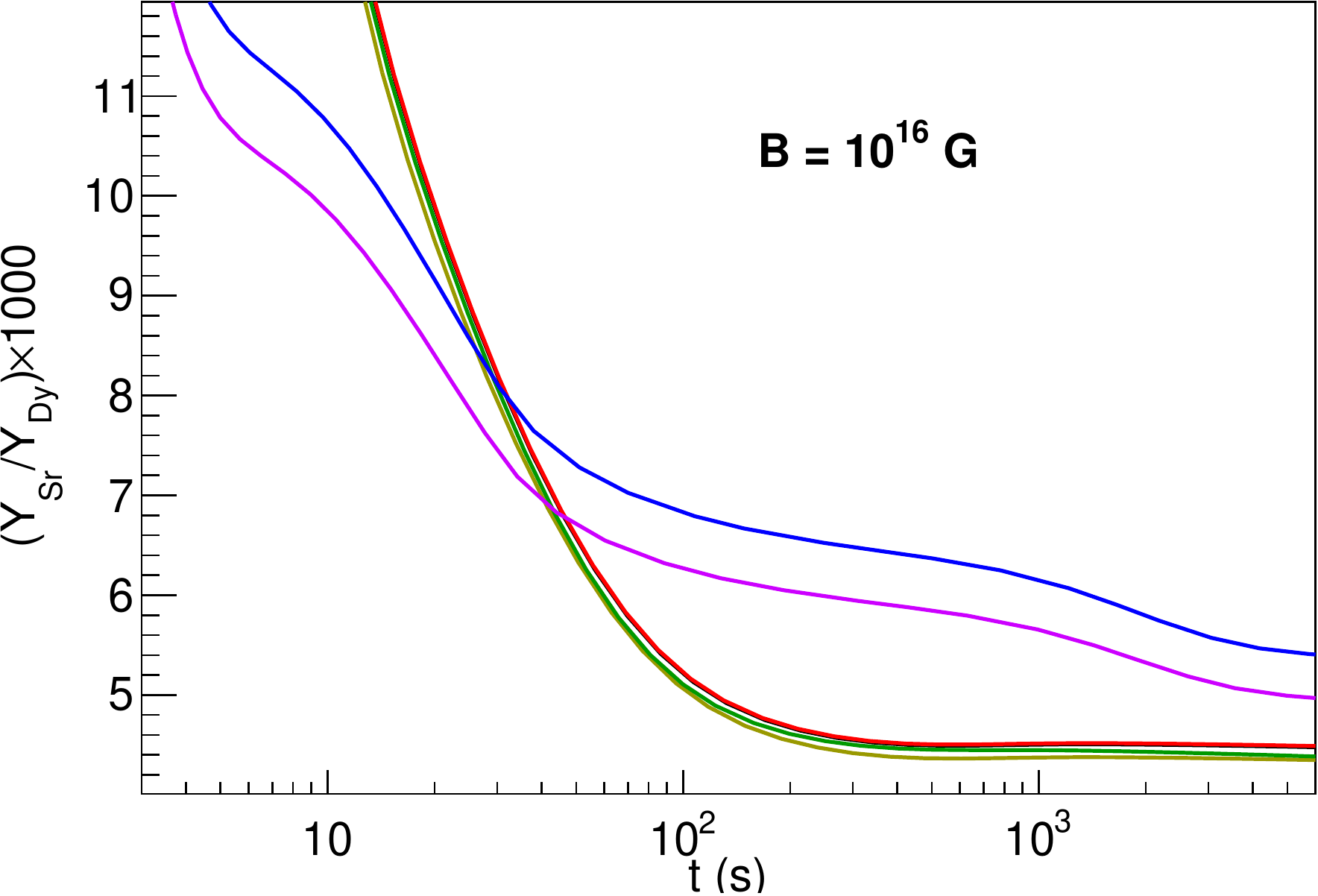}{0.49\textwidth}{(c)}
\fig{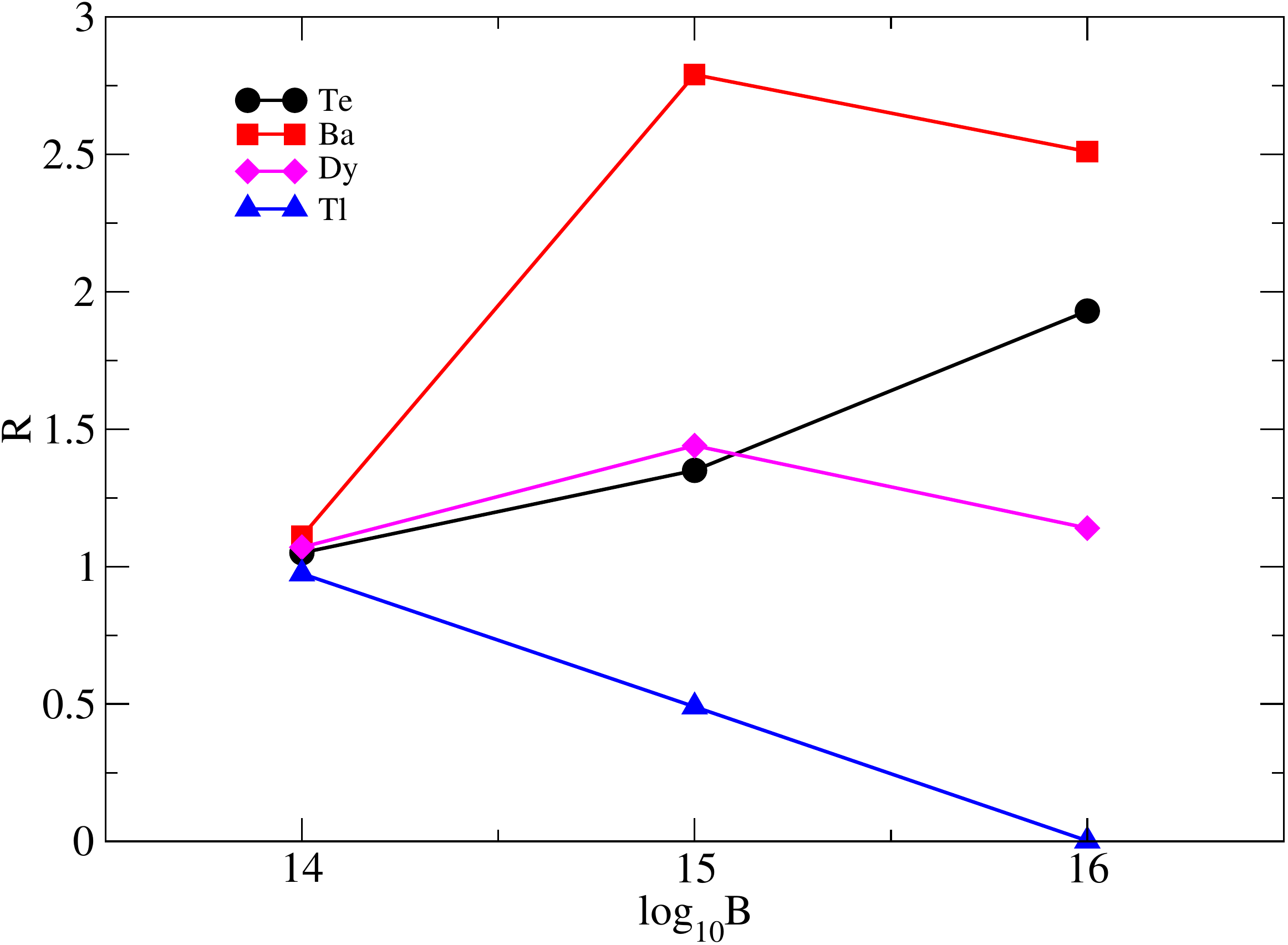}{0.49\textwidth}{(d)}
}
    \caption{Sr/Dy abundance ratios for the collapsar network
    calculation for all models in Table \ref{screen_mag_models} for
    (a) $B = 10^{14}$ G, (b) $B = 10^{15}$ G, and (c) $B = 10^{16}$ G. The colors are the same as those shown in Figure \ref{MHD_abun}.  Panel (d) shows 
    abundance double ratios given by Equation \ref{double_ratio} for four elements as a 
    function of magnetic field.}
    \label{sr_dy_fig}
\end{figure}
\subsection{Abundance Ratios}
The overall final abundance distribution can be characterized by various abundance ratios.  This is particularly helpful in that these
provide a characteristic number to gauge the relative contribution
from fission compared to the abundance buildup of light nuclei. 
This ratio is shown for three fields as a function of time in 
Figure \ref{sr_dy_fig} for all six models studied.  The zero-field cases
are represented by the unscreened and screened relativistic models.  The 
figure shows the abundance ratio for the cases in which
fission cycling is accounted for. 

In all cases, the value of the abundance ratio, $Y_{Sr}/Y_{Dy}$ drops 
rapidly as the r-process path moves to the heavier nuclei and into the
fissile nuclei, after which an equilibrium abundance of
Dy begins to be produced via fission.  The abundance ratio continues
to drop more gradually with time after $\sim$4 s, when the Dy 
continues to build more slowly, and an equilibrium abundance of Sr
is approached.   This evolution continues into the post-processing
of the r process.  It's also noticed the relativistic screening effect -- though small -- is
more prominent than effects from classical screening, resulting in a slight
reduction in the Sr/Dy ratio.  While this reduction is 
small compared to effects from the magnetic field on $\beta$ decays, it can be
seen in the figures.

For the lowest field, the effect of the enhanced rates is small because the field-enhanced rates -- consisting of decays to 
many Landau levels -- are similar to the non-enhanced rates.  If only
the LLL approximation is used (model EF$_{14}$), the evolution is
significantly different as the rates are grossly underestimated, resulting
in a very slow r-process evolution, and the Sr/Dy abundance ratio does
not drop until much later in the evolution.  For the highest field,
on the other hand, there is a smaller difference between the LLL approximation 
(model EF$_{16}$)
and the inclusion of all Landau levels (model FF$_{16}$) in the decay rates because
only a few Landau levels are populated in beta decays at this 
field. 

Shown in Figure \ref{sr_dy_fig}d are various abundance ratios 
Y$_{Sr}$/Y$_{X}$ (where $X$ indicates an arbitrary element)
at t=6000 s as a function of the magnetic field.  
Plotted in the figure is the relative elemental 
abundance double ratio, $R$, defined as:
\begin{equation}
\label{double_ratio}
    R\equiv \frac{\left(Y_{Sr}/Y_X\right)_B}{\left(Y_{Sr}/Y_X\right)_{B=0}}
\end{equation}
which shows the evolution of the elemental abundance ratios as the field
increases.
For low fields, all values are expected to converge at unity as seen in
the figure.  However, as fields increase, different physical processes
affect the ratios.

For the lowest $Z$ element (Te), which can be weakly populated by fission at 
all fields, a more rapid progression to the fission products can result
in a slightly increased production of Te.  
However, production of Sr via  neutron capture is enhanced by the strong magnetic field.  Also, the Sr decay rates are not as enhanced as much as those of Te. Thus
the Sr/Te ratio increases
with field.  For Ba and Dy, however, there is an increase, followed by
a decrease.  This is because the population of Ba and Dy by fission not
only depends on the rate of progression to the fissile nuclei, but also the 
final fission distribution.  As the r-process path progression to fission
for $B=10^{15}$ G is similar to that for $B=10^{14}$ G (as will be described in
the next section), the production of the Ba and Dy progenitors is
faster as the field increases up to $B=10^{15}$ G.  However, above this field,
the $\beta^-$ decay rates are fast enough such that the r-process
path itself -- being dynamic in nature -- shifts sufficiently such that the
distribution of fissile nuclei changes,
and the fission product distribution changes
somewhat.  One might imagine the peaks of the fission distribution in
Figure \ref{fission_dist} shifting to lower mass, thus raising
or lowering abundances of
the progenitors of Ba and Dy.  Clearly, the fission model used
in this work is
too simplistic to make a more than qualitative conclusion, but
the interplay
between the fission product distribution and the magnetic
fields compels further
investigation.

The element Tl is also fascinating. It is seen that the Sr/Tl
ratio decreases
with field.   Tl lies above the fission products in mass and
Z.  However, it also lies
just above the A=195 peak in the r-process distribution.  
Recall from Figure \ref{MHD_abun} that the
r-process
distribution shifts slightly to higher mass as the field
increases, shifting
the $A=195$ abundance peak as well.  This shift, in turn
increases the Tl 
abundance dramatically, thus reducing the Sr/Tl abundance
ratio. 

This effect of the magnetic field on the shape of the final r-process 
abundance distribution, and hence, the Sr/X abundance ratio is
compelling as an r-process from a single collapsar site can be characterized by
the abundance distribution, and the magnetic field may be constrained by
the abundance ratios.  Obviously, a more thorough evaluation incorporating a more realistic fission model is 
necessary \citep{beun08,mumpower18,suzuki18,vassh2019}, 
but the 
effect on the shape of the abundance distribution can still be made.

While not explicitly evaluated here, it is noted that if 
magnetic fields as high as $10^{15}$--$10^{16}$ G exist in 
r-process sites, neutron capture rates as well as charged-particle
reactions may be affected significantly via changes in nuclear 
distribution functions. The field effect on reaction rates has 
been studied for one important reaction in big bang 
nucleosynthesis, $^7$Be($n$,$p$)$^7$Li \citep{kawasaki12}. That 
reaction rate is affected only in cases of large magnetic fields
which can be excluded from observations of primordial abundances. 
However, the field effect through modified distribution functions 
can potentially change the neutron capture reaction rates with
non-flat $(\sigma v)(E)$ curves at low energies under strong
magnetic fields.

The ratios studied here may be of particular interest to 
astronomers in evaluating elemental abundance ratios
in stars enriched in single sites. These ratios
are generally low compared to solar r-process abundance values \citep{arlandini99} owing to the fact that the single neutron-rich trajectory
presented here results in a large abundance of massive elements.  The range of
observed values from the SAGA database \citep{suda08} are also large compared to the values here. This may likely result from a both a detection limit as well
as from the fact that if collapsar jets
contribute to the galactic r-process abundance distribution, they contribute in combination with
other sites.
\subsection{Fission Cycling Time}
The fission cycling time has also been explored, and the effects
of a strong magnetic field on the overall fission cycling has been
explored from the standpoint of the total time to progress from light nuclei to the fissile nuclei.    Naively, one would expect that the
fission cycling time would decrease with magnetic field as the nucleosynthetic 
progression speeds up.

Here, the fission cycling time is thought of as the time
to progress from the low-mass nuclei in an r-process path to the fissile
nuclei.  The low-mass nuclei are defined to be those in the Zr isotopic
chain ($Z=40$), and the high-mass nuclei are defined to be those in the fissile region ($Z=98$).  While this is a somewhat arbitrary choice, and
while fission cycling is more complex than this, such a method 
provides a figure-of-merit for the speed at which nuclei
can cycle through the r-process to the fissile nuclei.

Using the $\beta^-$ decay lifetimes, $\tau_{\beta,i}$, for nuclei along the r-process 
path, the fission cycling time, $\tau_f$, is then defined as:
\begin{equation}
\label{sum_fission}
    \tau_{f} \equiv \sum\limits_{z=40}^{98}\tau_{\beta,z}
\end{equation}
where the sum is over the most abundant isotope of each element between Zr ($Z=40$) and Cf ($Z=98$) along the r-process path
at a specific point in time.  It then remains to choose a point in time
at which the r-process path is chosen.  Two methods are utilized
to characterize the r-process path.

With the first method, the r-process path is chosen at the point in the evolution when the region containing fissile nuclei is first reached in the
r process.  Here, the r-process path is chosen at the epoch when the
first Cf nuclei are produced.  This is defined to be the point in time when
$Y_{Cf}\ge 10^{-20}$.  Because this point depends on 
the magnetic field, the r-process path at this epoch is 
unique for each magnetic field. In addition, the temperature and density of the environment are also different at this point, and thus the electron chemical potentials
vary in each case.  Here, the total fission time is defined by the
term $\tau(B)$, and the path so-chosen is referred to as the
``dynamic'' r-process path.

A second method is adopted for comparison.  With this method the r-process path,
temperature, and density is chosen to be fixed and independent of the
external field; the chosen isotopes are
the same for each choice of field. The $\beta$-decay lifetimes are then computed
for this path as a function of the magnetic field.  In this case, the r-process
path is chosen to be defined by the isotope with maximum abundance for
each element at the point in time when $Y_{Cf}\ge 10^{-20}$ for a specific
field of $10^{14}$ G.  At this point in the r-process
evolution, the temperature and density are $T_9=1.76$ and 
$\rho=377.9$ g cm$^{-3}$, respectively.  Here, the total lifetime is defined by
the term $\tau_s$, and the path is referred to as the ``static'' path.

\begin{figure}
\gridline{
\fig{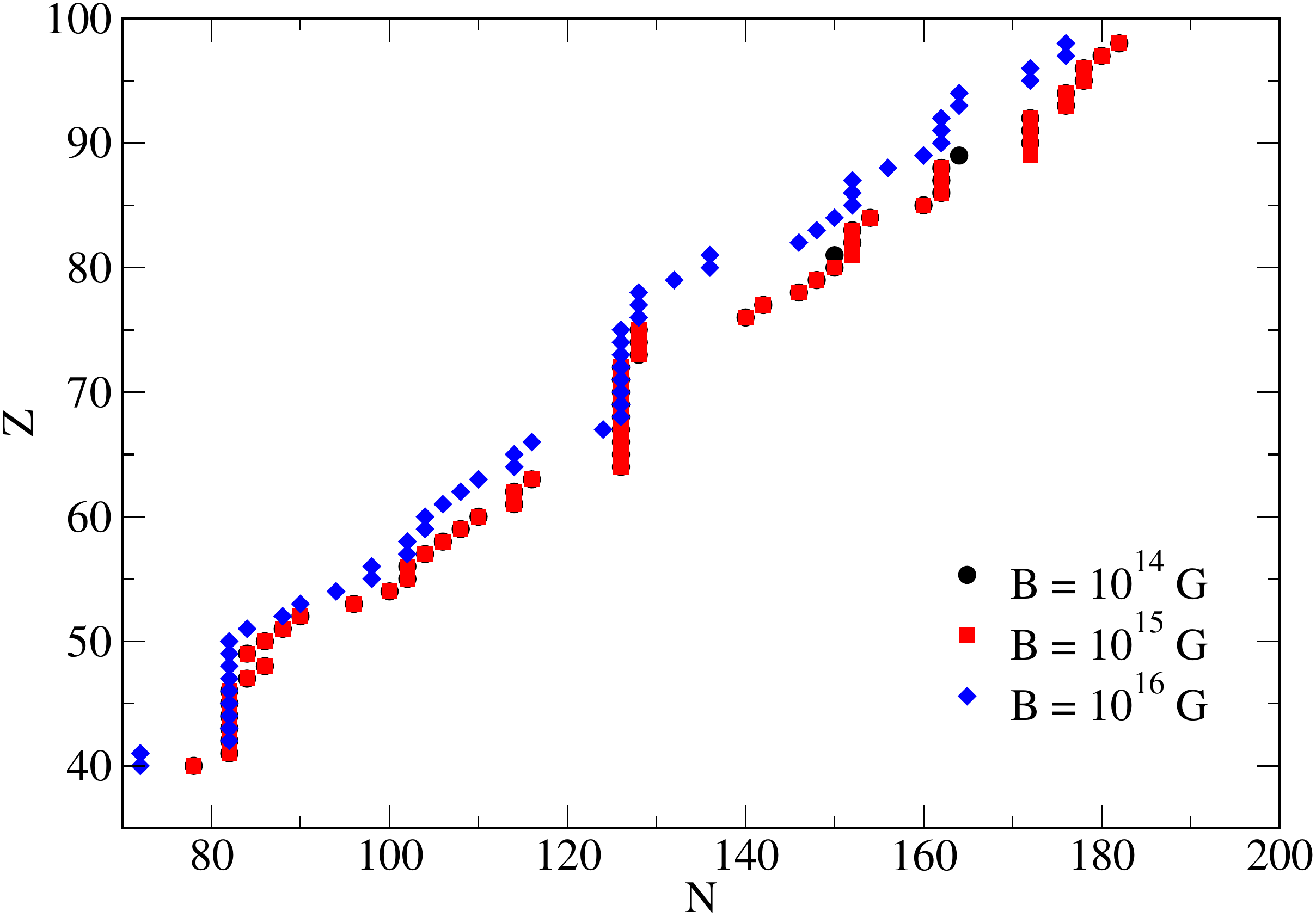}{0.49\textwidth}{(a)}
\fig{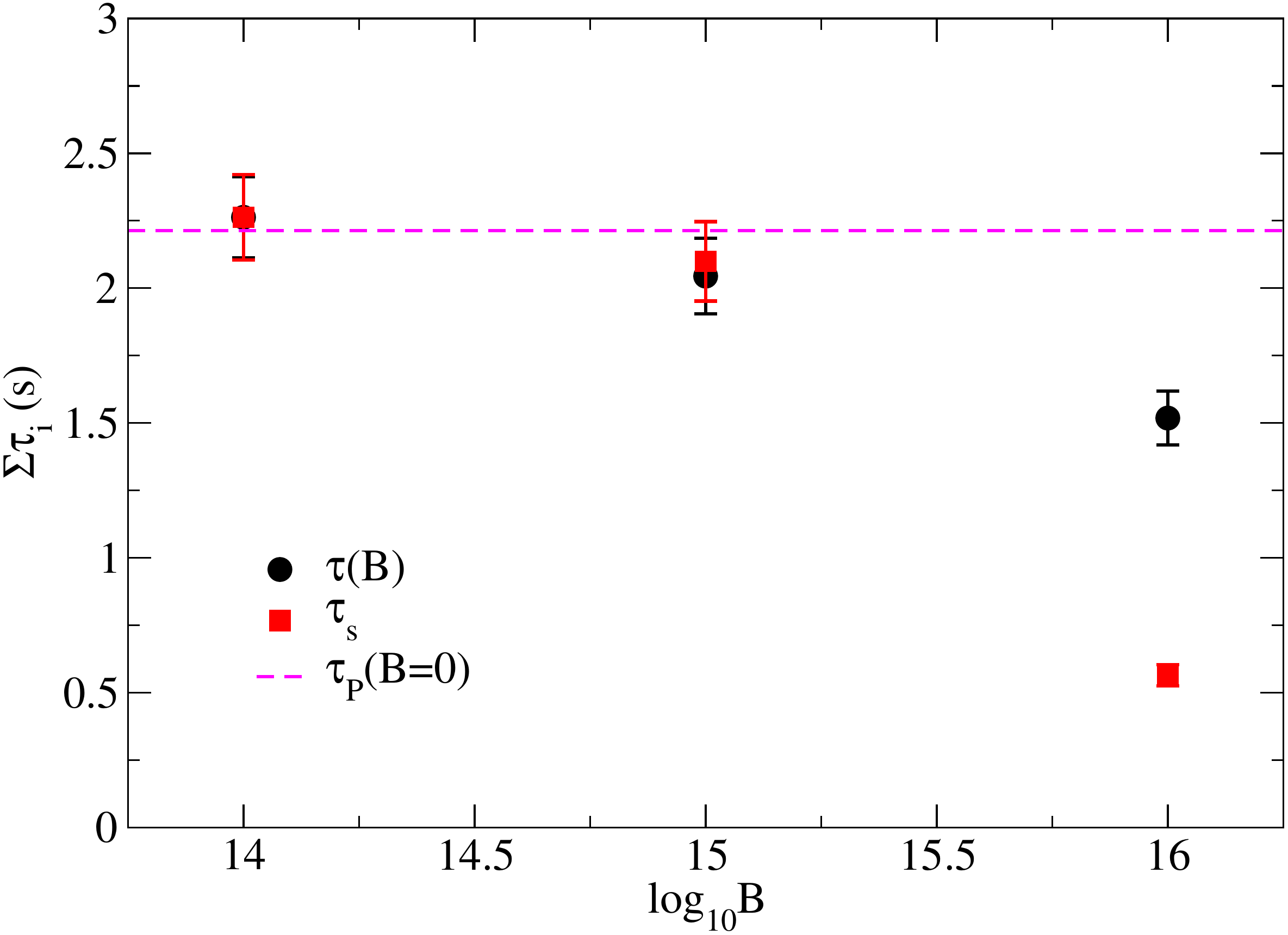}{0.49\textwidth}{(b)}
}
    \caption{(a) Dynamic r-process paths used to determine fission cycling times
    as described in the text.  The dynamic path at 10$^{14}$ G is also the
    static path used in this evaluation.(b) Fission cycling times as a function
    of magnetic
    field (in units of G) for the dynamic path, static path, and the zero-field case.}
    \label{fission_fig}
\end{figure}

In order to compute $\tau(B)$, the r-process path must then
be defined for each field, including a field of $10^{14}$ G, which is also used
to define the r-process path used to compute $\tau_s$.  The paths defined this
way are shown in Figure \ref{fission_fig}. For fields of
10$^{14}$ G and 10$^{15}$ G, the paths are very similar.  The dynamic path
corresponding to a field of 10$^{16}$ G is significantly closer to the 
valley of stability because of the significantly faster $\beta$ decay
rates.

The computed fission-cycling times are also shown in Figure
\ref{fission_fig}b for both definitions of r-process path and for the 
zero-field case.  In either case of the path definition, the fission cycling
time decreases with field.  The fission cycling time for a field of 
10$^{14}$ G roughly corresponds to a fission cycling time with
zero field.  (There is a small difference due to an imposed decay rate calculation
accuracy of 1\%, which accounts for the difference in both calculations.)
For the static path, the difference is more pronounced at higher fields
because the static path is defined to be farther from stability than the dynamic
path for the
10$^{16}$ G field.  The difference between the static path and the dynamic path
case at a field of 10$^{15}$ G is small as the dynamic path at
10$^{15}$ G is very similar to the static path.

It is apparent from this result that fission cycling is faster at higher fields
and thus becomes more prominent.  The production of fissile nuclei increases
during the r-processing time.  This results in more fission products, but also
an additional neutron abundance in the r-process environment.  The initial very
low $Y_e$ in the collapsar model is particularly conducive to 
producing a significant abundance of fissile nuclei.
\section{Discussion}
Plasma effects on nuclear fusion and weak interactions in hot, highly-magnetized
plasmas were 
evaluated, and the example of r-processing in a collapsar MHD jet site were
examined.  Two primary effects
were analyzed. The first is the effect of Coulomb screening on fusion reactions
of charged particles.
Because the r-process is dominated by neutron captures, screening has a small
effect on the overall
evolution and final abundance distribution of the r process.  However,
charged-particle reactions 
in the early stages (e.g., ($\alpha$,n) and ($\alpha$,$\gamma$) reactions) may be affected.  Coulomb
screening 
is affected by both the temperature and the magnetic field of the environment.
While the
default classical weak screening commonly used in astrophysics codes was found
to have virtually
no effect on the final r-process abundance distribution, relativistic effects
from 
high temperatures and high magnetic fields were found to have a slight effect on
the r-process evolution.

The second effect studied is the effect of high magnetic fields on nuclear weak
interaction
rates.   As fields increase in strength, electron momentum transverse to the
field direction is
quantized into Landau levels.  This alters the Fermi-Dirac distribution,
resulting in
a shift in the electron spectrum.  While the magnetic field was found to have a
small effect
on Coulomb screening, strong fields may have a larger effect on nucleosynthesis
when
applied to weak interaction rates.  This is because -- particularly in the case
of 
the finite $\beta$-decay spectrum -- only a limited number of Landau levels can
be occupied by the
emitted charged lepton, as indicated in Figures \ref{beta_spec_evolution} and
\ref{beta_spec}.  
For very high fields, $\sqrt{eB}\sim Q$, only a couple of Landau levels are
available to the
emitted electron or positron.  The electron energy spectra have strong peaks
where the electron
longitudinal momentum is zero.  The integrated spectrum, which is proportional
to the 
decay rate, is thus much larger than that for the zero-field case.  Large fields
can affect the r-process evolution.

A simple MHD collapsar jet model was adapted from the
hydrodynamics calculations of \cite{nakamura12} as an
illustrative model.  In
this model, static fields of various strengths were assumed. 
Various effects of thermal
and field effects were studied individually in a systematic
manner to gauge the effects
of individual environmental parameters.  While the temperature
was treated dynamically
following a single trajectory, which was assumed to decay
adiabatically after 2.8 s, the
magnetic field in this case was assumed constant.  

One interesting result of magnetic-field effects on the r process
studied was that $\beta^-$ decay rates increase with field strength.  Because
of this, the r-process path, which
changes dynamically in time, may shift somewhat closer to stability for very
strong
fields.  This has multiple effects.  First, the point at which the r-process
path crosses
the magic numbers change, thus shifting the abundance peaks of the final
distribution.  This shape can be evaluated using elemental abundance ratios,
such as Y$_{Sr}$/Y$_{Tl}$.  Second, the
fissile nuclei produced in the r process will be different, resulting in
potentially different
fission rates and distributions.  This could possibly be studied using abundance
ratios
such as Y$_{Sr}$/Y$_{Ba}$, Y$_{Sr}$/Y$_{Dy}$, or something similar.  Finally,
the fission cycling
time decreases somewhat with increasing field, resulting in an increase in
fission products as well as a slight addition of neutrons to the r-process
environment.

While the results presented require more precise evaluations, it
is interesting to note that -- in a highly-magnetized r process site -- the elemental abundance ratios can constrain the magnetic field of the
site and vice-versa.  This might be of interest to astronomers
in evaluating stellar abundance ratios in objects thought to contain
single-site abundances.  The characteristic abundance ratios 
with an MHD/collapsar model -- even at zero field -- may 
characterize the contribution to r-process elements in a star.  While 
the fields presented in this paper are quite large -- commensurate
with a collapsar, MHD, or possibly NS merger -- if such fields
can be sustained in an r-process site, they would be manifest 
in the isotopic ratios of the site.

Further, it is noted that fission in the collapsar model and effects
from the magnetic field may change the contribution to
currently observed elements in Galactic chemical evolution (GCE) 
models.  This will be studied in a subsequent paper.

The limitations of the model presented here are noted.  These include
primarily the static field assumption and the simplified 
fission model used.  If the static field is assumed to be the maximum
field in the site, then the results could be thought of as upper limits.  Also, the simplified fission model was used as the
primary evaluation of this paper was on the effects of strong
magnetic fields in nucleosynthesis sites.  The progenitor
nuclei examined in the r-process site in this paper -- being
quite far from stability -- were treated in this much simpler
matter.  Future work will concentrate on a more thorough treatment
of fission in the collapsar/MHD site and its effects on GCE.  In 
addition, a dynamic treatment of the magnetic field will be
examined.
\acknowledgments
M.A.F. is supported by National Science Foundation Grant No. PHY-1712832 and by NASA Grant No. 80NSSC20K0498. A.B.B. is supported in part by the U.S. National Science Foundation Grant No. PHY-1806368. 
T.K. is supported in part by Grants-in-Aid for Scientific Research of JSPS (17K05459, 20K03958). K.M. is supported by JSPS 
KAKENHI Grant Number JP19J12892.  M.K. is supported by NSFC Research Fund for International Young Scientists (11850410441). Y.L. is supported by JSPS KAKENHI Grant Number 19J22167.   
M.A.F. and A.B.B. acknowledge support from the NAOJ Visiting Professor program.

\bibliographystyle{aasjournal}
\bibliography{screening_paper}

\end{document}